\documentclass[useAMS,usenatbib]{mnras}
\usepackage{graphicx,amsmath,url,natbib,amssymb}

\usepackage{multirow}
\usepackage[flushleft]{threeparttable}
\usepackage{xcolor,ulem}
\usepackage{calligra}

\newcommand{\ergs}{\mbox{ erg s}^{-1}}
\newcommand{\machj}{\mathcal{M}_j}   
\newcommand{\cc}{\mbox{ cm}^{-3}}
\newcommand{\kms}{\mbox{ km s}^{-1}}
\newcommand{\kpc}{\mbox{ kpc}}
\newcommand{\pc}{\mbox{ pc}}
\newcommand{\Msun}{M_\odot}

\newcommand{\rbar}{\tilde{r}}
\newcommand{\rcbar}{\tilde{r_c}}
\newcommand{\lbar}{\tilde{l}}
\newcommand{\Lbar}{\tilde{L_0}}
\newcommand{\tbar}{\tilde{t}}

\begin{document}

\title[Dynamics of relativistic MHD jets]{Simulating the dynamics and non-thermal emission of relativistic magnetised jets I. Dynamics}
\author[Mukherjee et al.]{Dipanjan Mukherjee$^1$$^2$$^3$\thanks{dipanjan@iucaa.in},
  Gianluigi Bodo$^3$,  Andrea Mignone$^2$, Paola Rossi$^3$  
  \newauthor
  \& Bhargav Vaidya $^4$ \\
$^{1}$ Inter-University Centre for Astronomy and Astrophysics, Post Bag 4, Pune - 411007 \\
$^{2}$ Dipartimento di Fisica Generale, Universita degli Studi di Torino , Via Pietro Giuria 1, 10125 Torino, Italy\\
$^{3}$ INAF/Osservatorio Astrofisico di Torino, Strada Osservatorio 20, I-10025 Pino Torinese, Italy \\
$^{4}$ Discipline of Astronomy, Astrophysics and Space Engineering, Indian Institute of Technology Indore, \\ 
       Khandwa Road, Simrol, 453552, India
}
\date{\today}
\pagerange{\pageref{firstpage}--\pageref{lastpage}} 
\pubyear{2019}
\maketitle

\begin{abstract}
We have performed magneto-hydrodynamic simulations of relativistic jets from supermassive blackholes over a few tens of kpc for a range of jet parameters. One of the primary aims were to investigate the effect of different MHD instabilities on the jet dynamics and their dependence on the choice of jet parameters. We find that two dominant MHD instabilities affect the dynamics of the jet, small scale Kelvin- Helmholtz (KH) modes and large scale kink modes, whose evolution depend on internal jet parameters like the Lorentz factor, the ratio of the density and pressure to the external medium and the magnetisation and hence consequently on the jet power. Low power jets are susceptible to both instabilities, kink modes for jets with higher central magnetic field and KH modes for lower magnetisation. Moderate power jets do not show appreciable growth of kink modes, but KH modes develop for lower magnetisation. Higher power jets are generally stable to both instabilities. Such instabilities decelerate and decollimate the jet while inducing turbulence in the cocoon, with consequences on the magnetic field structure. We model the dynamics of the jets following a generalised treatment of the Begelman-Cioffi relations which we present here. We find that the dynamics of stable jets match well with simplified analytic models of expansion of non self-similar FRII jets, whereas jets with prominent MHD instabilities show a nearly self-similar evolution of the morphology as the energy is more evenly distributed between the jet head and the cocoon. 
\end{abstract}


\section{Introduction}\label{sec:intro}

Relativistic jets are one of the major drivers of galaxy evolution \citep{fabian12a}. Jets deposit energy over a large range of spatial scales, from the galactic core of a few kpc \citep{wagner11a,mukherjee16a,mukherjee17a,morganti13a,morganti20a} to the circum-galactic media, some extending to Mpc in length \cite{dabhade17a,dabhade19a,dabhade20a}. Understanding the evolution and dynamics of such jets is thus crucial in unraveling how galaxies evolve over cosmic time.

Since the discovery of radio emission from jet driven lobes \citep{jennison53a}, there has been significant observational and theoretical investigations to understand the nature of these extragalactic objects \citep[see e.g. ][ for reviews]{begelman84a,worrall09a,blandford19a}. While it is now  common understanding that  non-thermal processes such as synchrotron and inverse-Compton contribute to the multi-wavelength emission from the jets \citep{worrall09a,worrall06a}, there still remain several open questions on how the evolution and dynamics of the jet affect the above emission processes.

Several early works have attempted to describe the jet dynamics and subsequently explain the observed emission through semi-analytic modeling of the jet expansion such as \citet{begelman89a}, \citet{falle91a}, \citet{kaiser97a}, \citet{komissarov98a}, \citet{bromberg09a}, \citet{bromberg11a}, \citet{turner15a}, \citet{harrison18a} and \citet{hardcastle18a} to name a few. With the development of numerical schemes to simulate relativistic flows, several papers have investigated the dynamics of relativistic jets as they expand into the ambient medium \citep{marti97a,komissarov98a,komissarov99a,sheck02a,perucho07a,rossi08a,mignone10a,perucho14a,rossi17a,perucho19a}. In the present paper and other subsequent follow up publications in future, we intend to give a broad interpretation of the dynamics and emission properties of relativistic, magnetised jets, considering in detail the effects of instabilities and the role played by the magnetic field on jet propagation (paper I). This first paper, which focuses on the dynamics, provides a basis and a reference for interpreting the radiative properties, that will be investigated in the following papers.

MHD instabilities can play a significant role in determining the dynamics and evolution of the jet. The two major instabilities that can affect the jet are the current driven modes \citep{nakamura07a,mignone10a,mignone13a,mizuno14a,bromberg16a} and Kelvin-Helmholtz modes \citep{bodo89a,birkinshaw91b,bodo96a,perucho04a,perucho10a,bodo13a,bodo19a}. The growth of such instabilities and their efficiency in disrupting the jet column depends on several factors intrinsic to the properties of the jet such as its velocity, magnetisation and opening angle as well as the density profile of the external medium \citep{porth15a,tchekhovskoy16a}. The pressure in the cocoon surrounding the jet can also initiate the onset of instabilities due to higher sound speeds that facilitate the growth of perturbations \citep{hardee98a,rosen99a}.

Jets with higher velocities, stronger magnetisation and colder plasma have slower growth of Kelvin-Helmholtz modes \citep{rosen99a,perucho04a,bodo13a}. Strongly magnetised, collimated jets are however susceptible to the current driven modes \citep{bodo13a,bromberg16a,tchekhovskoy16a}. Thus the relative efficiency of the different modes depend on internal parameters of the jet. Many of the above works, especially those involving semi-analytic linear analysis \citep{bodo89a,perucho04a,bodo13a,bodo19a} rely on idealistic approximations to keep the problem tractable. In a realistic scenario of a jet traversing through an ambient density whose radial profile is defined by the gravitational potential of the host galaxy, several of the above modes can occur simultaneously. 

Simulations of relativistic jets expanding into an ambient medium have been carried out in several earlier papers \citep[such as ][]{marti97a,komissarov98a,sheck02a,perucho07a,rossi08a,mignone10a,perucho14a,english16a,perucho19a}. However, very few of the above explore in a systematic way the impact of different jet parameters on the development of various MHD instabilities and their effect on the jet dynamics. In the present paper we perform a suite of relativistic magneto-hydrodynamic simulations to explore the dynamics and evolution of the jet and its cocoon over a few tens of kpc for a varying range of initial jet parameters such as the jet's power, velocity, magnetisation and contrast of the pressure (or temperature) and  density with the ambient medium.

We investigate how the jet parameters impact the growth of different instabilities and their effect on the dynamics and morphology of the jet by comparing with an analytic extension of the jet evolution model proposed in \citet{begelman89a}. We also present the distribution and evolution of the magnetic field in the cocoon and its dependence on the onset of different MHD instabilities, which is important in predicting synchrotron emission from the jet lobes \citep{hardcastle13a,hardcastle14a,english16a}.  Some of the simulations have been performed with the new \textsc{lagrangian particle} module in the \textsc{PLUTO} code, as described in \citet{vaidya18a} that computes the spectral and spatial evolution of relativistic electrons in the jet. This enables one to make accurate predictions of synchrotron emission expected from such systems. In this paper we restrict ourselves to the discussions of dynamics of the jet the evolution based on the fluid parameters alone. In subsequent publications of this series, we will discuss the nature of the observable emission and its connection to the jet dynamics and MHD instabilities.

We structure the paper as described below. In Sec.~\ref{sec.setup} we describe the initialisation of the simulation parameters and the details of the numerical implementation. In Sec.~\ref{sec.results} we describe the results of the simulations and the impact of different parameters on the jet dynamics. In the sub-sections therein we describe the onset of different MHD instabilities for different jet parameters and the relative comparison of the different simulations with an analytic model of jet evolution. Finally in Sec.~\ref{sec.discuss} and Sec.~\ref{sec.conclusion} we discuss the implications of the results and summarise our findings.

\section{Simulation setup}\label{sec.setup}
\subsection{The problem}
We investigate the propagation of relativistic magnetised jets in a stratified ambient medium.  The relevant equations to be solved are the relativistic magnetohydrodynamic (RMHD) equations in a constant Minkowski metric for special relativistic flows \citep[see e.g.][]{mignone07,  rossi17a}. We assume a single-species relativistic perfect fluid (the Synge gas) described by the approximated Taub-Matthews equation of state \citep{mignone05b, mignone07a}. The ambient medium, better described in subsection \ref{sec:atmosphere}, is maintained in hydrostatic equilibrium by an external gravitational potential. No magnetic field is present in the initial configuration at $t=0$ and a toroidal magnetic field is injected along with the jets. The equations are solved in a 3D Cartesian geometry with the $z$ axis pointing along the jet direction. 

\subsection{Ambient atmosphere}
\label{sec:atmosphere}
\begin{figure}
	\centering
	\includegraphics[width = 7 cm, keepaspectratio] 
	{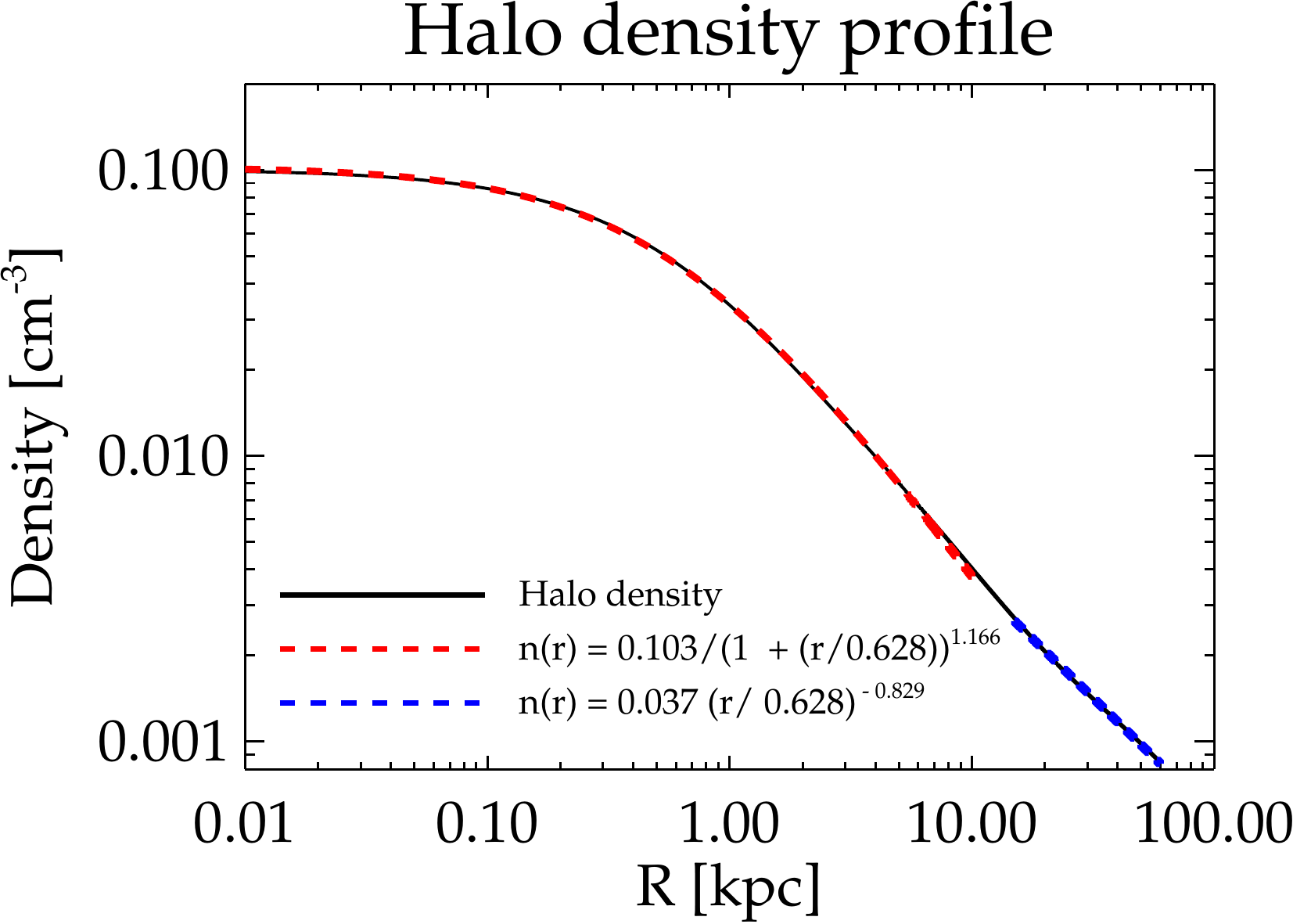}
	\includegraphics[width = 7 cm, keepaspectratio] 
	{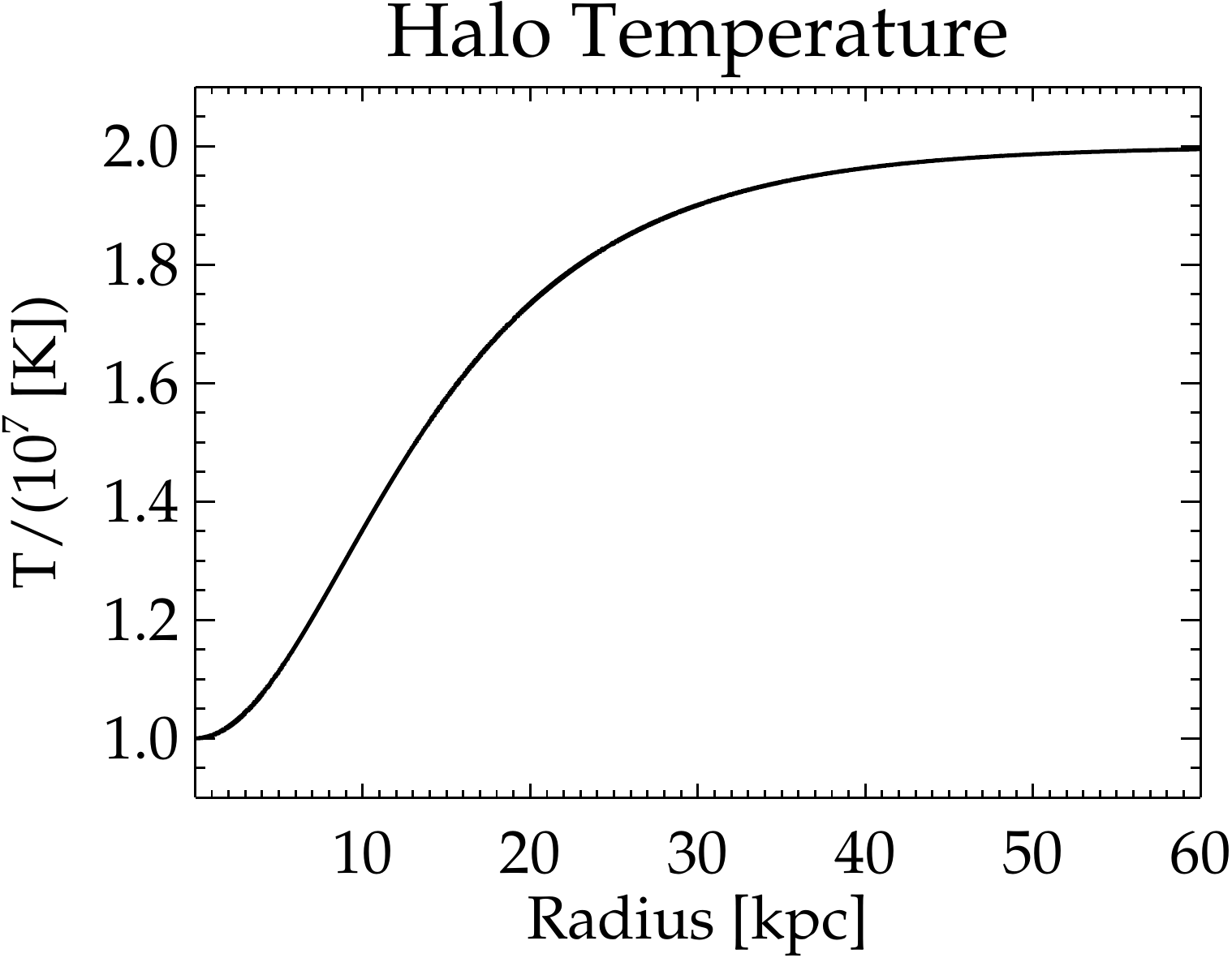}
	\caption{\small \textbf{Top}: Density profile of the ambient halo as a function of radius set to be in equilibrium with the external gravitational field. Fits to the density profile using simple analytical expressions (eq.~\ref{eq.frho1} and eq.~\ref{eq.frho3}) have been presented for two different regimes, $1-10$ kpc in red and $15-60$ kpc in blue. \textbf{Bottom:} The temperature profile assumed for the halo gas, using eq.~\ref{eq.temp}.}
	\label{fig.halo}
\end{figure}
We assume an external static gravitational field to keep the ambient halo gas in pressure equilibrium. We take a Hernquist potential \citep{hernquist90a} to represent the contribution of the stellar (baryonic) component of the galaxy:
\begin{equation}
\phi_{B} = -\frac{G M_B}{r+a_H}
\end{equation}
Here $G$ is the gravitational constant, $M_B = 2\times10^{11} \Msun$ is the stellar mass of the galaxy, typical of large ellipticals which host powerful radio jets \citep{best05a,sabater19a} and $a_H = 2$ kpc is the scale radius, which corresponds to a half-mass radius $r_{1/2} = \left(1 + \sqrt2\right) a_H = 4.8$ kpc and the half-light radius of $R_e = 1.8153a_H \simeq 3.63$ kpc \citep{hernquist90a}, typical of giant ellipticals \citep{kormendy09a}. The contribution of the dark matter component to the gravitational potential is modelled by a NFW profile \citep{navarro96a}:
\begin{align}
&\phi_{\rm DM} = \frac{-G M_{200}}{\left\lbrack \ln(1+\tilde{c}) + \tilde{c}/(1+\tilde{c}) \right\rbrack} \left(\frac{1}{r+d}\right) \ln\left(1+\frac{r}{r_s}\right) \\
&\mbox{ where } M_{200} = 200 \rho_{\rm cr} \frac{4 \pi}{3} \tilde{c}^3 r_s^3 \,\,\, ; \,\,\, r_s = r_{200}/\tilde{c} \nonumber
\end{align}
Here $r_{200}$ is the radius where the mean density of the dark-matter halo is 200 times the critical density of the universe, $\tilde{c}$ is the concentration parameter and  $\rho_{cr} = 3H^2/(8 \pi G) = 8.506 10^{-30} \mbox{g cm}^{-3}$ is the critical density of the universe at $z=0$ with the Hubble constant $H = 67 \mbox{km s}^{-1}\mbox{Mpc}^{-1}$ \citep{planck16a}. The NFW profile is modified with an arbitrarily chosen small core radius of $d=10^{-3}$ kpc to avoid the singularity at $r=0$. 
\\
For our simulations we assumed $\tilde{c}=10$, $d=10^{-3}$ kpc and $r_{200} = 1$ Mpc which gives a virial mass of $M_{200} = 1\times10^{14} \Msun \left(r_{200}/1 \mbox{Mpc}\right)^3$. The above are comparable to values inferred from observations of galaxy clusters \citep{croston08a}. Thus the galaxy parameters used represent a typical giant elliptical at the centre of a cluster.

The ambient atmosphere in several early type galaxies \citep{paggi17a} and centres of clusters \citep{leccardi08a} are usually found to have radially increasing gas temperatures. For our simulations we model the ambient halo to have a radially varying temperature profile as (as shown in Fig.~\ref{fig.halo}): 
\begin{equation}
T_a(r) = T_c + \left\lbrack1 - \frac{1}{\cosh(r/r_c)}\right\rbrack\left(T_H - T_c\right). \label{eq.temp}
\end{equation}
Here $T_c = 10^7$ K is the temperature at $r=0$ and $T_H$ is the temperature at radii beyond the scale radius $r_c$. For our simulations we assume $T_H = 2T_c$ and $r_c=10$ kpc. The density and pressure are then evaluated by considering the atmosphere to be in hydrostatic equilibrium with the external gravitational force, by solving:
\begin{align}
\frac{dp_a(r)}{dr} &= -\rho_a(r) \frac{d\phi(r)}{dr} \,\,\, ; \,\,\, p_a(r) = \frac{\rho_h(r)}{\mu m_a} k_B T_h(r) \nonumber \\
p_a(r) &= \left(n_0 k_B T_c\right) \exp \left\lbrack -\int^r_0 \left(\frac{\mu m_a}{k_B T_a(r)}\right) \frac{d\phi(r)}{dr} dr \right\rbrack \label{eq.halopres}
\end{align}
where $p_a$ and $\rho_a = \mu m_a n_h$ are the pressure and density of the ambient halo gas, $\phi = \phi_B + \phi_{DM}$ is the total gravitational potential, $\mu=0.6$ is the mean molecular weight for a fully ionised gas \citep{sutherland17a} with $m_a$ being the atomic weight, $n_0$ is the number density at $r=0$ and the temperature, $T_a(r)$, is given by eq.~\ref{eq.temp}. Equation~\ref{eq.halopres} is solved numerically to obtain a tabulated list of density and pressure as a function of radius, which is then interpolated on to the \textsc{pluto} domain at the initialisation step.

\subsection{Jet parameters}\label{sec.jetsetup}
The jet properties are defined by four non dimensional parameters: 
\begin{itemize}
\item \emph{The density contrast:} It is defined as
\begin{equation}
\eta_j = \frac{n_j(r_{\rm inj})}{n_h(r_{\rm inj})}
\end{equation} which gives the ratio of the number density of the jet plasma ($n_j$) to the number density of the ambient halo ($n_h$) at the radius of injection ($r_{\rm inj}$). The typical choices in the simulations range from $\sim 4\times10^{-5} - 10^{-4}$, similar to previous works \citep{sheck02a,rossi08a,perucho14a,wykes19a}.

\item \emph{The pressure contrast:} It is defined as 
\begin{equation}
\zeta_p = \frac{p_j(r_{\rm inj})}{p_h(r_{\rm inj})}
\end{equation} which sets the ratio of the pressure of the jet ($p_j$)  with respect to the pressure of the ambient halo at the injection radius. For all of our simulations we assume the jet to be in pressure equilibrium with the atmosphere at $t=0$, except for simulation G (see Table~\ref{tab.sims}) where the jet is over-pressured at launch with $\zeta_p = 5$. In the Appendix~\ref{sec.jetchi} we show that the values of pressure and density of the jet used in our simulations are consistent with that of an proton-electron jet.

\item \emph{Jet Lorentz factor:} The bulk Lorentz factor of the jet ($\gamma_b$), from which the magnitude of the jet speed is computed. In our simulations we choose a range of Lorentz factors ($3-10$) which are typical values inferred from doppler boosted luminosity estimates of Blazars \citep{cohen07a,lister09a} or VLBA studies \citep{jorstad05a}. The jet is primarily directed along the $z$ axis. The different components of the velocity vectors are then calculated by assuming the jet to be launched with an opening half-angle of $5^\circ$, as in \citet{mukherjee18b}. 

\item \emph{Jet radius:} We consider a jet radius of $R_j = 100$ pc for all simulations except for G, H and J, where the radius was increased to $R_j=200$ pc to obtain a higher jet power. For our simulations with a resolution of $15.6$ pc, this choice of jet radius ensures that the radius of the jet inlet is resolved by at least 6 computational cells and 12 cells for simulations G, H and J. The above values of jet radii are higher than those obtained from observations at heights similar to our injection zone. However, our choice was restricted due to limitations of computational resources and the need to sufficiently resolve the jet diameter to prevent spurious numerical artefacts and suppressed growth of instabilities and entrainment \citep[e.g.][]{rossi08a,english16a,english19a}.

\item \emph{Jet magnetisation:} The jet magnetisation parameter is defined as the ratio of the Poynting flux ($S_j$) to the jet enthalpy flux ($F_j$):
\begin{equation}
\sigma_B = \frac{|\mathbf{S}_j\cdot\hat{z}|}{|\mathbf{F}_j\cdot\hat{z}|} = \frac{|\left(\mathbf{B}_j\times(\mathbf{v}_j\times\mathbf{B}_j)\right)\cdot\hat{z}|}{4\pi  \left(\gamma^2 \rho_j h_j - \gamma \rho_j c^2\right)\left(|\mathbf{v}_j\cdot\hat{z}|\right)} \label{eq.sigma}
\end{equation}
where $\mathbf{B}_j$ is the magnetic field vector of the jet, $\mathbf{v}_j$ is the jet velocity, and $\rho_j h_j$ is the relativistic enthalpy density of the jet per unit volume. The contribution of the rest mass energy to the enthalpy flux is removed while computing the jet enthalpy flux $F_j$. The above is a more general definition of the magnetisation parameter. For a highly relativistic plasma where the enthalpy dominates over rest mass energy, eq.~\ref{eq.sigma} reduces to $\sigma_B = B^2/\left(4 \pi \gamma^2 \rho h\right)$, similar to the expressions used in earlier papers \citep[e.g.][]{rossi08a,nalewajko16a}. 

The fluxes are considered along the jet $z$ axis, i.e. the direction of launch of the jets. The relativistic enthalpy is computed for a Taub-Matthews equation of state \citep{mignone05b} as:
\begin{equation}
\rho_j h_j = \frac{5}{2} p_j + \sqrt{\frac{9}{4} p_j^2 + \left(\rho_j c^2\right)^2}\label{eq.enth} .
\end{equation}
Eq.~\ref{eq.sigma} can be used to derive the strength of the magnetic field of the jet. For a toroidal magnetic field in a jet directed along the $z$ axis, we derive the peak field strength as $B_0 = \sqrt{\left(4 \pi  F_j \sigma_B\right)/v_j}$,  which is used in eq.~\ref{eq.vecpot} to define the magnetic field profile in the jet at the injection zone. The values of $B_0$ listed in Table~\ref{tab.sims} are similar to the ranges of magnetic fields inferred from observational studies of kilo-parsec scale jets \citep{carilli96a,stawarz05a,kataoka05a,stawarz06a,wu17a}; as well field strengths inferred from smaller parsec scale jets \citep[e.g.][]{osullivan09a} when extrapolated to larger scales.
\end{itemize}

The  jet power $P_j$ is found by integrating the total enthalpy flux (without the rest mass energy) over the injection surface, including the contribution of the magnetic field. For a flow with a total enthalpy $w_t = \rho_jh_j + B^2/(\gamma^2 4\pi) + \left(\mathbf{v}\cdot\mathbf{B}\right)^2/(4 \pi)$, the enthalpy flux per unit area along the $z$ axis, excluding the rest mass energy, is \citep{mignone09a}
\begin{align}
F_z^T &= \left( w_t \gamma^2 -\gamma \rho_j c^2 \right) v_z \nonumber \\
    &-  \gamma\left(\frac{\mathbf{v}\cdot\mathbf{B}}{\sqrt{4 \pi}}\right) \left(\frac{B_z}{\gamma \sqrt{4 \pi}} + \gamma \left(\frac{\mathbf{v}\cdot\mathbf{B}}{\sqrt{4 \pi}}\right) v_z \right)   \nonumber\\
    &= \left( \gamma^2 \rho_j h_j  - \gamma \rho_j c^2 \right) v_z + \frac{B^2}{4 \pi} v_z - \left(\mathbf{v}\cdot\mathbf{B}\right) \frac{B_z}{4 \pi} \label{eq.flux} .
\end{align}
 In order to get the jet power  in physical units we need to fix the value of the jet radius $R_j$ and the number density of the ambient halo at the radius of injection $n_h(r_{\rm inj})$. As discussed earlier, we assume $R_j = 100$ pc in all cases except cases G, H and J, where $R_j = 200$ pc (see Table~\ref{tab.sims}). The number density of the ambient gas is $n_h(r_{\rm inj}) = 0.1 \cc$ in all cases except for simulation I, where $n_h(r_{\rm inj}) = 1 \cc$. 

The list of simulations performed with the different choice of parameters and other inferred quantities is summarised in Table~\ref{tab.sims}. Besides the above described parameters, we also present the jet Mach number defined following \cite{rossi08a} as 
\begin{equation}
\machj = \frac{\gamma_b v_j}{\left(\gamma_s c_s\right)}, \quad \gamma_s = \frac{1}{\sqrt{1-\left(c_s/c\right)^2}} \label{eq.mach}
\end{equation}
Here $c_s$ is the sound speed defined in eq.~\ref{eq.jetcs}. This would facilitate ready comparison with previous simulations where the non-magnetic hydrodynamic Mach number has been used as an input parameter \citep[e.g.][]{komissarov98a,hardee98a,rosen99a,rossi08a,mignone10a,massaglia16a}. In the last column we present the temperature parameter defined as $\Theta_j = p_j/(\rho_j c^2)$ as done in \citet{mignone07a}, which gives an approximate estimate of the adiabatic index of the gas.  
\begin{table*}
\caption{List of simulations and parameters}\label{tab.sims}

\centering
\begin{tabular}{| l | l | l | c | c | c | c  | c | c | c | c |}
\hline
Sim. 	     &  Physical domain                                  &  Grid point              &$\eta_j$             &$\gamma_b $  & $\sigma_B$    & $r_j$    & $P_j$ 		   & $B_0$   &$\machj$   &$\Theta_j$    \\
label	     &  ($\mbox{kpc}\times\mbox{kpc}\times\mbox{kpc}$)   &                          &                     &             &               & kpc      & (ergs$^{-1}$)	   & (m G)   &           &              \\ 
\hline
A            & $4.5 \times 4.5 \times 10$                        & $288\times288\times640$  & $4\times 10^{-5}$   & 3           & 0.01          &  0.1     & $1.57\times10^{44}$   & 0.054   &  11.5     & 0.039         \\
B            & $4.5 \times 4.5 \times 10$                        & $288\times288\times640$  & $4\times 10^{-5}$   & 3           & 0.1           &  0.1     & $1.65\times10^{44}$   & 0.171   &  11.5     & 0.039        \\
C            & $4.5 \times 4.5 \times 10$                        & $288\times288\times640$  & $4\times 10^{-5}$   & 3           & 0.2           &  0.1     & $1.73\times10^{44}$   & 0.241   &  11.5     & 0.039        \\
\hline                                                              
D            & $4.5 \times 4.5 \times 10$                        & $288\times288\times640$  & $1\times 10^{-4}$   & 5           & 0.01          &  0.1     & $1.11\times10^{45}$   & 0.152   &  30.9     & 0.015        \\
E$^a$        & $6   \times 6   \times 18$                        & $384\times384\times1152$ & $1\times 10^{-4}$   & 5           & 0.05          &  0.1     & $1.15\times10^{45}$   & 0.304   &  30.9     & 0.015        \\
F            & $4.5 \times 4.5 \times 10$                        & $288\times288\times640$  & $1\times 10^{-4}$   & 5           & 0.1           &  0.1     & $1.17\times10^{45}$   & 0.48    &  30.9     & 0.015        \\
\hline                                                              
G$^b$        & $4.5 \times 4.5 \times 10$                        & $288\times288\times640$  & $1\times10^{-4}$    & 6           & 0.2           &  0.2     & $8.29\times10^{45}$   & 0.907   &  17.49    & 0.077        \\
H            & $4.5 \times 4.5 \times 10$                        & $288\times288\times640$  & $1\times10^{-4}$    & 10          & 0.2           &  0.2     & $1.64\times10^{46}$   & 1.363   &  62.77    & 0.015        \\
I$^c$        & $4.5 \times 4.5 \times 10$                        & $288\times288\times640$  & $1\times10^{-4}$    & 5           & 0.1           &  0.1     & $1.17\times10^{46}$   & 1.36    &  30.9     & 0.015        \\
J            & $6 \times 6 \times 40$                            & $384\times384\times2560$ & $1\times10^{-4}$    & 10          & 0.1           &  0.2     & $1.51\times10^{46}$   & 0.964   &  62.77    & 0.015        \\
\hline
\end{tabular} 

\begin{tablenotes}
\small
\item {$^a$}  Simulation E is a two sided jet with the injection zone located at the centre of the domain. 
\item {$^b$}  Over-pressured jet $\zeta_p = 5$. For the rest $\zeta_p = 1$. 
\item  $^c$ $n_h(r_{\rm inj}) = 1 \cc$. For other simulation $n_h(r_{\rm inj}) = 0.1 \cc$.
\end{tablenotes}

\flushleft

        Parameters: \\
\begin{tabular}{|l | l|}
    $\eta_j$:            & Ratio of jet density to ambient density. \\
    $\gamma_b$:          & Jet Lorentz factor. \\
	$\sigma_B$:          & Jet magnetisation parameter, the ratio of jet Poynting flux to  enthalpy flux. \\
	$r_j$:               & Jet radius\\
	$P_j$:               & Jet power computed from eq.~\ref{eq.flux}. \\
        $B_0$:               & Maximum strength of toroidal magnetic field in milli-Gauss \\
	$\machj$:            & Jet Mach number defined in eq.~\ref{eq.mach}  \\
    $\Theta_j$:          & The temperature parameter for the jet equation of state: $\Theta_j = p_j/(\rho_j c^2)$ \\
        
\end{tabular}
\end{table*}

\subsection{Numerical implementation}

We perform the simulations with the \textsc{pluto} code \citep{mignone07}, utilising the relativistic magnetohydrodynamic module (RMHD). We employ the piece-wise parabolic reconstruction scheme \citep[\textsc{ppm}: ][]{colella84a}, with a second-order Runge-Kutta method for time integration and the HLLD Riemann solver \cite{mignone09a}.   The magnetic field components, defined on the face-centres of a staggered mesh, are updated using the constrained transport (CT) method \citep{balsara99a, gardiner05a}. The electromotive force is defined on the zone edges of a computational cell, and reconstructed with the upwind constrained transport technique \citep[\textsc{uct hll} scheme of \textsc{pluto}:][]{londrillo04a} by solving a 2-D Riemann problem. For better numerical stability, in some simulations we employed  a more diffusive Riemann solver (\textsc{hll}) and limiter (\textsc{min-mod}) for cells identified as strongly shocked in the central region where the jet is injected ($Z < \pm 1$ kpc). A computational cell was identified to be shocked if $\delta p/p_{\rm min} > 4$, where $\delta p$ is the sum of the difference in pressure between neighbouring cells in each direction and $p_{\rm min}$ is the minimum pressure of all surrounding cell. An outflow boundary condition was applied on all sides of the computational box with the jet injected from a volume inside the computational box. 

\begin{figure}
	\centering
	\includegraphics[width = 7 cm, keepaspectratio] 
	{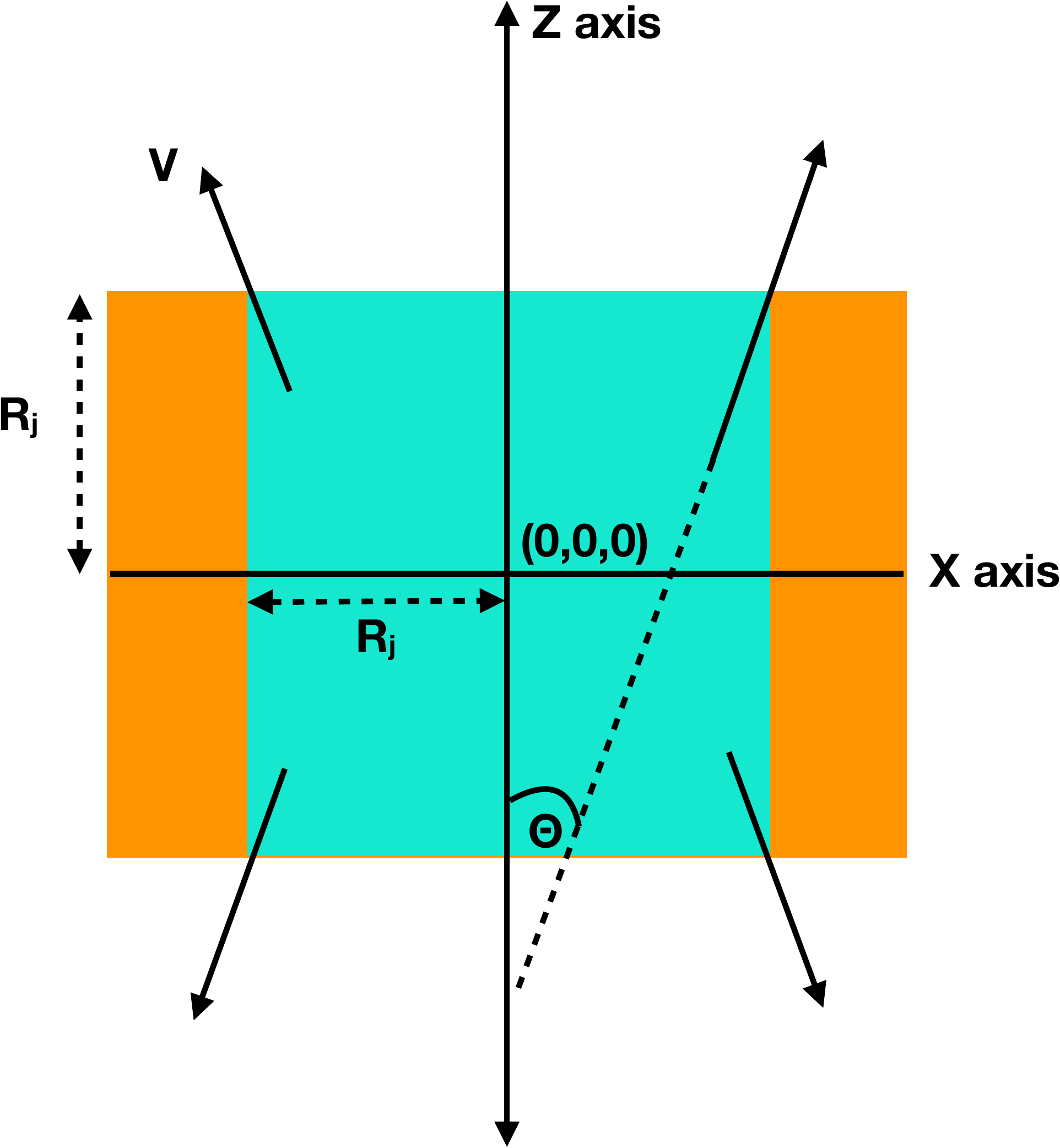}
	\caption{\small A cartoon demonstrating the $X-Z$ plane of the jet injection, centred at the origin. The shaded regions (in blue and orange) represent the injection zone where the fluid variables are not updated. The injection zone is initialised with jet parameters in the central region (shaded in blue) with lateral extent up to $R_j$. The injection zone is extended beyond $R_j$ by a few computational cells where fluid variables are set to values of the ambient medium. The velocity vectors lie along a cone which makes a half-opening angle of $\theta = 5^\circ$ with the $Z$ axis. }
	\label{fig.injectcartoon}
\end{figure}
The jet is injected along both positive and negative $Z$ axis from an injection region centred at $(0,0,0)$, as shown in Fig.~\ref{fig.injectcartoon}.  The vertical extent of the injection zone is set at $z=\pm R_j$, while the horizontal extent is chosen to have a few computational cells larger than the jet radius. In the jet injection zone the fluxes of the Riemann solvers are set to zero and hence the fluid variables ($\rho$, $p$, $\mathbf{v}$) remain unchanged.  For most of the simulations, the computational box has a short extension of $\sim 1$ kpc along the negative $z$ axis. This avoids the use of a reflecting boundary condition, as has been traditionally used in typical jet simulations \cite{mignone10a,massaglia16a,perucho19a}, which may result in spurious features at the lower boundary. For simulation E (see Table~\ref{tab.sims}), the injection zone was centred at the middle of the total computational domain, and the evolution of both jet lobes were followed in full. The extent of the computational domain and the grid resolution are detailed in Table \ref{tab.sims}. The grid resolution is chosen in a way such  that the number of points on the jet radius is always larger than 6.

The density and pressure of the jet in the injection zone are tapered radially with a smoothing function: $Q = Q_0 / \left(\cosh\left\lbrack\left(R/R_j\right)^6\right\rbrack\right)$, $R$ being the cylindrical radius, to avoid sharp discontinuities at $R=R_j$. The velocity components were strictly truncated at the jet radius ($R=R_j$) so that there is no energy flux beyond $R_j$. This ensures that the injected jet energy flux is not greater than the intended value calculated by integrating eq.~\ref{eq.flux} over the injection surface bounded by $R=R_j$. Besides the bulk velocity defined by $\gamma_b$, we additionally imposed small perturbations on the transverse components to induce pinching, helical and fluting mode instabilities as in \citet{rossi08a}
\begin{equation}
\left(v_x, v_y\right) = \frac{\tilde{A}}{24} \sum_{m=0}^{2} \sum_{l=1}^{8} \cos(m\phi + \omega_l t + b_l)(cos\phi,\sin\phi)
\end{equation}
where $\phi = \tan^{-1}(y/x)$, $\omega_l = c_s(1/2, 1, 2, 3)$ for $l \in (1,4)$ and $\omega_l = c_s(0.03, 0.06, 0.12, 0.25)$ for $l \in (5,8)$. Here $c_s$ is the relativistic sound speed in the jet, which for a Taub-Matthews equation of state is defined as \citep{mignone05b}
\begin{equation}
c_s^2 = \left(\frac{p_j}{3 \rho_j h_j}\right)\left( \frac{5\rho_j h_j - 8p_j}{\rho_j h_j - p_j}\right)\label{eq.jetcs}
\end{equation}
where $\rho_j h_j$ is computed from eq.~\ref{eq.enth}.
The perturbation amplitude is defined to be
\begin{equation}
\tilde{A} = \frac{1}{\sqrt{2}\gamma_b}\frac{\sqrt{\left(1+\epsilon\right)^2 - 1}}{\left(1 + \epsilon\right)}
\end{equation}
which gives the Lorentz factor of the perturbed velocity field to be $\gamma = \gamma_b(1+\epsilon)$. We choose $\epsilon = 0.005$ for our simulations to induce very mild perturbations in the jet flow.

The magnetic field components were assigned from a vector potential defined by
\begin{align}
A_z &= -\int_0^\infty B_0 f\left(\frac{R}{R_j}\right) dR \label{eq.vecpot} \\
\mbox{ where } f\left(\frac{R}{R_j}\right) &= \frac{R}{R_j \left(\cosh\left\lbrack\left(R/R_j\right)^6\right\rbrack\right)} \mbox{ for } R \le R_j \nonumber \\
                                                     &= 0 \mbox{ for } R>R_j \label{eq.bprofile}
\end{align}
Eq.~\ref{eq.vecpot} is numerically integrated to radii much larger than the jet radius to obtain a tabulated list of vector potential as a function of cylindrical radius, which is then interpolated on to the \textsc{pluto} domain. This gives a toroidal magnetic field of peak strength $B_0$, as defined by the choice of the magnetisation parameter $\sigma_B$ in eq.~\ref{eq.sigma}. Thus the radial profile of the jet magnetic field is:
\begin{align}
B_\theta (R) &= B_0 \left(\frac{R}{R_j}\right) \frac{1}{\cosh\left\lbrack\left(R/R_j\right)^6\right\rbrack} \mbox{ for } R \le R_j \nonumber \\
 & = 0  \mbox{ for } R>R_j
\end{align}

  The staggered magnetic field components were not updated inside the jet injection zone except  at the faces of the outer surfaces of the injection domain. Similarly, the components of the EMF were also not updated within the injection zone, except for the edges of the injection domain. The sign of the toroidal component of the magnetic field and $z$ component of the velocity were reversed for injection of jet along the negative $z$ axis.

\section{Results}\label{sec.results}
We have performed a series of simulations to investigate the difference in the dynamics of the jet for different powers, magnetisation, jet pressure contrast with respect to the ambient gas and density of the ambient medium. The main focus of these studies has been to understand the impact of these parameters on the evolution of the jet's morphology, the deceleration of the jet and the impact of instabilities such as kink and Kelvin-Helmholtz modes. In this section we summarise the results of the different simulations and compare analytical models that predict the evolution of the jet kinematics.

\subsection{Dynamics of jet}
\begin{figure*}
	\centering
	\includegraphics[width = 6.2cm, keepaspectratio] 
	{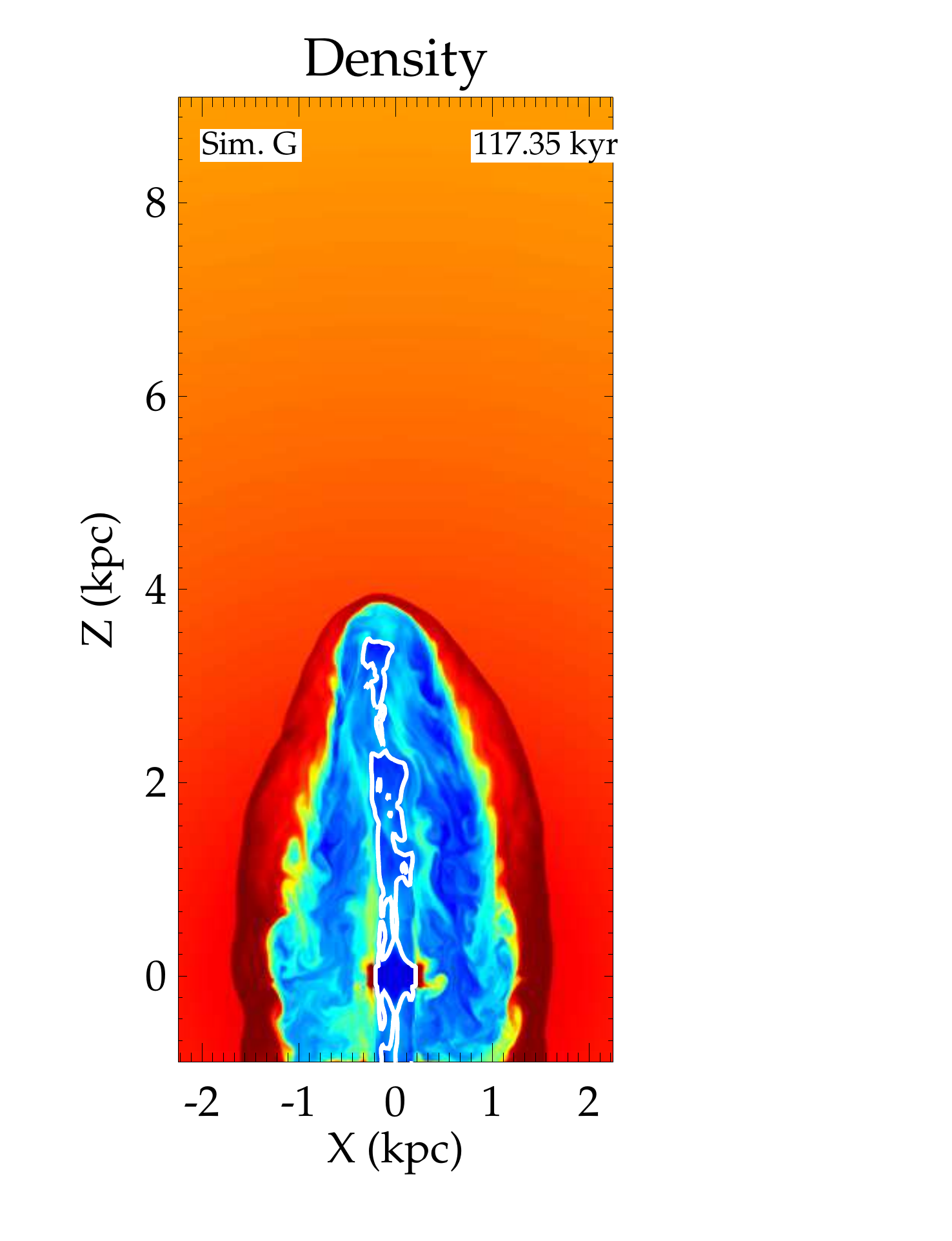}\vspace{-0cm}\hspace{-3.4cm}
	\includegraphics[width = 6.2cm, keepaspectratio] 
	{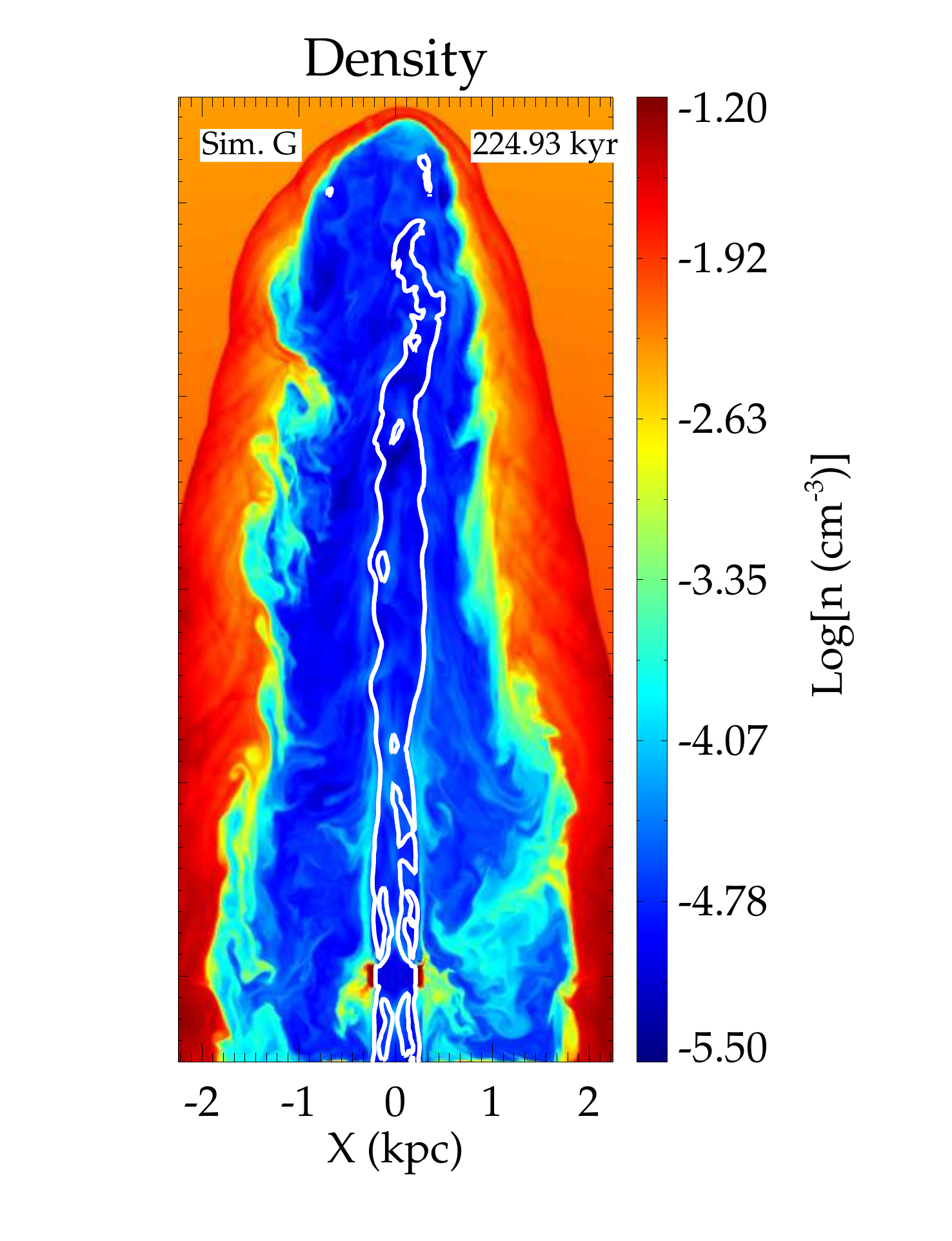}\vspace{-0cm}\hspace{-1cm}
	\includegraphics[width = 6.2cm, keepaspectratio] 
	{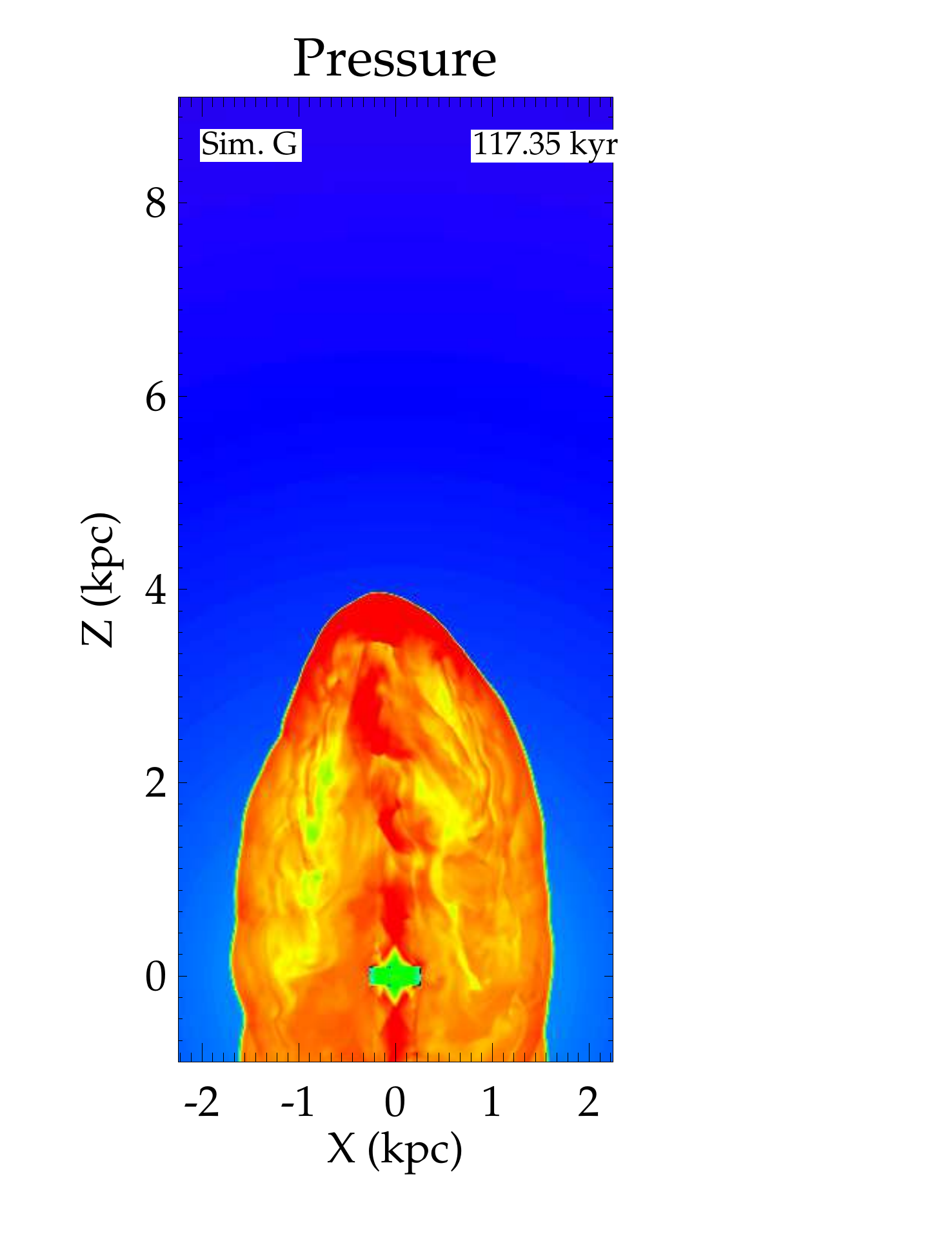}\vspace{-0cm}\hspace{-3.4cm}
	\includegraphics[width = 6.2cm, keepaspectratio] 
	{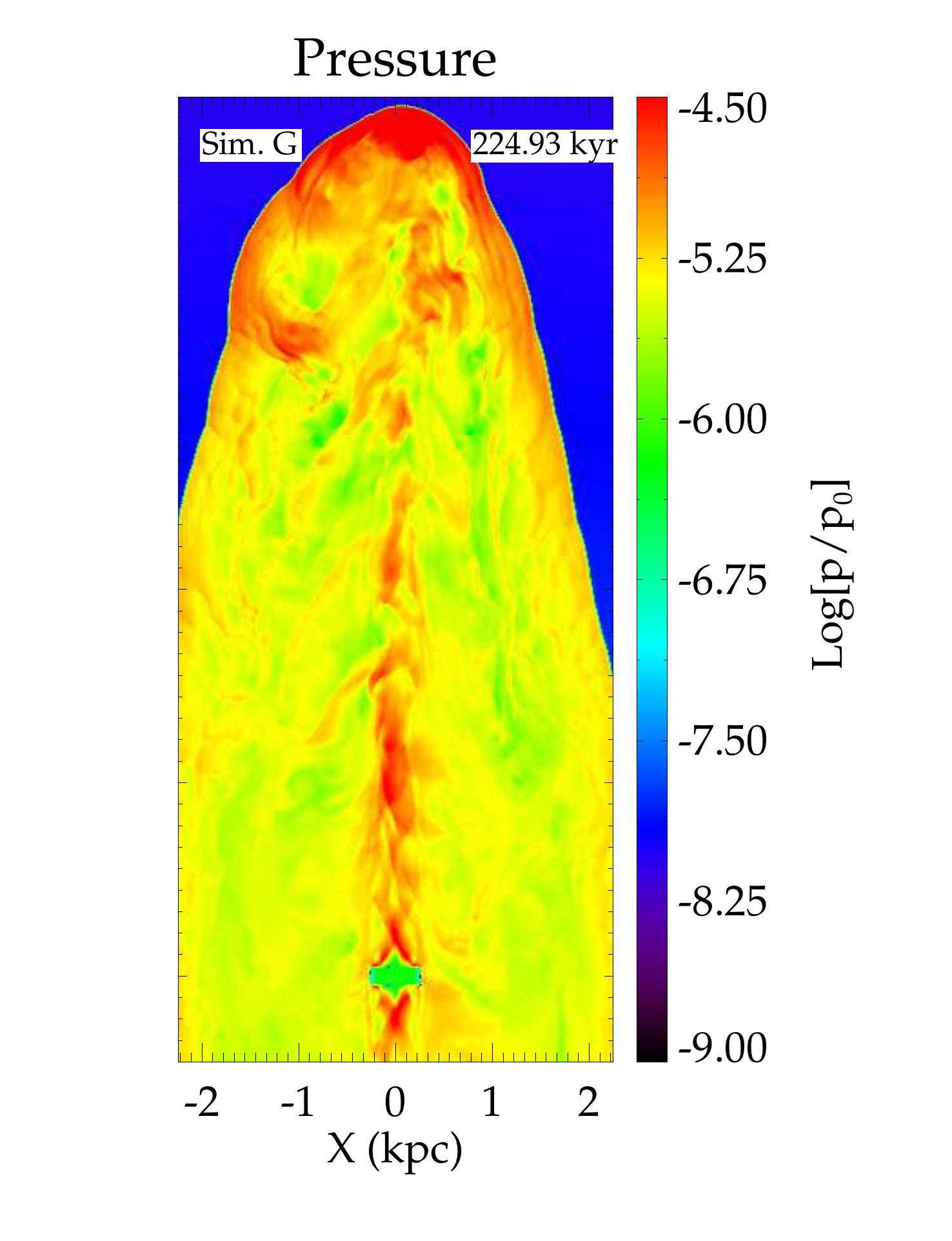}\vspace{-0cm}
	\includegraphics[width = 6.2cm, keepaspectratio] 
	{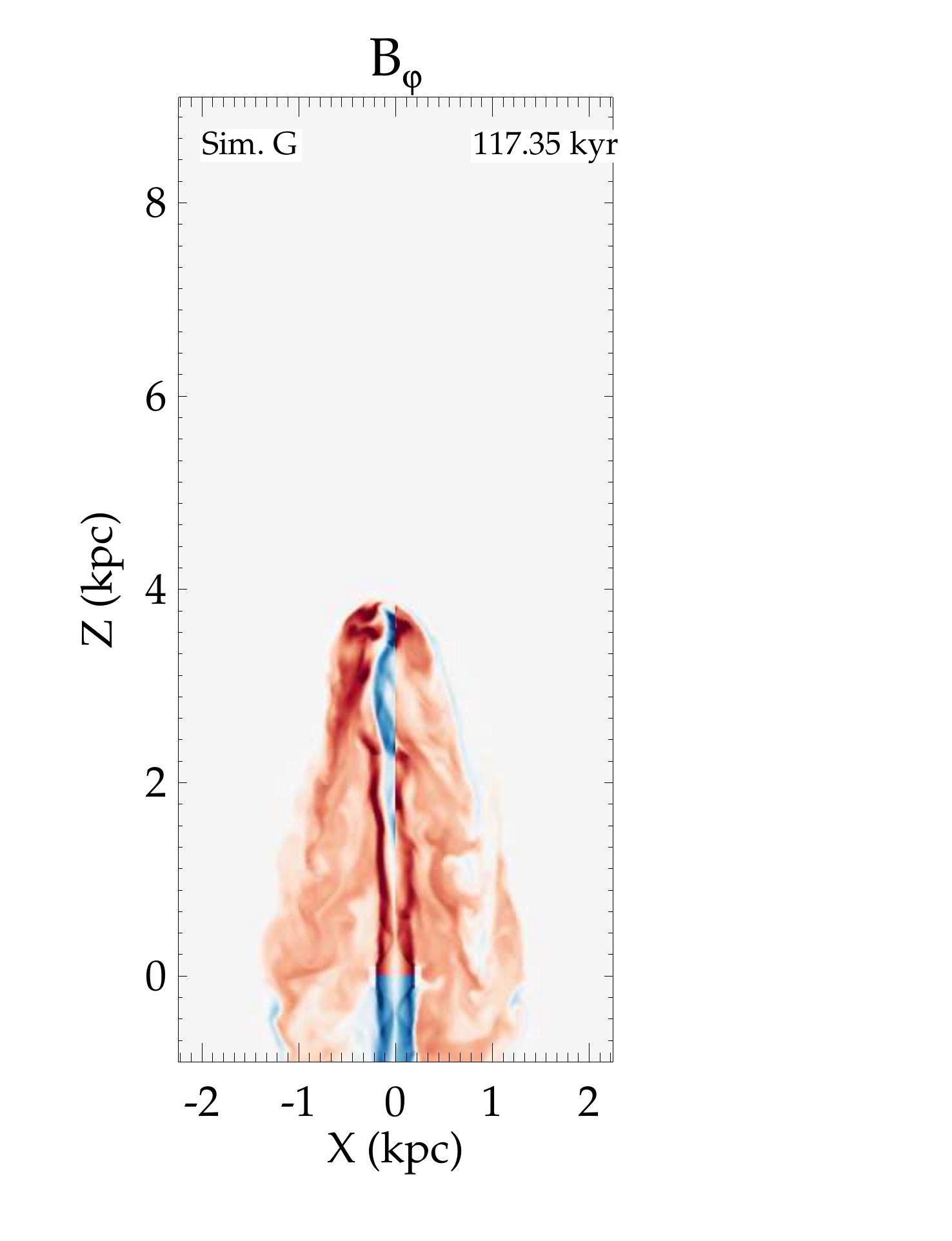}\vspace{-0cm}\hspace{-3.4cm}
	\includegraphics[width = 6.2cm, keepaspectratio] 
	{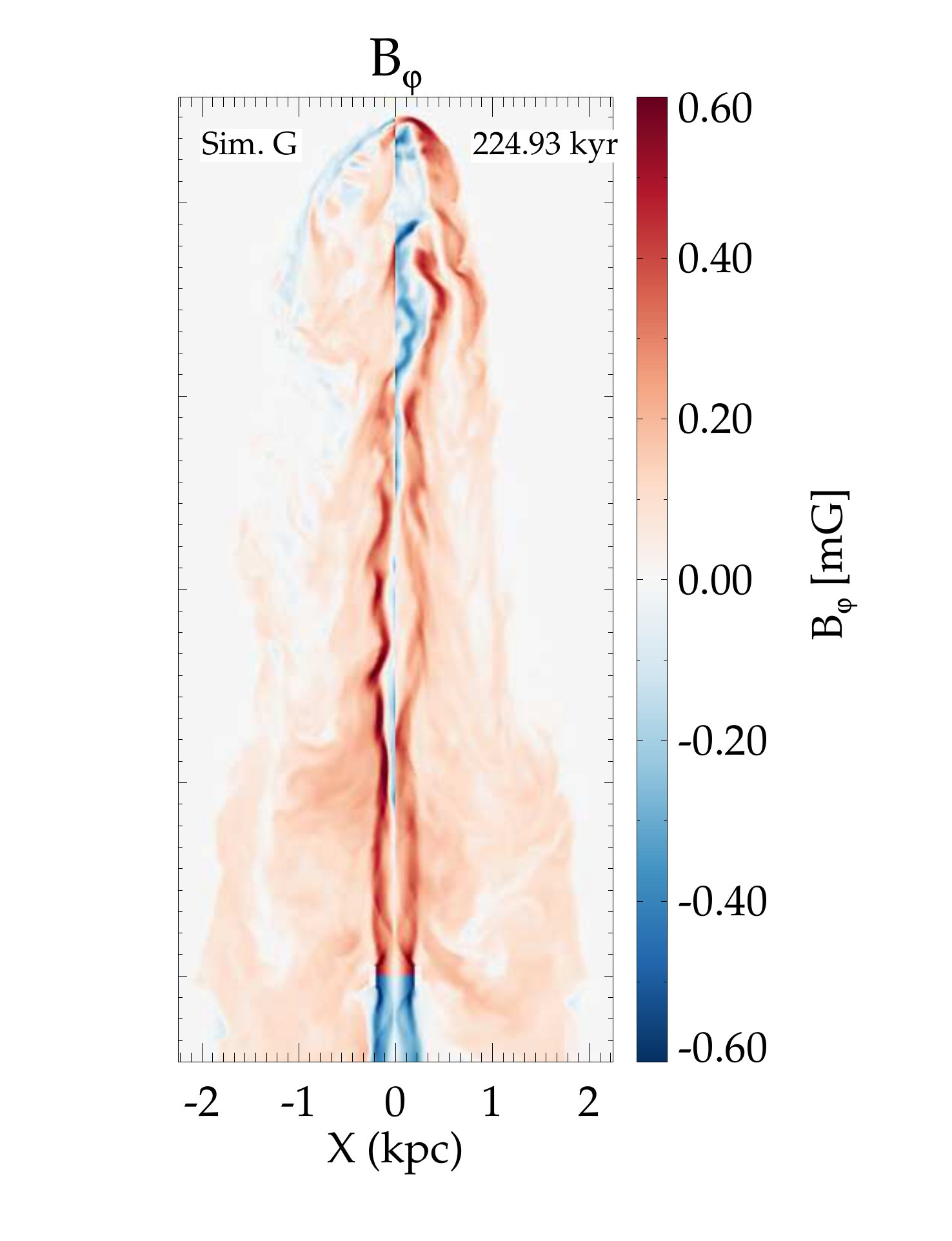}\vspace{-0cm}\hspace{-1cm}
	\includegraphics[width = 6.2cm, keepaspectratio] 
	{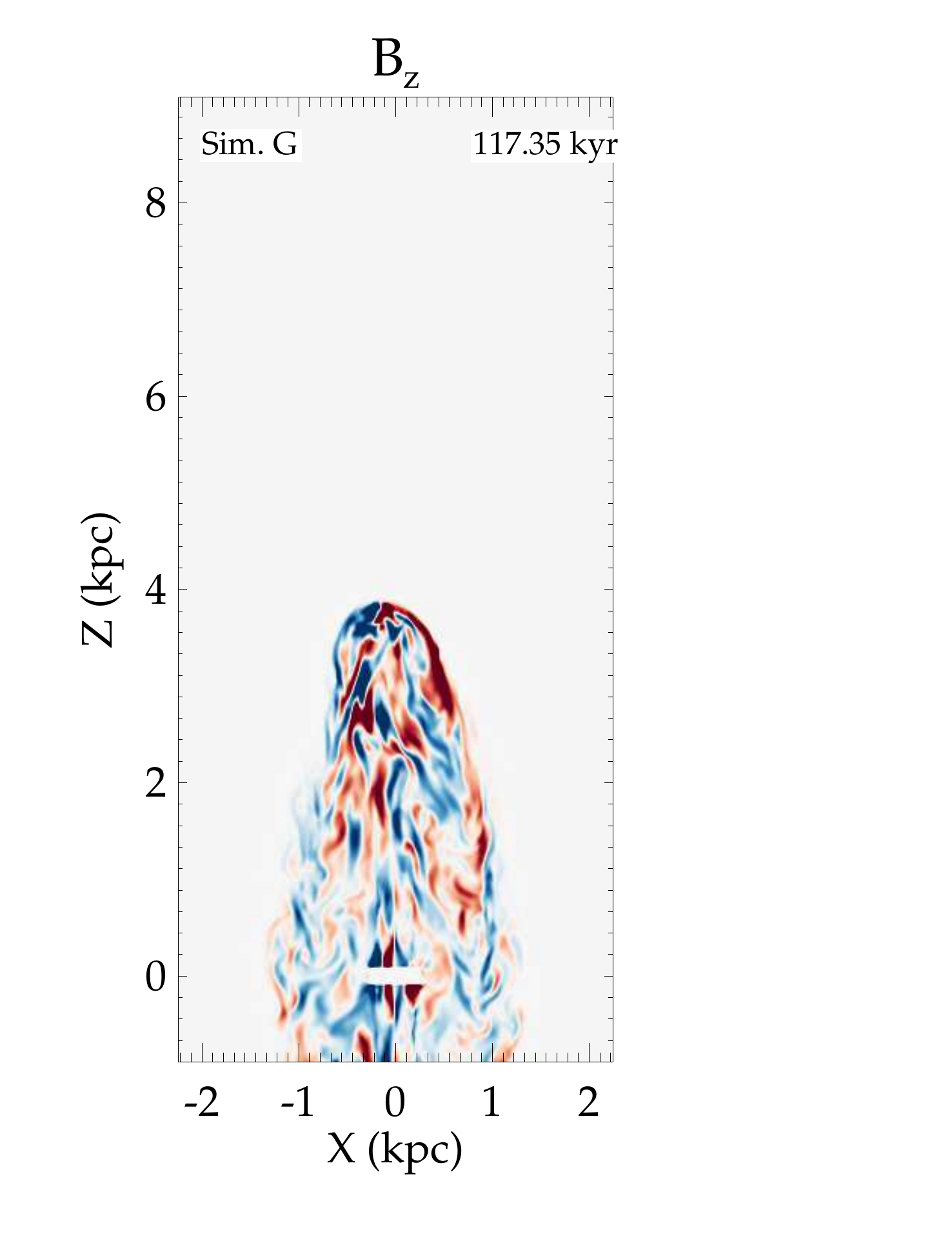}\vspace{-0cm}\hspace{-3.4cm}
	\includegraphics[width = 6.2cm, keepaspectratio] 
	{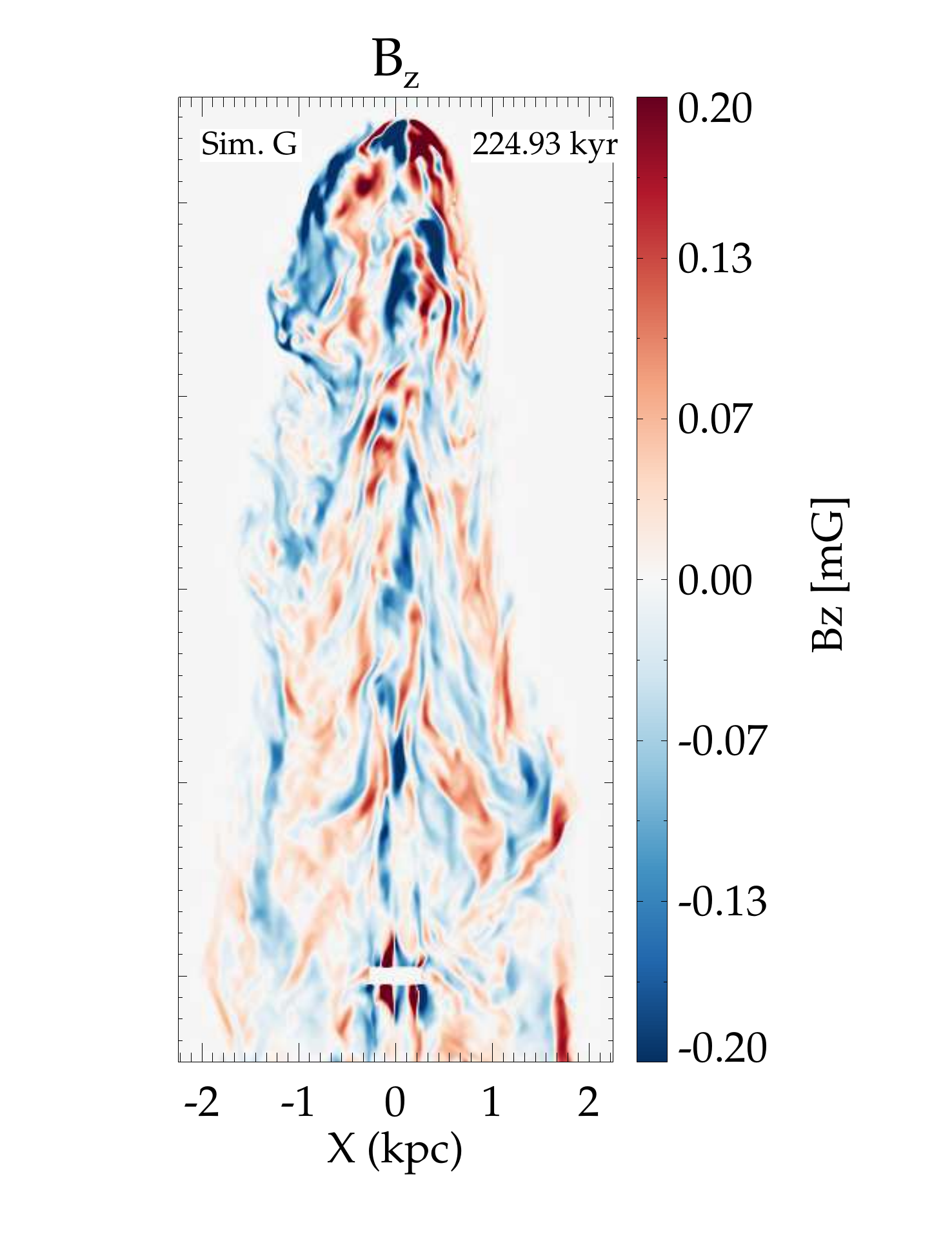}\vspace{-0cm}
	\caption{\small The top left panel shows slices in the $X-Z$ plane of the logarithm of the number density at two different times for Simulation G (see Table~\ref{tab.sims} for list of simulations). The white lines represent contours of Lorentz factor $\gamma = 2$, representing the bulk relativistic flow. The logarithm of pressure slices are on the top right panel. The bottom panels show the toroidal (left) and vertical component of the magnetic field in milli-Gauss.  }
	\label{fig.simG}
\end{figure*}
In Fig.~\ref{fig.simG} we present the density and pressure at two different times for simulation G (see Table~\ref{tab.sims}), which represents a typical  powerful FRII jet \citep[as per the classification of ][]{fanaroff74a}. The density slices show an internal cavity bounded by a contact discontinuity and forward shock \citep[typical of over-pressured outflows as shown in][]{komissarov98a,kaiser97a}.  The jet moves at bulk relativistic velocities near the axis, represented by the contour of $\gamma = 2$ in white. The jet terminates at a hot-spot with enhanced pressure due to the strong shock with the ambient gas. The internal cavity has low density ($\sim 10^{-4} \cc$) plasma resulting from the mixing of thermal gas due to Kelvin-Helmholtz instabilities at the contact discontinuity with the jet backflow that originates from the forward shock at the jet-head. 

Within the axis of the jet there are several sites of enhanced pressure, arising out of recollimation shocks \citep{norman82a,komissarov98a,nalewajko09a,nawaz14a,fuentes18a,bodo18a}. In the bottom panels we show the $Y$ and $Z$ components of the magnetic field. It is evident from Fig.~\ref{fig.simG} that the jet is not collimated along the $Z$ axis, showing both small scale distortions as well as bending near the jet head spread over $\sim 1$ kpc. Such distortions arise from both small scale instabilities resulting shearing of the jet axis driven by high order Kelvin-Helmholtz modes, as well as kink type $m=1$ mode instabilities \citep{mignone10a,bodo13a,mizuno14a,bodo19a,bromberg19a}. It is to be noted that although we inject a purely toroidal magnetic field, the jet magnetic field develops a vertical component as it propagates. This results in a helical topology of the resultant magnetic field along the jet axis, although dominated by the toroidal component. We shall elaborate more on the effect of instabilities on the jet dynamics in the following sections.

\subsection{Effect of magnetisation on jet stability}
Two different kinds of fluid instabilities affect the dynamics of the jets in our simulations. Weakly magnetised jets have a faster onset of Kelvin-Helmholtz (KH) instabilities which deform the jet cross section with short wavelengths modes and promote mixing between the jet and the surrounding medium. With a stronger toroidal magnetic field, the magnetic tension opposes jet deformation and  stabilises the KH modes \citep{mignone10a}. However, stronger magnetisation can also instigate the onset of current driven instabilities, of which the most relevant is  the $m=1$ mode, which will result in large scale deformations and bending of the jet from its initial axis \citep{bodo13a}. The relative growth rates of the different modes depend on the magnetic pitch parameter, the jet velocity and magnetisation \citep{bodo13a}. In the following sections we discuss the effect of magnetisation on the evolution of the jets in different power regimes.

\subsubsection{Low power jets: Kink modes}\label{sec.kink}
\begin{figure*}
	\centering
	\includegraphics[width = 9cm, keepaspectratio] 
	{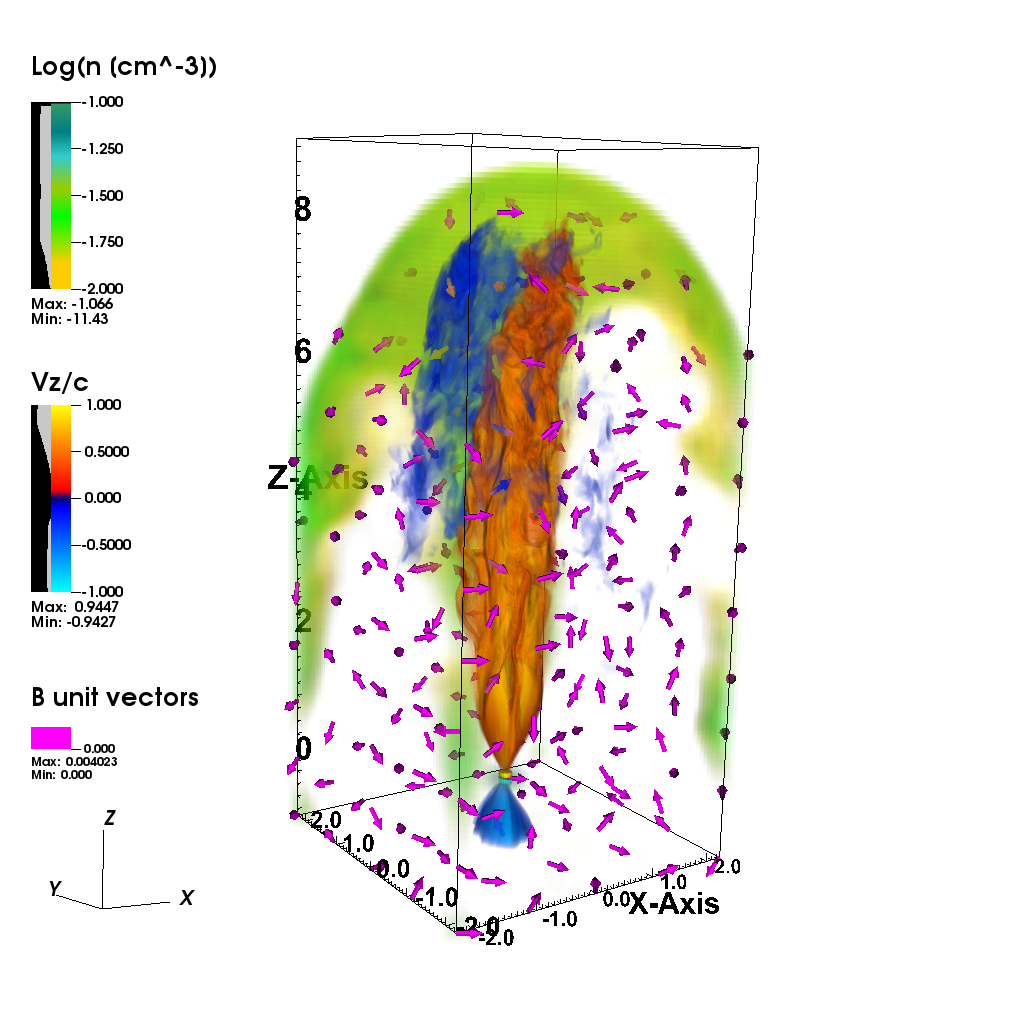}\hspace{-1.8cm}
	\includegraphics[width = 9cm, keepaspectratio] 
	{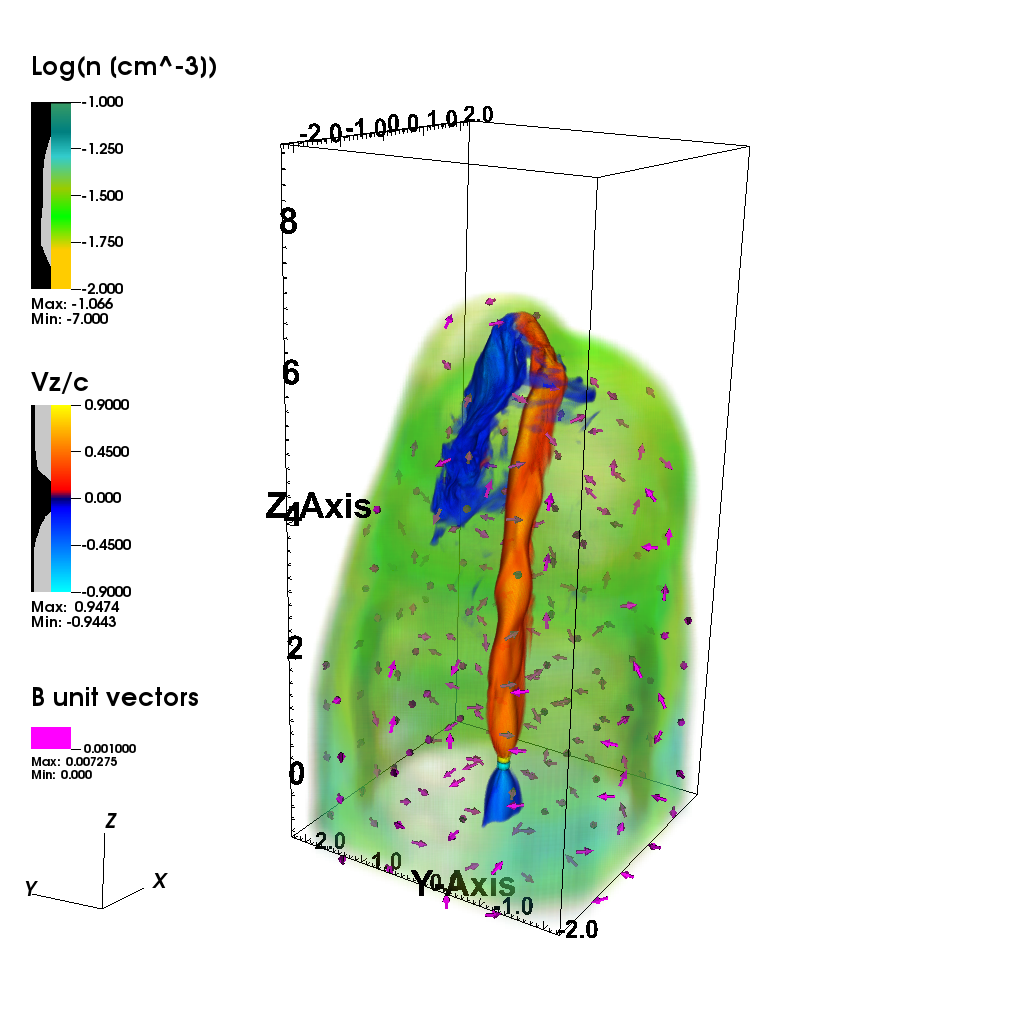}\hspace{-1.8cm}
	\caption{\small 3D volume rendering of the velocity in orange-blue palette with the density of the jet and cocoon in the yellow-green palette for simulations A (left) and B (right). The magnetic field vectors are plotted in magenta with their length scaled to their magnitude.}
	\label{fig.visitimgs}
\end{figure*}
Simulations A,B,C have similar jet power ($\sim 10^{44} \ergs$) and injection speed ($\gamma\sim3$) while differing in jet magnetisation with $\sigma_B=0.01,0.1,0.2$ respectively. Figure~\ref{fig.visitimgs} shows the 3D volume rendering of the jet speed and density for simulations A and B. The $Z$ component of the velocity (normalised to $c$) is presented in a blue-red palette with the red-orange depicting positive velocities and velocities directed along the negative $Z$ axis in blue. The  spine of the jet in simulation B (right panel in Fig.~\ref{fig.visitimgs}) shows clear bends and twists indicative of kink mode instabilities. At the top, the jet head bends sharply, almost perpendicular to its original axis, before bending backwards to eventually form the backflow. The morphology of the jet head is thus very different from that of usual jets where the relativistic flow terminates in a shock, at a mach disc, symmetric around the jet axis before flowing backwards in the cocoon \citep{kaiser97a,marti97a,komissarov98a,rosen99a}. 

The cocoon of the jet can be discerned from the volume rendering of the density presented in green. The morphology of the cocoon is highly asymmetric, with local bubble shaped protrusions. These correspond to the locally expanding bow shock where the jet was temporarily directed before bending to a different direction. Over the course of its growth, the swings of the jet-head results in a broader spread of the jet energy over a much larger solid angle. This results in the formation of the cocoon with an over-all cylindrical shape, as opposed to a narrow conical shape expected for stable jets. The instabilities decelerate the jet, reducing its advance speed as discussed later in Sec.~\ref{sec.GBCcompare}.

Simulation A with lower magnetisation (Table~\ref{tab.sims}) on the other hand do not show the onset of the kink modes on similar time scales. The jet forms a conical cocoon with stable spine along the launch axis. The central spine broadens and shows evidence of shear, as expected for low magnetic fields (discussed more in the next section). The magnetic field vectors in simulation B are less ordered compared to that in A. The randomness of the field topology arises from the stronger interaction of the jet with the  ambient gas due to the kink modes, which also enhances turbulent motions in the cocoon.


\subsubsection{Moderate power jets: small scale Kelvin-Helmholtz modes}\label{sec.moderate}
\begin{figure}
	\centering
	\includegraphics[width = 9cm, keepaspectratio] 
	{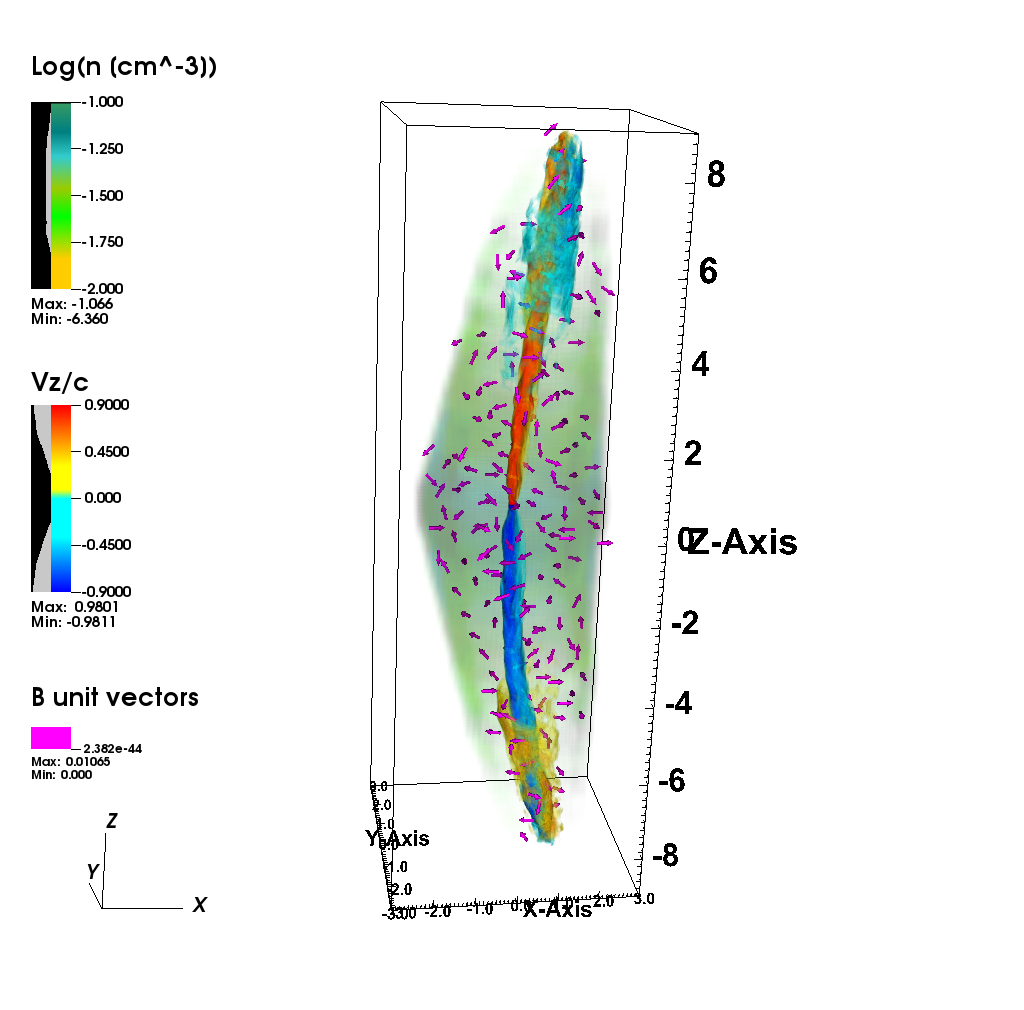}\hspace{-1.8cm}
	\caption{\small 3D volume rendering of the jet and cocoon, as in Fig.~\ref{fig.visitimgs}, for simulation E.}
	\label{fig.visitimgs2}
\end{figure}
\begin{figure*}
	\centering
	\includegraphics[width = 6.2cm, keepaspectratio] 
	{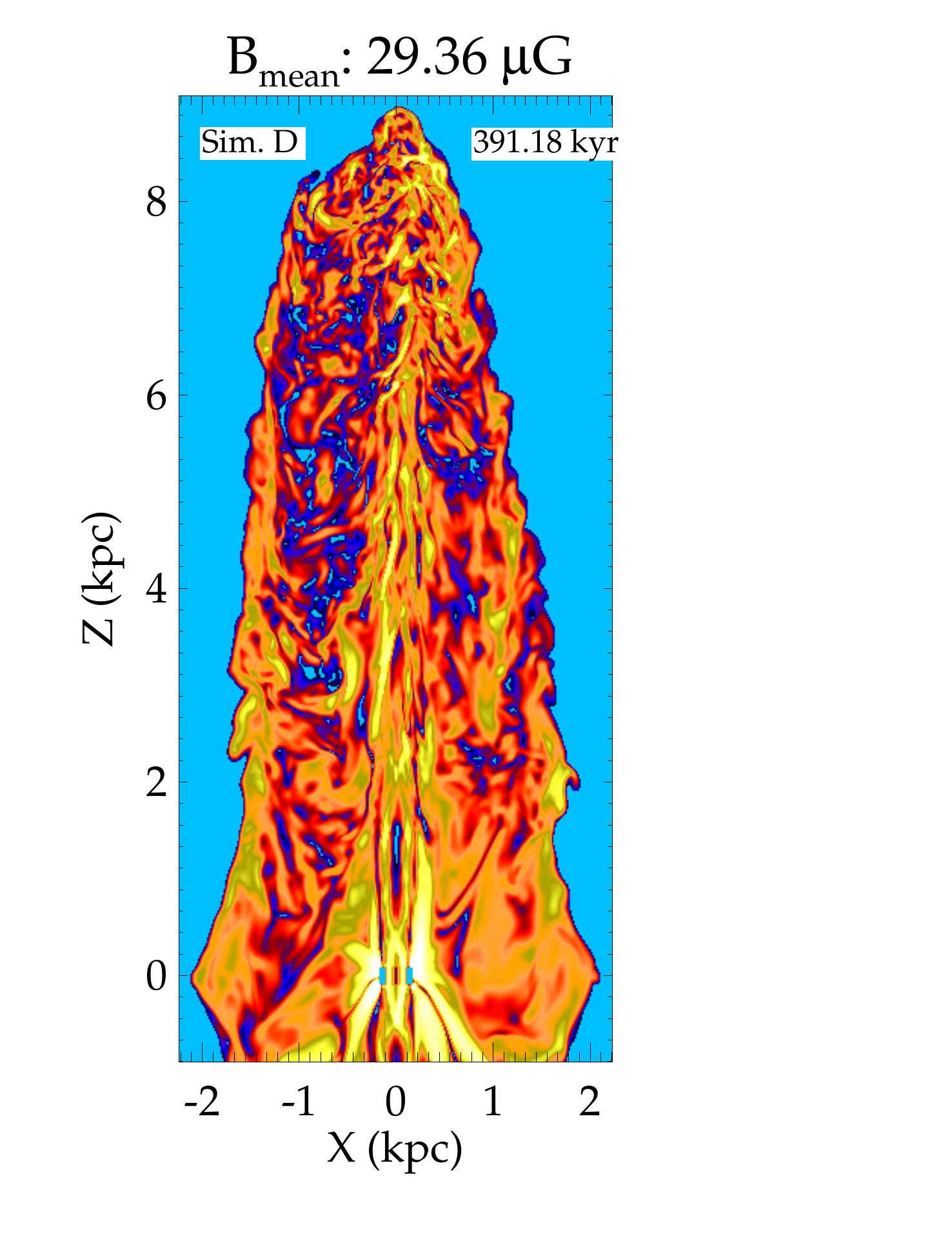}\vspace{-0cm}\hspace{-3.4cm}
	\includegraphics[width = 6.2cm, keepaspectratio] 
	{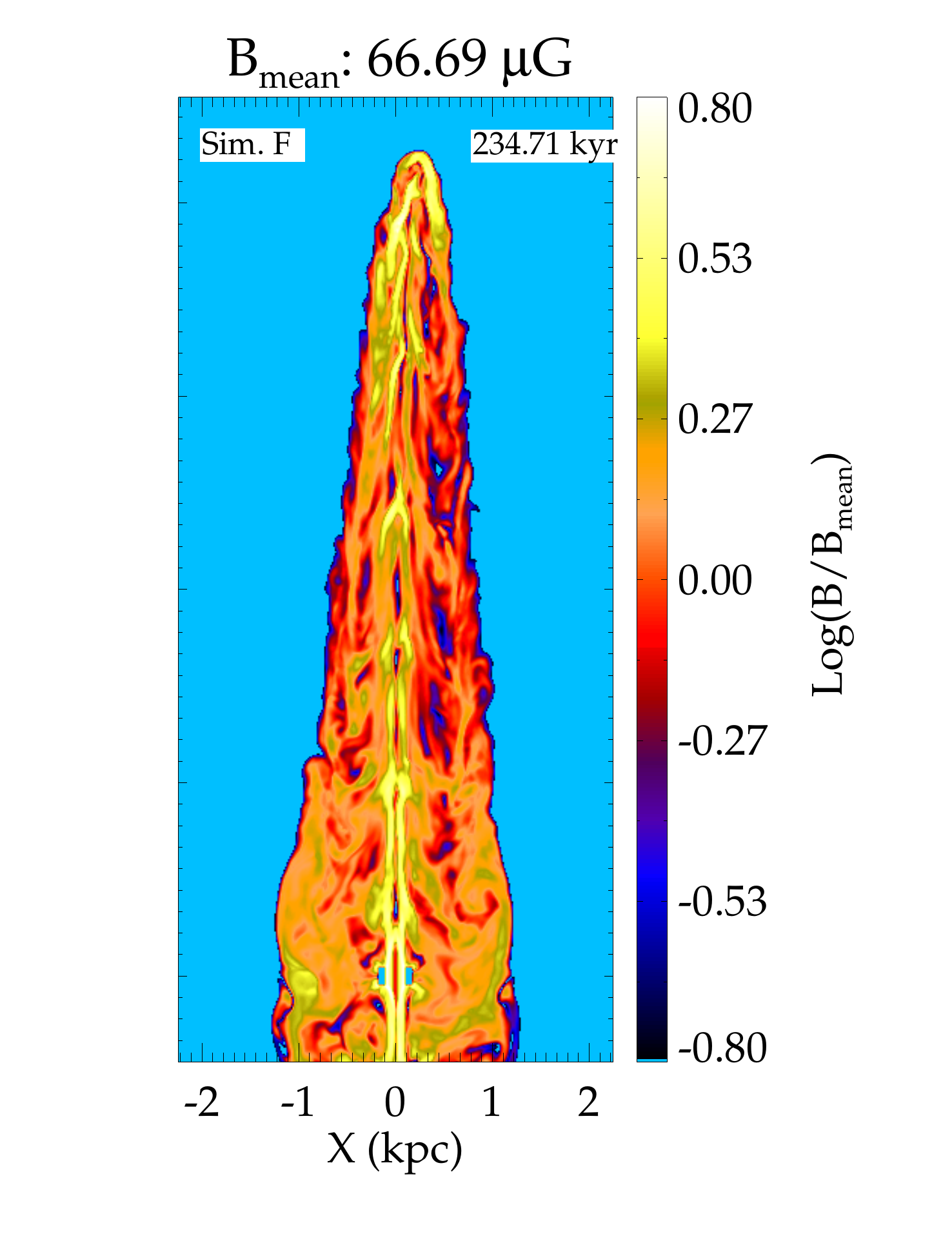}\vspace{-0cm}\hspace{-0.8cm}
	\includegraphics[width = 6.2cm, keepaspectratio] 
	{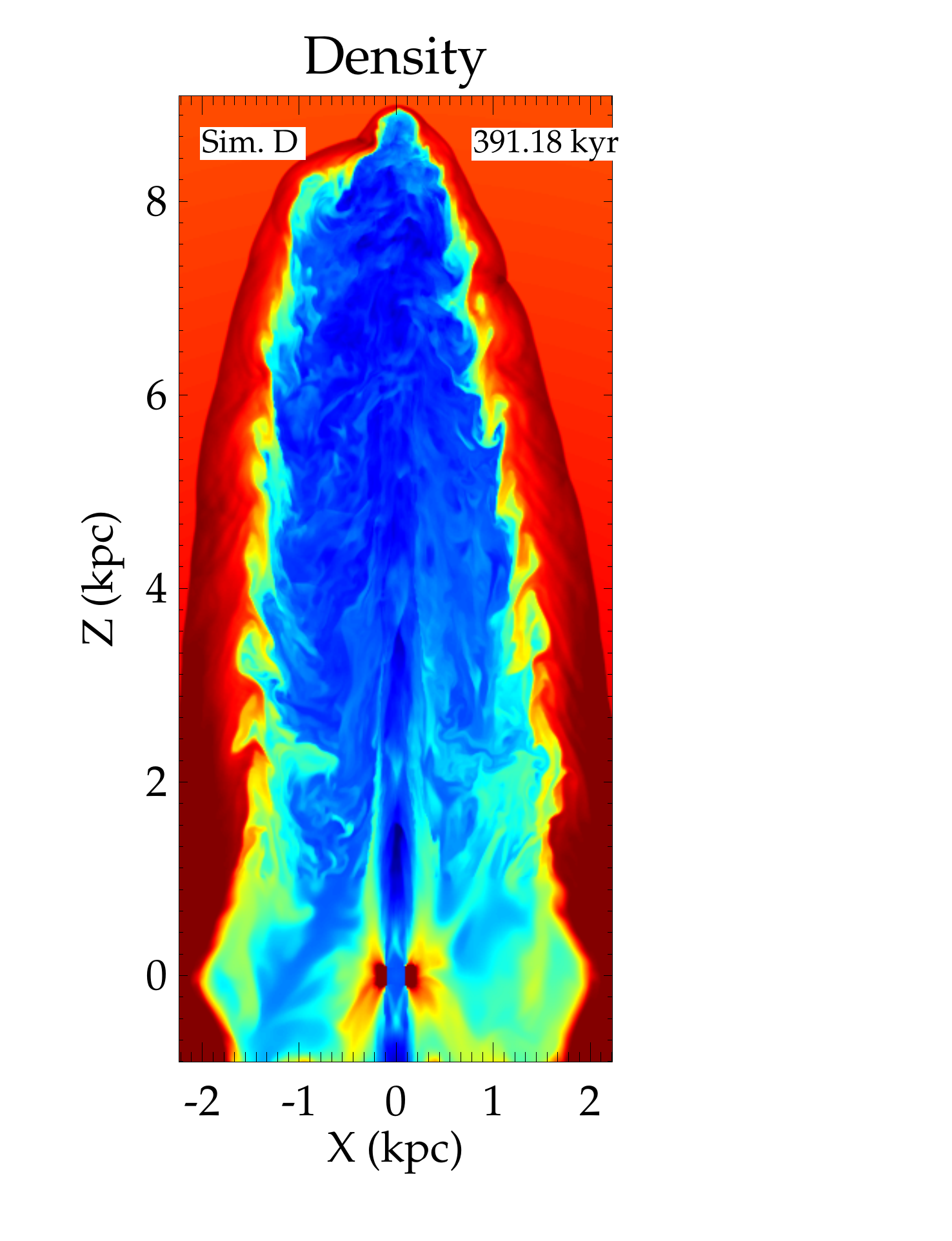}\vspace{-0cm}\hspace{-3.4cm}
	\includegraphics[width = 6.2cm, keepaspectratio] 
	{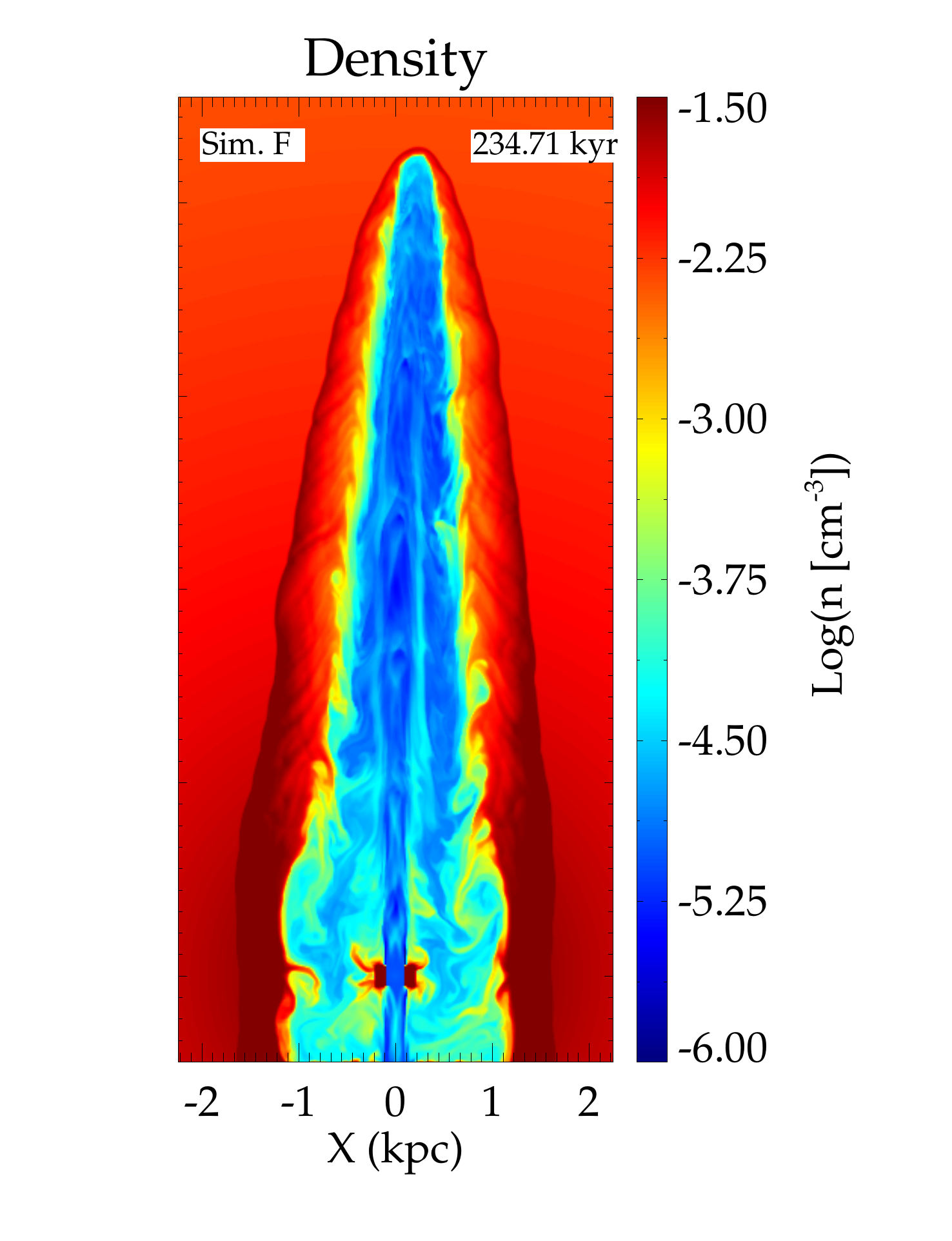}\vspace{-0cm}
	\caption{\small \textbf{Top}: Plots of magnetic field and density for simulations D ($\sigma_B=0.01$) and F ($\sigma_B=0.1$) to show difference in morphology due to high m modes arising from Kelvin-Helmholtz instabilities}
	\label{fig.sigcompare}
\end{figure*}
\begin{figure}
	\centering
	\includegraphics[width = 5.8cm, keepaspectratio] 
	{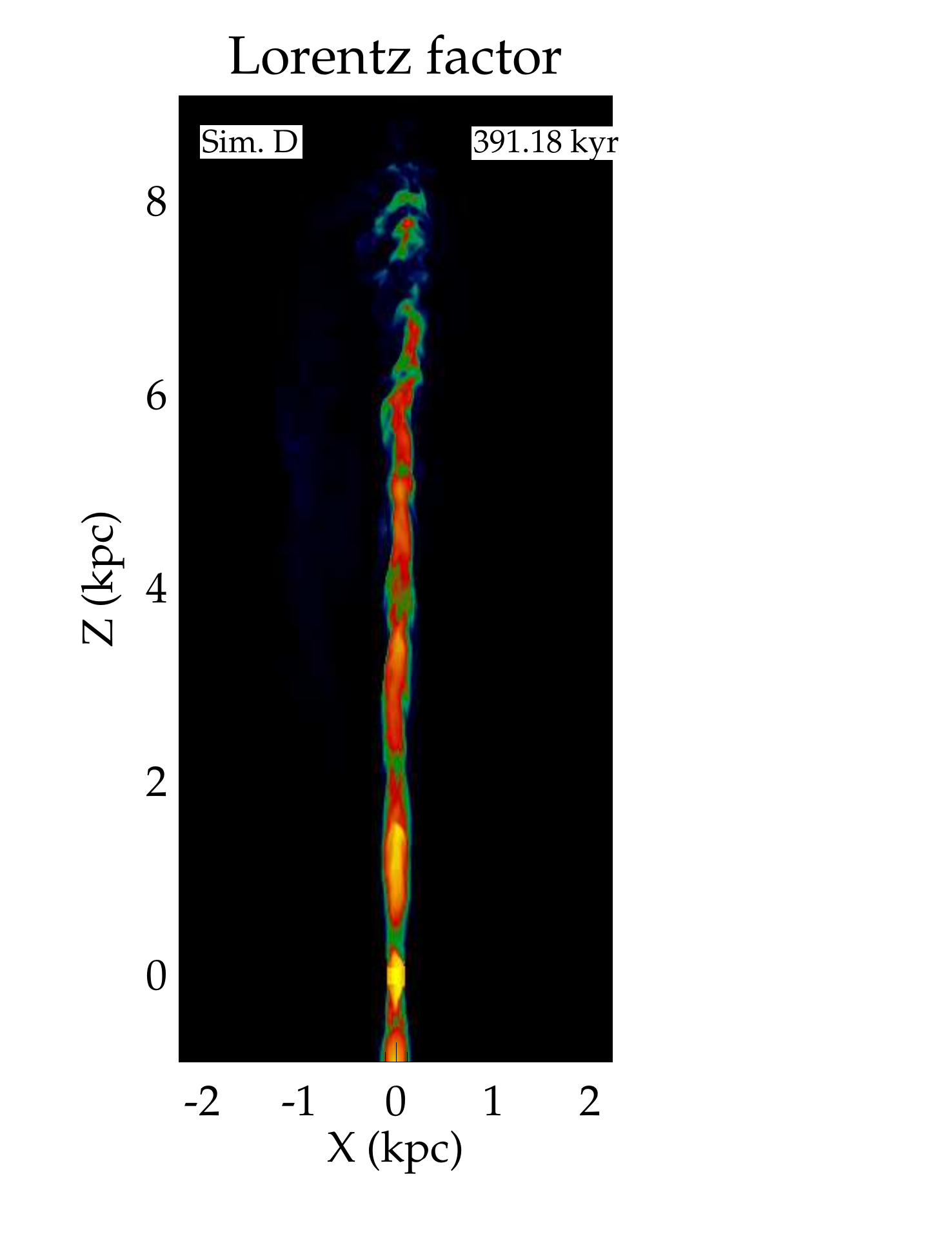}\vspace{-0cm}\hspace{-3.3cm}
	\includegraphics[width = 5.8cm, keepaspectratio] 
	{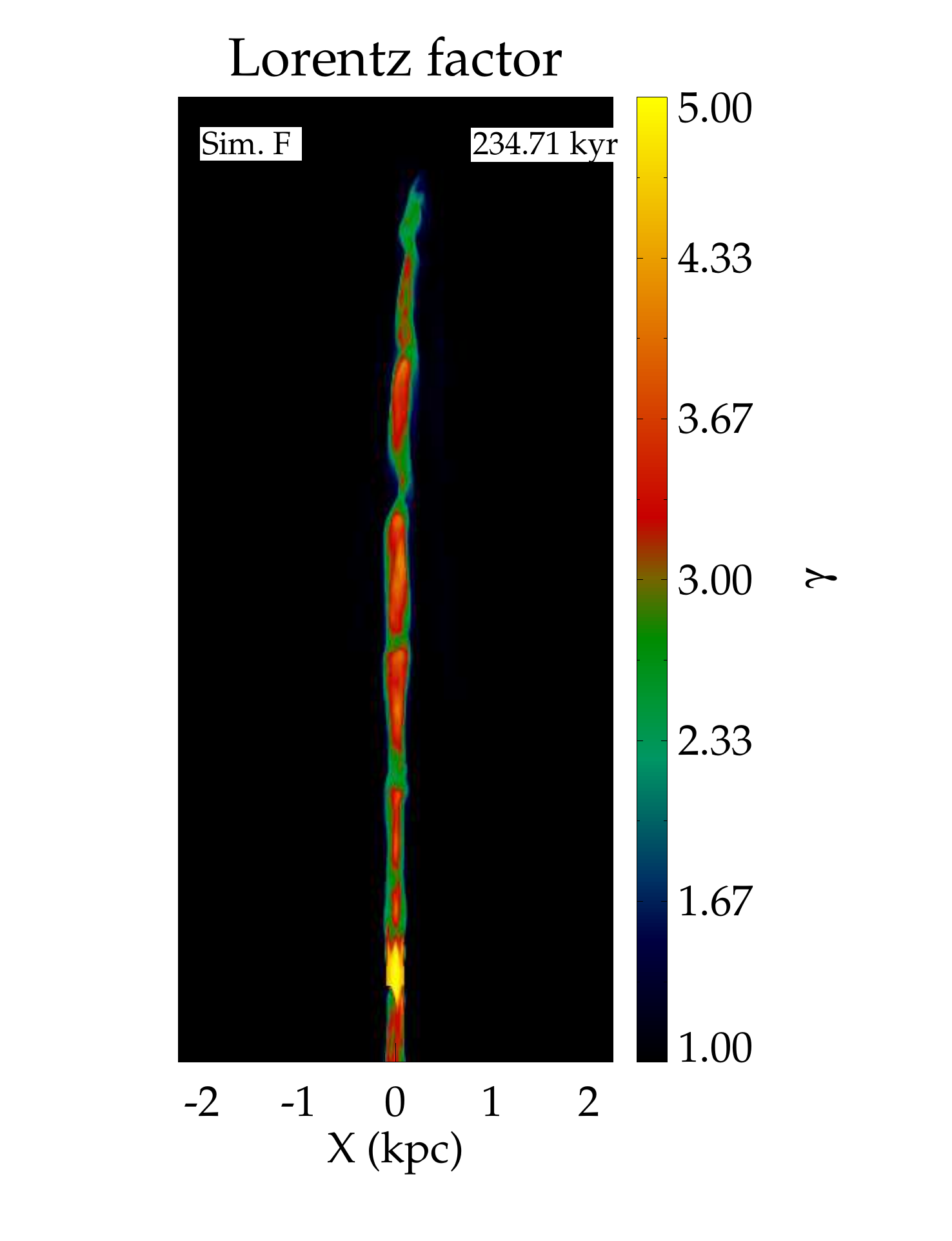}\vspace{-0cm}\hspace{-0.8cm}
	\caption{\small Plots of the Lorentz factor in the $X-Z$ plane for simulations D and F. Simulation D shows deceleration at the top with irregular distribution of flow implying onset of decollimation. Simulation F shows a steady cylindrical spine along the $Z$.}
	\label{fig.gammacompare}
\end{figure}
Simulations D,E,F have moderate jet powers of $\sim 10^{45} \ergs$,  Lorentz gamma of $\gamma \sim 5$ but differing jet magnetisation with $\sigma_B = 0.01, 0.05, 0.1$. These jets do not show strong growth of kink modes within the simulation run times, as was seen for lower power jets. Simulation E shows mild bending away from the axis (as shown in Fig.~\ref{fig.visitimgs2}), but much less pronounced as compared to simulation B. Simulation D however, shows intermittent turbulent distribution of magnetic field  resulting from the development of small scale  Kelvin-Helmholtz (KH) instabilities  at the jet-cocoon interface. These instabilities develop over small scales and are absent in simulation F with higher magnetisation. The higher strength of the toroidal magnetic field prevents deformation of the inner jet spine through the increased magnetic tension  and suppresses the disruptive KH modes \citep{mignone10a,bodo13a}.

In Fig.~\ref{fig.sigcompare} we show the magnitude of the magnetic field normalised to its mean value, for simulations D and F, and their corresponding density slices. Firstly we notice that simulation D has a much wider cocoon, with an asymmetrical head. The development of KH modes results in a stronger deceleration of the jet head, as is evident from a comparison of the times at which the two jets reach a similar length ($t=391.18$ kyr for case D compared to $t=234.71$ kyr for case F). The cocoon in case D had therefore a longer time to expand in the lateral direction.  Simulation D shows onset of deceleration beyond $\sim 6$ kpc with irregular flow axis, as seen in plots of the Lorentz factor in Fig.~\ref{fig.gammacompare}. In simulation F the jet remains collimated with a regular cylindrical axis as seen in the plots of the Lorentz factor and density. The Lorentz factor shows intermediate dips following recollimation shocks whose locations are also seen in the density images in Fig.~\ref{fig.sigcompare}.

Both the magnetic field and density plots show more structures varying over smaller scales for simulation D than those in simulation F. Simulation F shows a distinct spine along its axis with enhanced magnetic field, accentuated by islands from recollimation shocks. Simulation D lacks such a clear morphology, with the  magnetic field near the jet spine being more turbulent. The field in the cocoon of simulation D shows intermittent structures over small scales, whereas simulation F has fields ordered over longer scales. 

\begin{figure*}
	\centering
	\includegraphics[width = 6.2cm, keepaspectratio] 
	{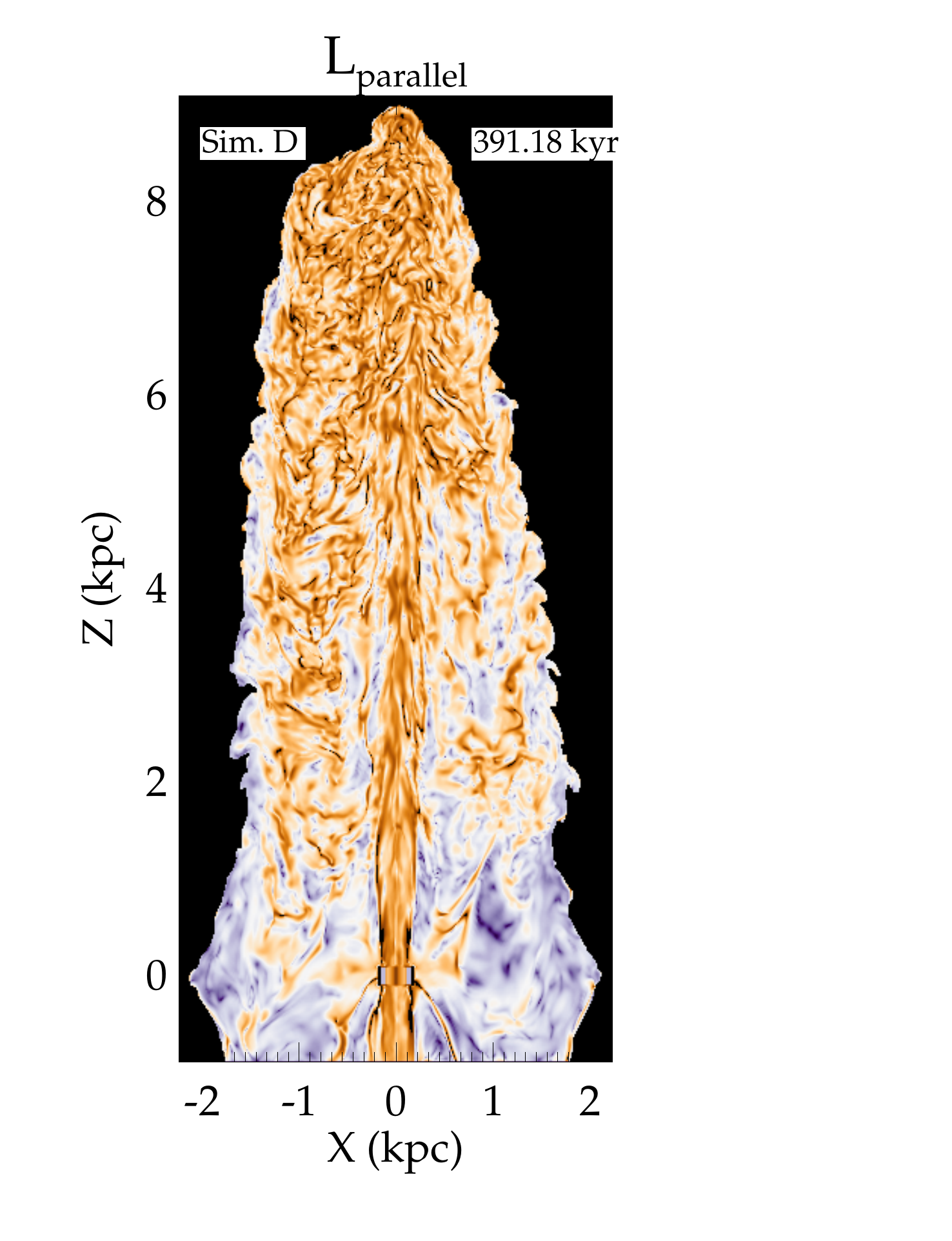}\vspace{-0cm}\hspace{-3.4cm}
	\includegraphics[width = 6.2cm, keepaspectratio] 
	{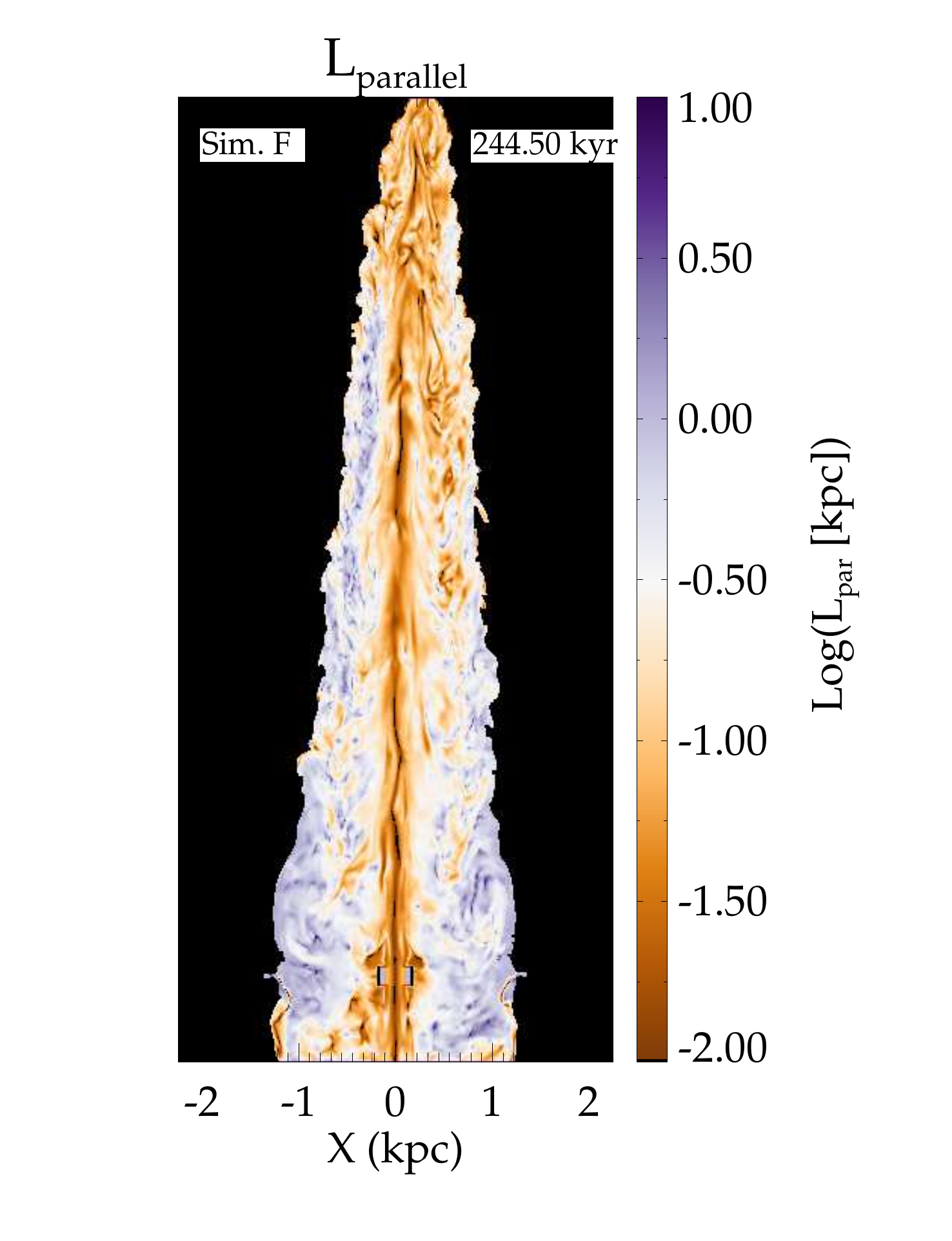}\vspace{-0cm}
	\includegraphics[width = 8 cm, keepaspectratio] 
	{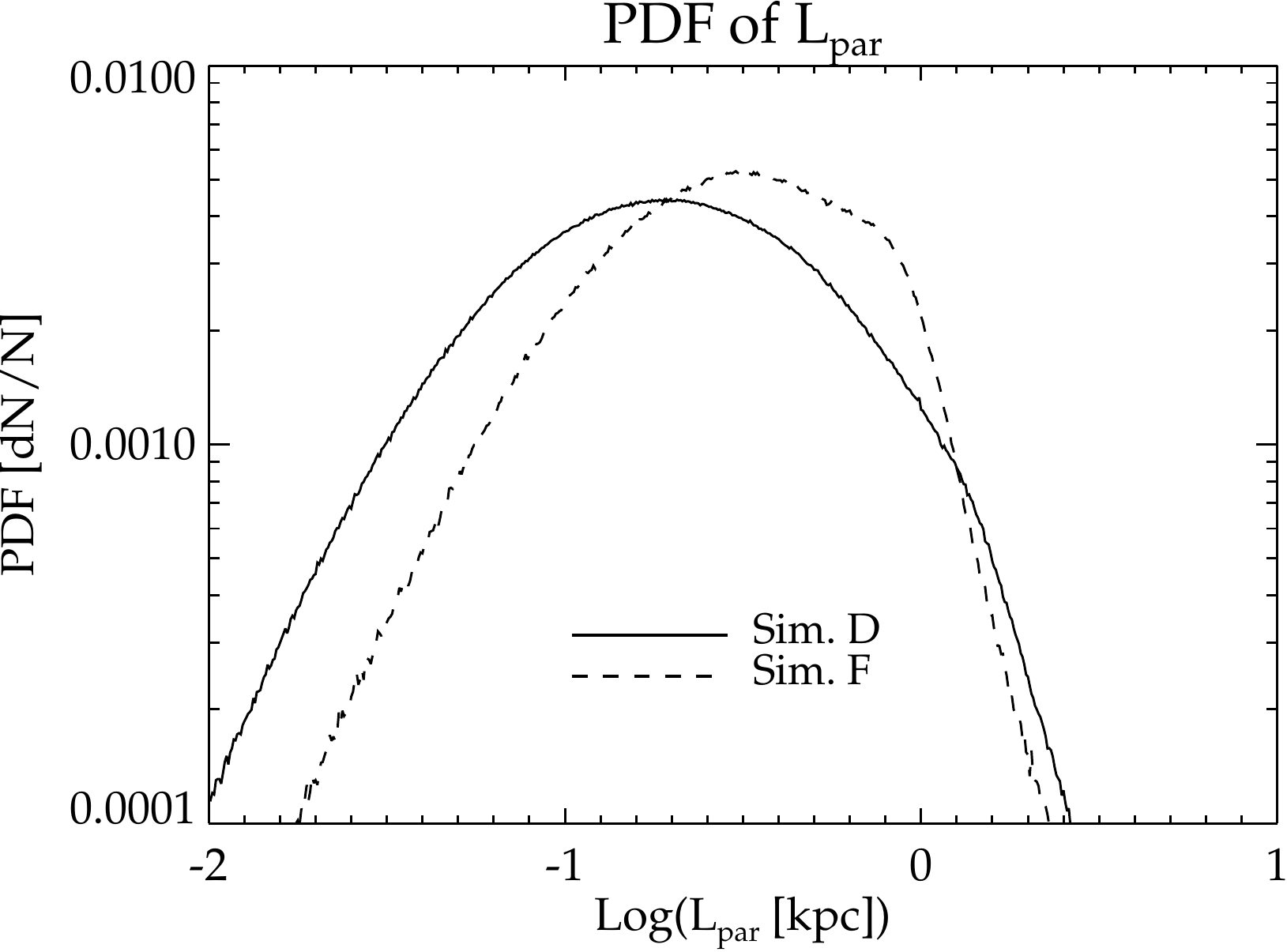}
	\caption{\small \textbf{Left:} 2-D slices in the $X-Z$ plane showing the length scale parallel to the magnetic field (eq.~\ref{eq.lpar}) for simulation D and F. The quantity plotted is $\log(l_\parallel/(1 \kpc))$. Simulation D with smaller magnetisation is dominated by smaller scale lengths. \textbf{Right:} The distribution (PDF) of length scales in the cocoon after excluding regions with jet tracer $>0.9$ and $z < 1$ kpc. }
	\label{fig.lpar}
\end{figure*}
KH instabilities result in the growth of unstable modes at different spatial scales with the shorter wavelengths having faster growth rates. This is demonstrated in Fig.~\ref{fig.lpar} where we plot the length scales parallel to the magnetic field defined as \citep{schekochihin04a,bodo11a}:
\begin{equation}
l_\parallel = \left\lbrack \frac{|\textbf{B}|^4}{|(\textbf{B}\cdot\nabla)\textbf{B}|^2} \right\rbrack^{1/2} \label{eq.lpar}
\end{equation}
The two left panels of Fig.~\ref{fig.lpar} show the distribution of $\log(l_\parallel/(1 \kpc))$ in the X-Z plane for simulations D and F.  The cocoon and jet-axis of simulation D is seen to be dominated by small length scales of $\sim 10-100$ pc or $\sim \Delta x - 10\Delta x$, $\Delta x$ being the grid resolution, which for our simulations is $\sim 15.6$ pc. For simulation F the jet-axis and jet-head have smaller length scales, whereas the cocoon has ordered fields with typical length scales $\gtrsim 1$ kpc. Since simulation F does not suffer from KH modes, the backflow has well ordered magnetic fields. The smaller length scales inside the jet-axis likely arise from recollimation shocks at different intervals from the injection region. 

In the right panel of Fig.~\ref{fig.lpar} is the volume weighted probability distribution function (PDF) of the length scales computed from eq.~\ref{eq.lpar}. The PDF excludes the jet axis, defined as regions with jet tracer $> 0.9$; and also excludes regions with $z<1$ kpc to remove artefacts that may arise from the lower-boundary. It can be seen that simulation D has a higher value of the PDF for length scales $\lesssim 100$ pc. The PDF of simulation F is higher for length scales $100 \mbox{ pc} < l_\parallel < 1$ kpc. The fractional volume occupied by length scales in the range $\Delta x - 10 \Delta x$ is $\sim 0.42$ for simulation D and $\sim 0.24$ for simulation F, whereas for $l_\parallel$ in the range $10 \Delta x - 100 \Delta x$ simulation D has $\sim 0.56$ by volume and simulation F has contributions from $\sim 0.74$ of the volume. Thus regions with small scale fields dominate the unstable simulation D by over 2 times in terms of relative fraction of the total volume of the cocoon as compared to the stable simulation F.

\begin{figure}
	\centering
	\includegraphics[width = 6.2cm, keepaspectratio] 
	{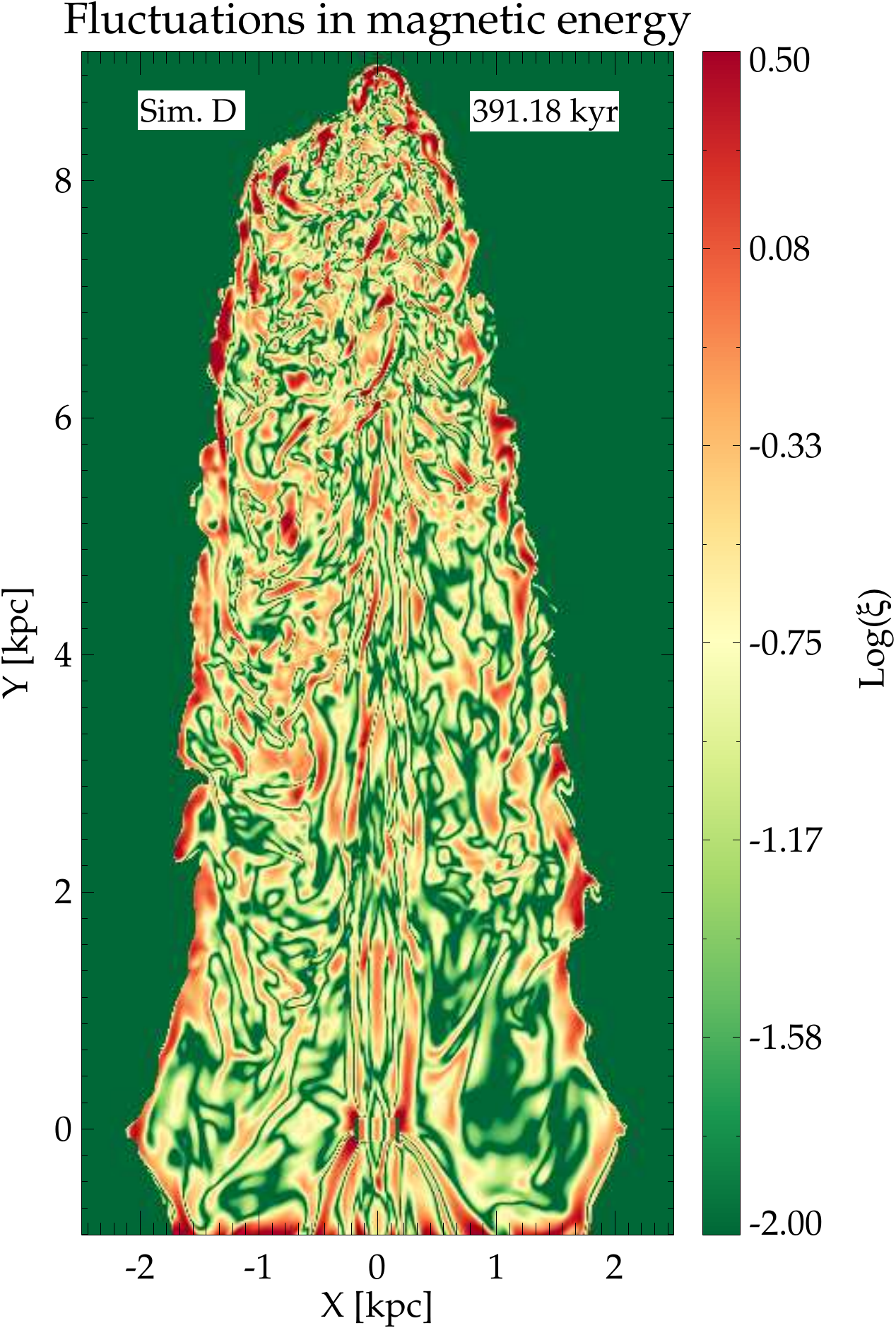}
	\caption{\small A plot of $\xi$ (eq.~\ref{eq.xi}), showing the fluctuating magnetic field energy density on varying intermittently on short length scales ($\sim 100$ pc) in the cocoon of simulation D.  }
	\label{fig.bavg}
\end{figure}
 To further show the developement of small scale intermittent magnetic field distribution in the cocoon of simulation D due to the onset of KH instabilities, we present in Fig.~\ref{fig.bavg} the plot of the relative strength of the fluctuating magnetic field energy density. We define this as:
\begin{align}
&\xi = \frac{(B - \bar{B})^2}{\bar{B}^2}, \quad \mbox{ where:} \nonumber\\
&\bar{B}(x,y,z) = \int \int \int G(x,x',y,y',z,z')  B(x',y',z') dx'dy' dz',  \nonumber \\
&G(x,x',y,y',z,z') = \frac{1}{(2\pi)^{3/2} (2\sigma_K)^2} \exp\left(\frac{ -\sum_{i=1}^3 (x_i-x_i^\prime)^2}{2 (\sigma_K)^2} \right) . \label{eq.xi}
\end{align}
Here $\bar{B}$ is the local average magnetic field computed by a convolving the local field with a Gaussian kernel with a width ($\sigma_K$) equal to the diameter of the jet ($\sigma_K = 2r_j$). The indices $i$ in eq.~\ref{eq.xi} refer to the three spatial dimensions ($x,y,z$). We see that the energy density of the fluctuating component of the field varies over small length scales, as also demonstrated earlier in Fig.~\ref{fig.lpar}. In certain areas the fluctuating fields are a few times stronger than the local mean.

\subsubsection{High power jets}\label{sec.highpower}
\begin{figure*}
	\centering
	\includegraphics[width = 16cm, keepaspectratio] 
	{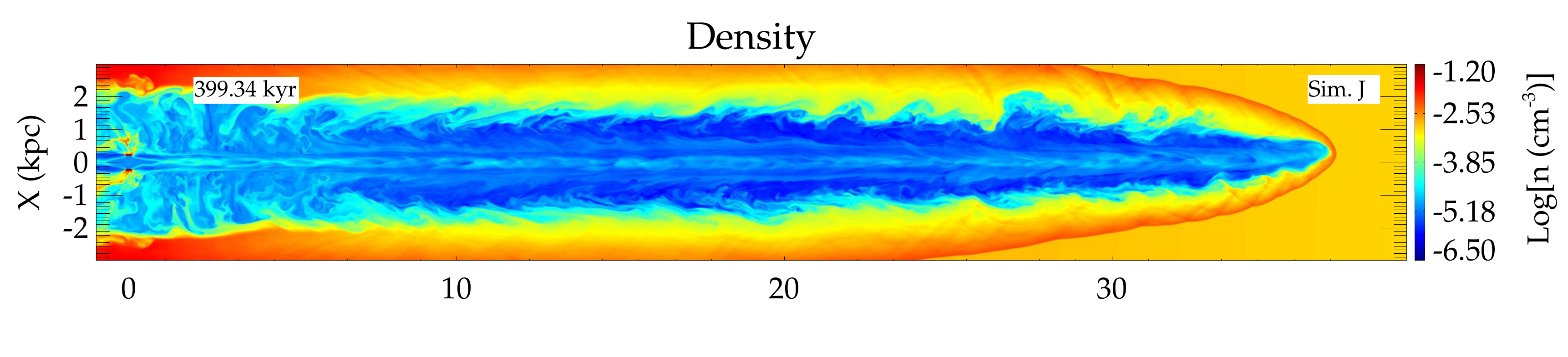}\vspace{-0.5cm}
	\includegraphics[width = 16cm, keepaspectratio] 
	{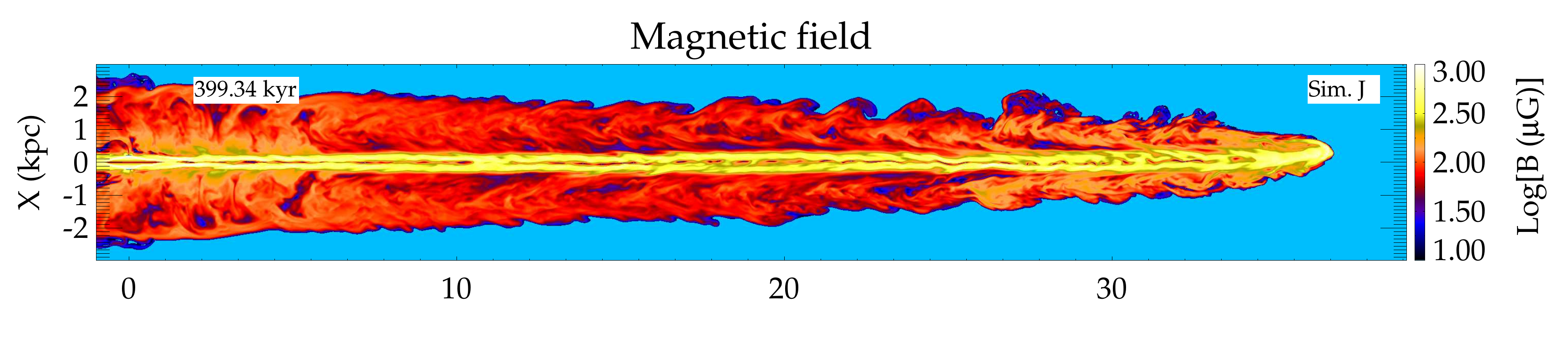}\vspace{-0.5cm}
	\includegraphics[width = 16cm, keepaspectratio] 
	{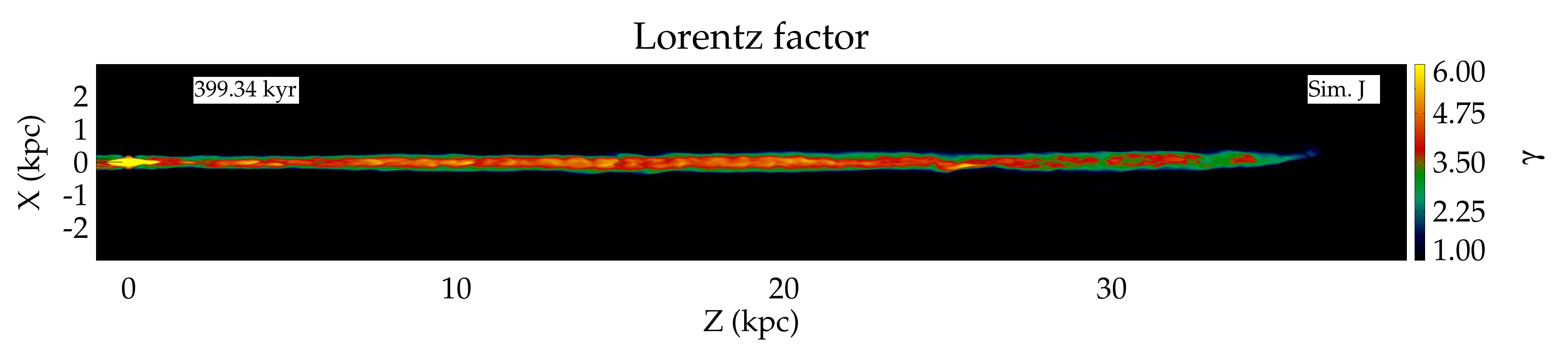}\vspace{-0cm}
	\caption{\small The density (top), magnetic field (middle) and Lorentz factor in the $X-Z$ plane for simulation J. The jet with power $\sim 10^{46}\ergs$ and initial Lorentz factor of $\gamma = 10$ at injection remains fairly stable up to $\sim 40$ kpc. }
	\label{fig.simJ}
\end{figure*}
Simulations G,H,I,J have higher jet powers $\sim 10^{46} \ergs$, with higher Lorentz gamma $\gamma \in (5-10)$. These simulations do not show strong growth of unstable modes as found earlier. Jets in simulations H and J were launched with higher velocity ($\gamma = 10$) and comparable magnetisation ($\sigma_B=0.1,0.2$ respectively) to that of simulation F. Similar to F, the jets evolve without any appreciable onset of instability. Simulation J was followed up to $\sim 40$ kpc and was found to be stable with a collimated spine, as shown in Fig.~\ref{fig.simJ}. The difference in magnetisation between simulations H and J did not have any significant qualitative difference. The absence of instabilities likely results from slower growth rates of instabilties in jets with higher Lorentz factors \citep{rosen99a,bodo13a}, which is discussed in more detail later in Sec.~\ref{sec.unstablemodes}.

\begin{figure}
	\centering
\hspace{-0.8cm}
	\includegraphics[width = 6cm, keepaspectratio] 
	{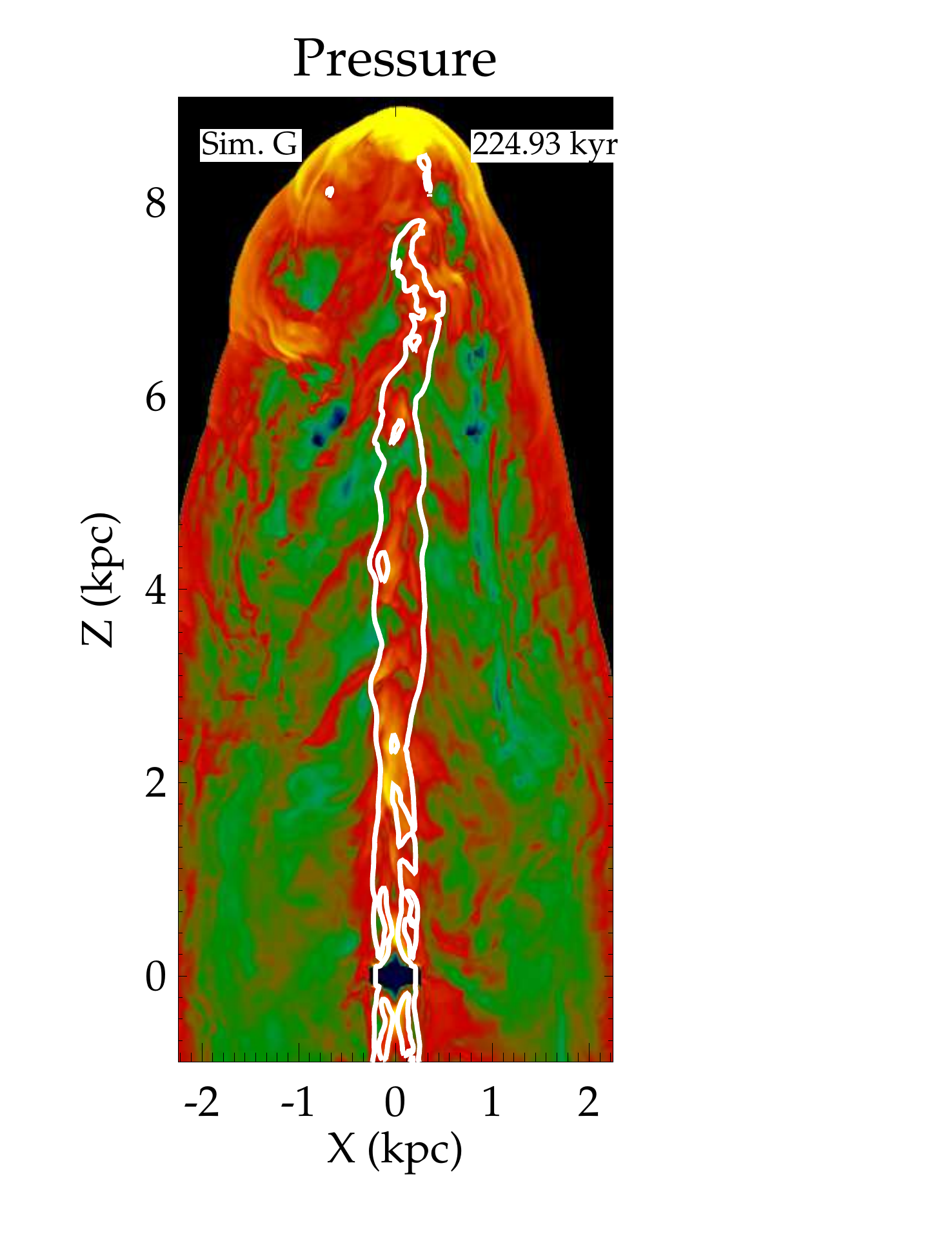}\vspace{-0cm}\hspace{-3.25cm}
	\includegraphics[width = 6cm, keepaspectratio] 
	{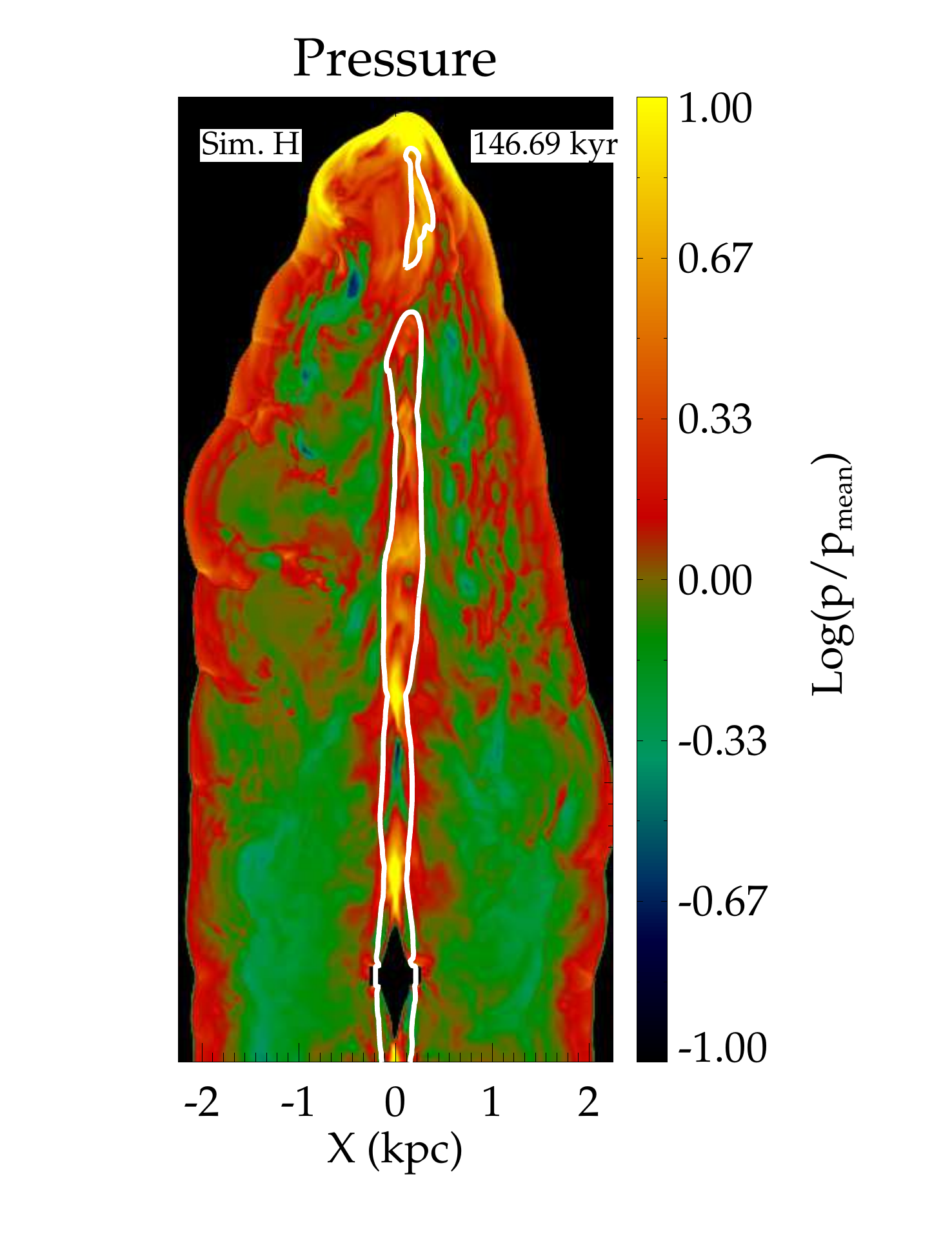}\vspace{-0cm}	\caption{\small Plots of the pressure normalised to its mean value in the $X-Z$ plane for simulations G and H. The white contours denote constant Lorentz factor with a value $\gamma=2$. Simulation G with a higher initial pressure but lower Lorentz factor has irregular jet axis (traced by the $\gamma=2$ contour), bending of the jet and more pronounced internal structures, implying faster growth of unstable modes (Kelvn-Helmholtz). Simulation H has a more regular jet axis and cocoon than that in G.}
	\label{fig.meanpres}
\end{figure}
Simulation G, which has a hotter jet with an initial pressure 5 times that of H (see Table~\ref{tab.sims}) shows some added structures and shear of the jet axis, and bending of the jet head, than in simulation H, as shown in Fig.~\ref{fig.meanpres}. This is similar to the results of \citet{rosen99a}, where hotter jets were found to have more structures due to faster growth rates of unstable modes. However, these are not as disruptive as in the low power jets. Simulation I was carried out in an ambient medium with a central density of $n_0 = 1 \cc$, 10 times the value of other simulations. However, within the domain of our simulation we did not see any appreciable deceleration compared to simulations G and H.

\subsection{The Generalised Begelman-Cioffi (GBC) model}\label{sec.GBCmodel}
There are several approximate analytical models that describe the evolution of the jet as a function of time or radius \citep{begelman89a,falle91a,kaiser97a,turner15a,perucho11a,bromberg11a,harrison18a}. One of the commonly used models was derived by \citet{begelman89a} where the time evolution of the jet length and mean cocoon pressure of a jet propagating into a homogeneous environment of constant density was derived. The solutions do not necessarily assume a self-similar evolution of the jet, which is often considered as a fundamental assumption in several analytical models \begin{NoHyper}\citep[e.g.][]{falle91a,kaiser97a,turner15a}\end{NoHyper}. Later works \citep{sheck02a,perucho07a} extended the Begelman-Cioffi model to account for a jet that steadily decelerates while expanding into an external medium whose density decreases as a power-law. In other works, \citet{bromberg11a} and \citet{harrison18a} have developed a semi-analytical model of the jet evolution by duly accounting for the structure of the recollimation shock that shapes the jet radius. However, the possible deceleration of the jet due to MHD instabilities were not accounted for. The effect of kink mode instabilities on the dyanamics of highly magnetised jets have been studied in \citet{bromberg16a} and \citet{tchekhovskoy16a}, an extension of the semi-analytic results of \citet{bromberg11a}. However, the jet magnetisations in the simulations presented in this work are much lower than those in \citet{bromberg16a}.

In this section we present a more generalised formulation of the Begelman Cioffi model (hereafter GBC), to compare with the results from the numerical simulations. We assume a simplified model of a jet evolution by evaluating the jet-head velocity following momentum flux balance. We consider a deceleration factor to account for the effect of MHD instabilities. The detailed derivations of the equations are outlined in Appendix~\ref{append.gbc}. We compare the approximate analytical results with the jet dynamics from the simulations by evaluating advance speed of the jet head. The model is simplistic in nature, although an update on the original \citet{begelman89a}. We do not consider the detailed nature of the recollimation shock structure, as done in \citet{bromberg11a}. Instead, we focus on matching the bulk energetics to approximately model the evolution of the cocoon and jet, which may be a better approach for jets with complicated morphologies resulting from 3D MHD instabilities.

\begin{figure}[!h]
	\centering
	\includegraphics[width = 7 cm, keepaspectratio] 
	{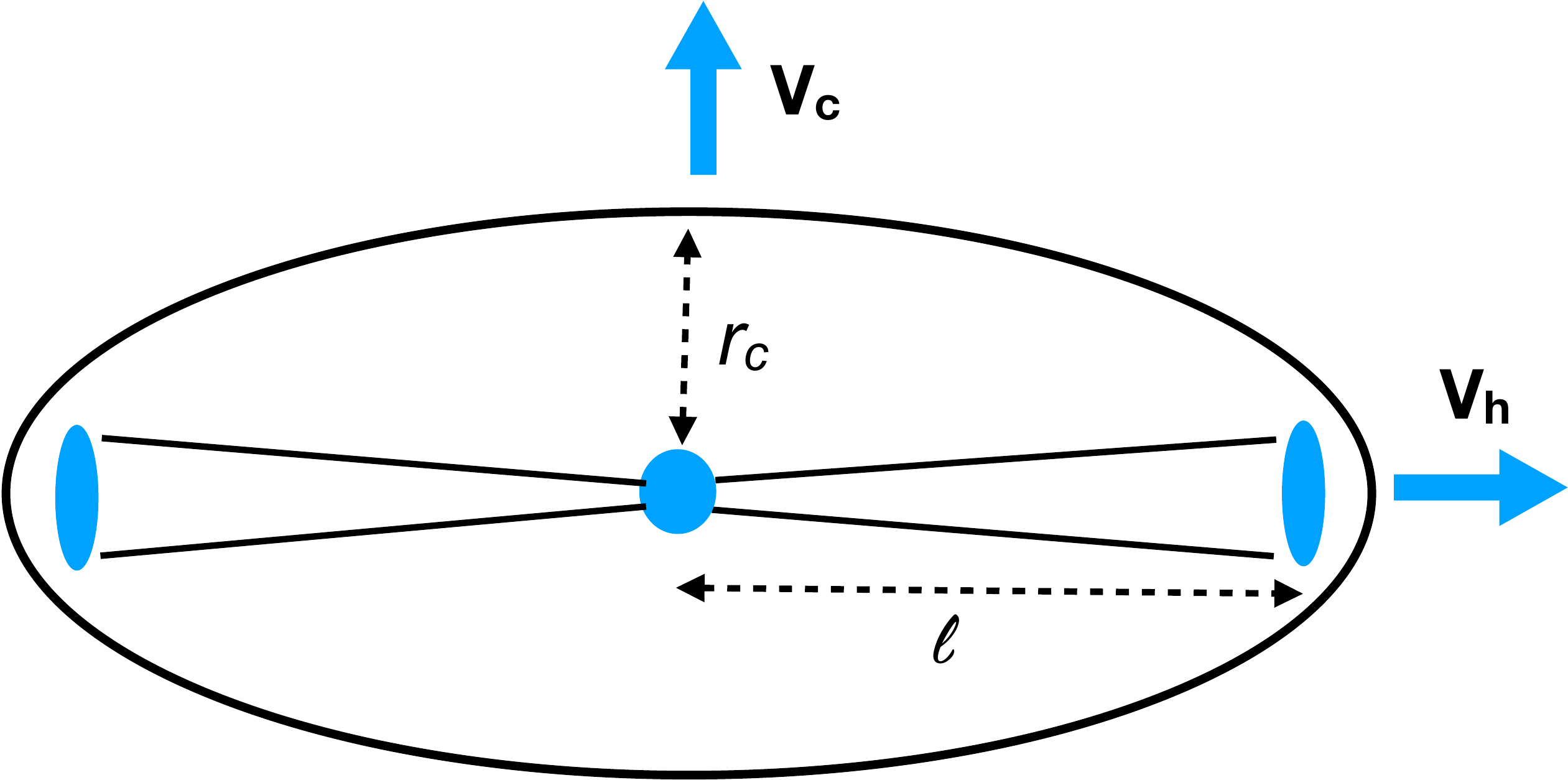}
	\caption{\small A schematic figure of a jet with an ellipsoidal cocoon whose evolution for the Generalised Begelman-Cioffi model discussed in Appendix~\ref{append.gbc} and Sec.~\ref{sec.GBCmodel}. The jet head, at a distance $l$, advances along the jet axis with speed $v_h$. The cocoon expands laterally in the transverse direction with speed $v_c$. The length of the cocoon along the semi-minor axis is considered to be the cocoon length $r_c$. }
	\label{fig.gbcschematic}
\end{figure}
By equating the (relativistic) momentum flux of the jet and the ambient medium, the advance speed of the jet ($v_h$) at the bow shock, can be expressed in terms of the pre-shock speed and density contrast with the ambient medium as \citep{marti97a,bromberg11a}:
\begin{equation}
v_h^M = \frac{\gamma_j\sqrt{\eta_R}}{1 + \gamma_j\sqrt{\eta_R}} v_j, \quad \eta_R = \frac{\rho_j h_j}{\rho_a h_a} \label{eq.vhmarti}.
\end{equation} 
Here $\eta_R$ is the ratio of the relativistic enthalpy of the jet with respect to the ambient medium. Assuming an ideal equation of state with adiabatic index $\Gamma$ for simplicity, the enthalpy of the ambient gas is
\begin{equation}
\rho_a h_a = \rho_a c^2 \left\lbrack 1 + \frac{1}{\Gamma -1}\left(\frac{c_{sa}}{c}\right)^2\right\rbrack \simeq \rho_a c^2
\end{equation}
where $c_{sa}$ is the sound speed of the ambient medium, which for $T_a \sim 10^7$ is $c_{sa} \simeq 372 \kms \ll	 c$. Thus
\begin{align}
\eta_R &= \left(\frac{\rho_j}{\rho_a} \right)\left\lbrack 1 + \frac{\Gamma p_j}{(\Gamma -1) \rho_j c^2} \right\rbrack \nonumber \\
       &= \eta_j f(\rbar)^{-1} \left \lbrack 1 + \frac{\Gamma p_j}{(\Gamma -1) \rho_j c^2} \right \rbrack,
\end{align}
where $\eta_j$ is the density contrast of the jet with respect to the ambient medium at $r=0$ (as in Table~\ref{tab.sims}) and $f(\rbar)$ is radial dependence of the ambient density profile. Typically, the density contrast of the jet with the ambient medium is small for light jets. For our simulations, $\eta_j f(\rbar)^{-1} \lesssim 2.8\times10^{-3}$ for $r < 10 \kpc$. Thus the jet-head velocity can be approximated as 
\begin{align}
v_h^M &\sim \gamma_j\sqrt{\eta_R} v_j \nonumber\\
    &= \gamma_j v_j \eta_j^{1/2} f(\rbar)^{-1/2} \left\lbrack 1 + \frac{\Gamma p_j}{(\Gamma -1) \rho_j c^2} \right\rbrack^{1/2} \label{eq.vh1} 
  \end{align}
From eq.~\ref{eq.vh1} it is evident that for a jet propagating into a medium with a decreasing density profile, the jet head velocity may increase with time for a constant pre-shock jet velocity. However, at large radii, the jet density may become comparable to the ambient density, in which case the above approximation of $\eta_j f(\rbar)^{-1} \ll 1$ is no longer valid, and the jet will propagate with a constant speed as $v_h^M \simeq v_j$.

The time evolution of the jet head can be found by integrating eq.~\ref{eq.vh1}. However, additional factors such as MHD instabilities or broadening of the hotspot area can lower the jet speed with time. We thus consider the actual jet head velocity to be modified by a deceleration factor $g(\tbar)$, $\tbar = t/\tau$ with $\tau$ a scale deceleration time, which accounts for a secular reduction in the advance speed of the jet with time. 

Thus the jet will evolve as 
\begin{equation}
v_h = \frac{dl}{dt}  = \frac{v_h^M}{\left(1+\frac{t}{\tau}\right)^n}, \label{eq.vh2}
\end{equation}
such that $v_h \simeq v_h^M$ for $t \ll \tau$ (no deceleration) and $v_h \simeq v_h^M t^{1-n}$ for $t \gg \tau$. For the above assumptions, eq.~\ref{eq.vh1} can be integrated under various limits to find the time evolution of the jet head (eq.~\ref{eq.lbar1}--eq.~\ref{eq.lbar4} in Appendix~\ref{append.gbc}).

The energy from the jet is spread over the entire cocoon, which tends to have nearly homogeneous pressure (as seen in Fig.~\ref{fig.simG}), except for the jet head which has values higher by more than an order of magnitude than the mean cocoon pressure. Assuming kinetic energy of the motions inside the cocoon from backflows and turbulence to be sub-dominant as compared to the thermal energy, the mean pressure   ($p_c$) of an ellipsoidal cocoon (see Fig.~\ref{fig.gbcschematic}) can be expressed in terms of the total energy injected by the jet up to a given time as
\begin{equation}
p_c = (\Gamma - 1) \frac{P_j t}{(4/3) \pi a^3 \rbar^2_c \lbar} \label{eq.pc0}
\end{equation}
where the cocoon radius ($r_c$) and jet length ($l$) have been normalised to the density scale length $a$. The rate of expansion of the cocoon radius ($v_c = dr_c/dt$) can be then obtained by equating the ram pressure experienced by the ambient medium to the cocoon pressure $p_c = \rho_a(\rcbar) v_c^2$ (as in eq.~\ref{eq.pc0b}). The mean pressure of the cocoon can then be derived for different limits of $l/a$ and $t/\tau$ as presented in eq.~\ref{eq.pc1} -- eq.~\ref{eq.pc4}.

\subsection{Comparison with GBC model}\label{sec.GBCcompare}
\subsubsection{Jet length and morphology}
\begin{figure}
	\centering
	\includegraphics[width = 6. cm, keepaspectratio] 
	{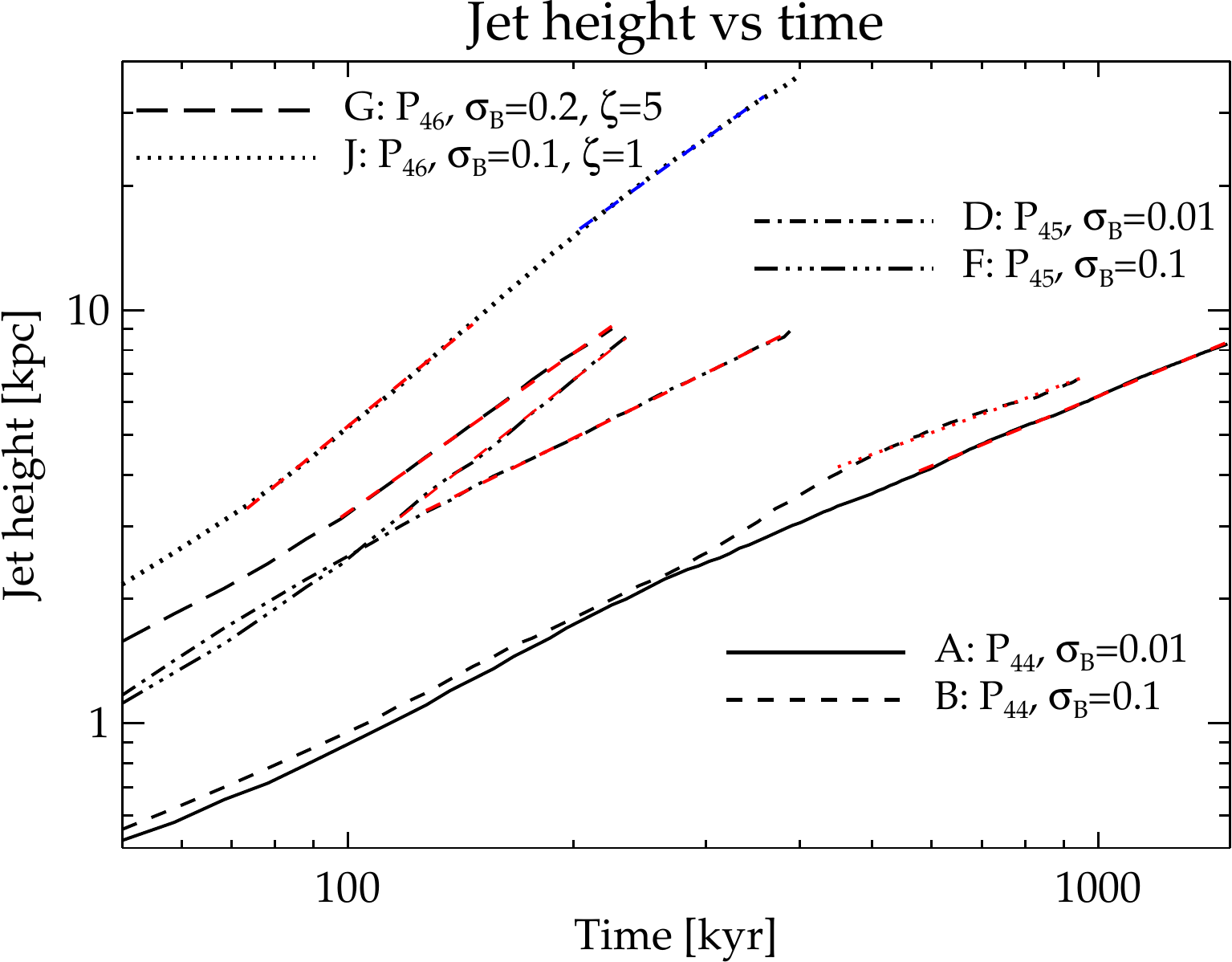}
	\includegraphics[width = 6.1 cm, keepaspectratio] 
	{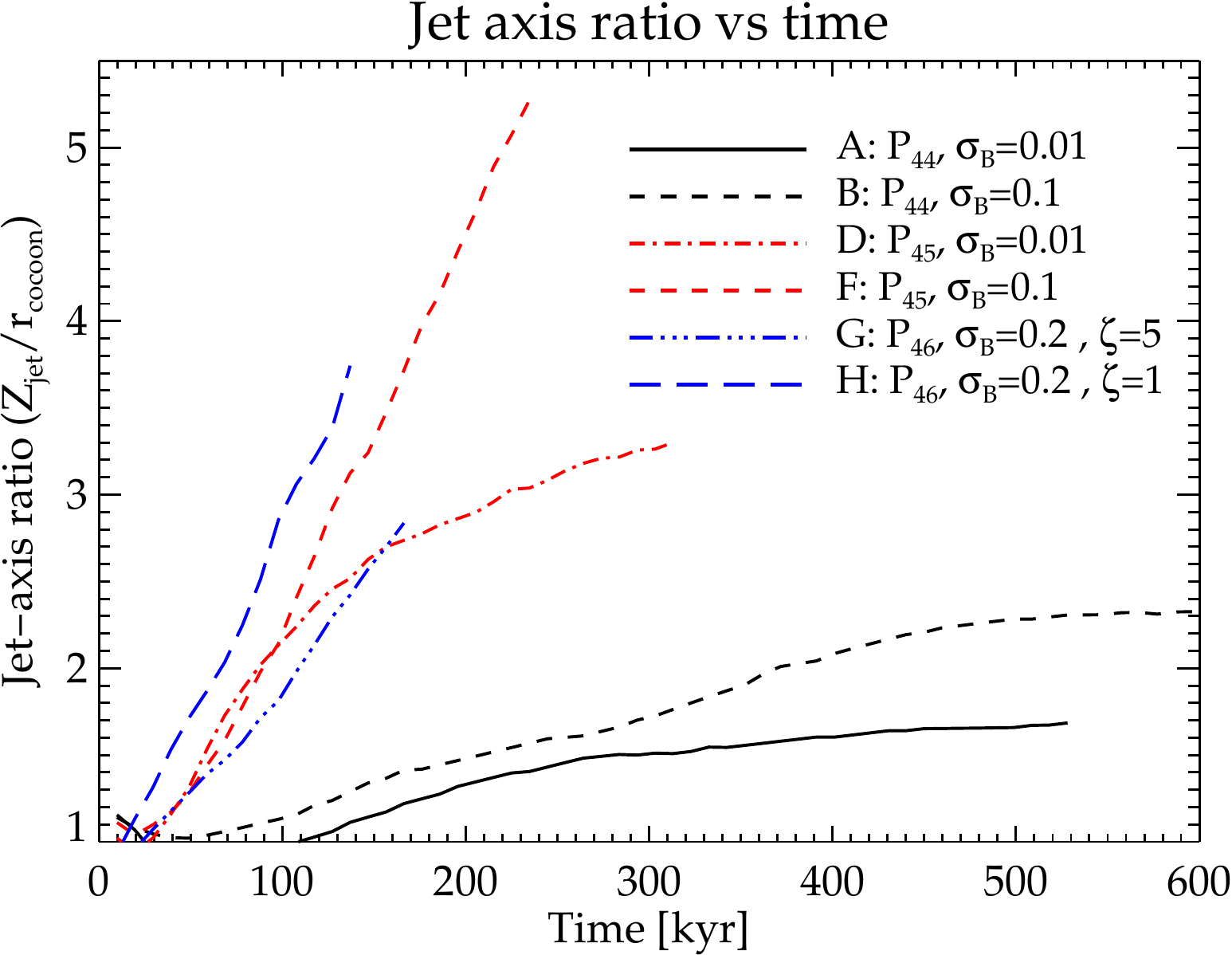}
	\includegraphics[width = 6 cm, keepaspectratio] 
	{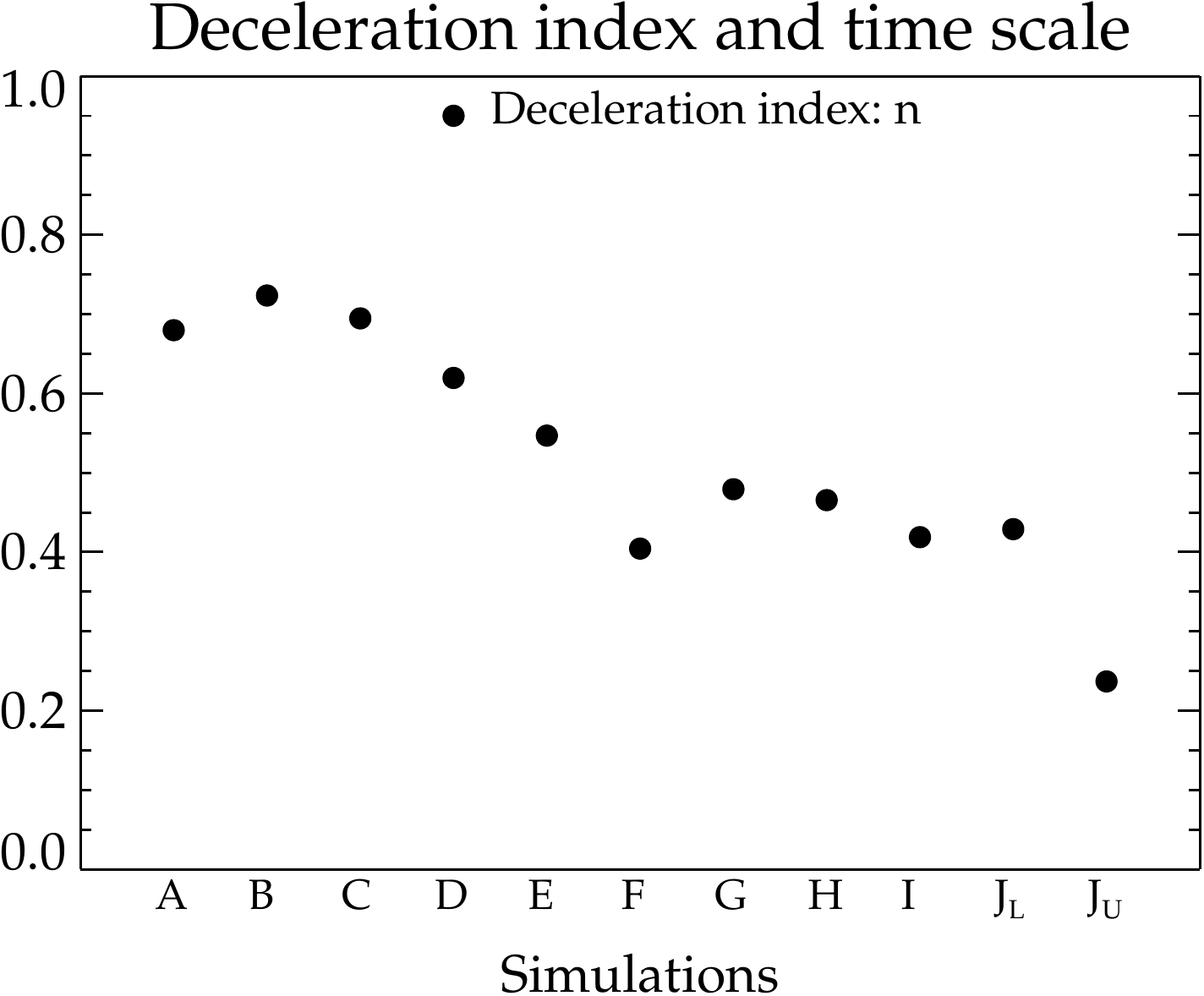}
	\caption{\small \textbf{Top}: Evolution of jet height with time for some representative simulations. The red dashed line overplotted shows the power-law fit function (see Sec.~\ref{sec.GBCcompare}). For the simulation J, the blue line denotes the fit function with  $\alpha=0.829$ for the jet's evolution beyond 15 kpc (as in eq.~\ref{eq.frho3}). See Table~\ref{tab.sims} for detailed description of parameters for different runs. The jet power for $P_j=10^{45}\ergs$ is abbreviated as $P_{45}$ and so forth. \textbf{Middle:} Plot of the axis ratio ($l/r_c$) with time for some  simulations. The axial length of the cocoon is computed from eq.~\ref{eq.rc}. \textbf{Bottom:} The deceleration coefficient evaluated from eq.~\ref{eq.lbar4} using the results of the fit function in the top panel. For simulation J fits to heights $\lesssim 10$ kpc and $\gtrsim 15$ kpc have been presented separately as J$_{\rm L}$ and J$_{\rm U}$.}
	\label{fig.jetheight}
\end{figure}
From the simulations we compute the maximum length of the jet as a function of time. In the top panel of fig.~\ref{fig.jetheight} we present the evolution of the jet height for some representative simulations. The jet length beyond 2 kpc was fit with a function power-law in time. From the fit parameters we derive the deceleration index $n$ and the deceleration time scale $\tau$ in eq.~\ref{eq.L0bar} and eq.~\ref{eq.lbar4} given in Appendix~\ref{append.gbc}. 

In the middle panel of Fig.~\ref{fig.jetheight} we present the axis ratio defined as the ratio of jet length ($l$) to effective lateral radius $r_c$ computed from
\begin{equation}
r_c = \left(\frac{3fV_c}{4\pi l_j}\right)^{1/2} \label{eq.rc}.
\end{equation}
Here $V_c$ is the volume of the cocoon, computed from the simulations by summing the volume with jet tracer $> 10^{-7}$. The factor $f$ has a value $f=2$ for simulations with half-sided jets injected close to the lower boundary. For simulation E where both lobes of the jets are followed, the value is $f=1$. The radius $r_c$ represents the lateral radius of an ellipsoid with the volume of the cocoon, which is a close approximation to the shape of the cocoon. From the time evolution of the axis ratio we find that for jets of power $\gtrsim 10^{46}\ergs$ the axis ratio steadily increases with time due to the faster expansion along the jet axis as compared with the lateral extent. 

For simulations showing instabilities however (simulations A, B and D), the rate of increase of the axis ratio slows down with time. For simulations A and B, the axis ratio is nearly steady with time, indicating an approximate self-similar evolution of the cocoon. This is also supported by the deceleration index being close to $\sim 0.67$, for which the GBC predicts a self-similar expansion of the jet (for $\alpha = 1.166$), as explained at the end of Appendix~\ref{append.gbc}. The jets showing onset of instabilities have a slower advance speed and the bending of jet-head results in a more uniform spread of the energy in the cocoon. This results in an approximate self-similar expansion of the cocoon \citep{komissarov98a,sheck02a,perucho19a}.

\begin{figure}
	\centering
	\includegraphics[width = 6.5 cm, keepaspectratio] 
	{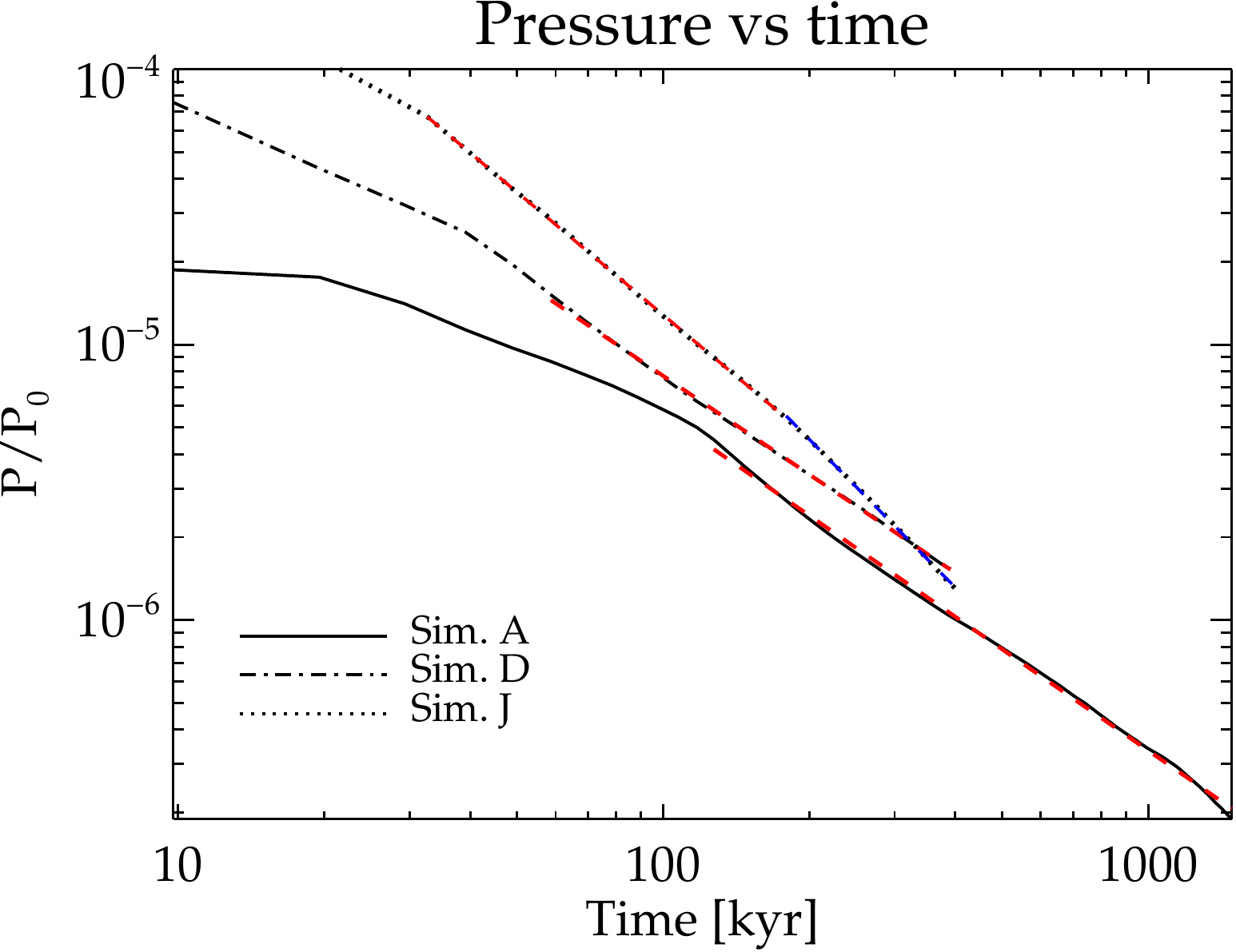}
	\includegraphics[width = 6.5 cm, keepaspectratio] 
	{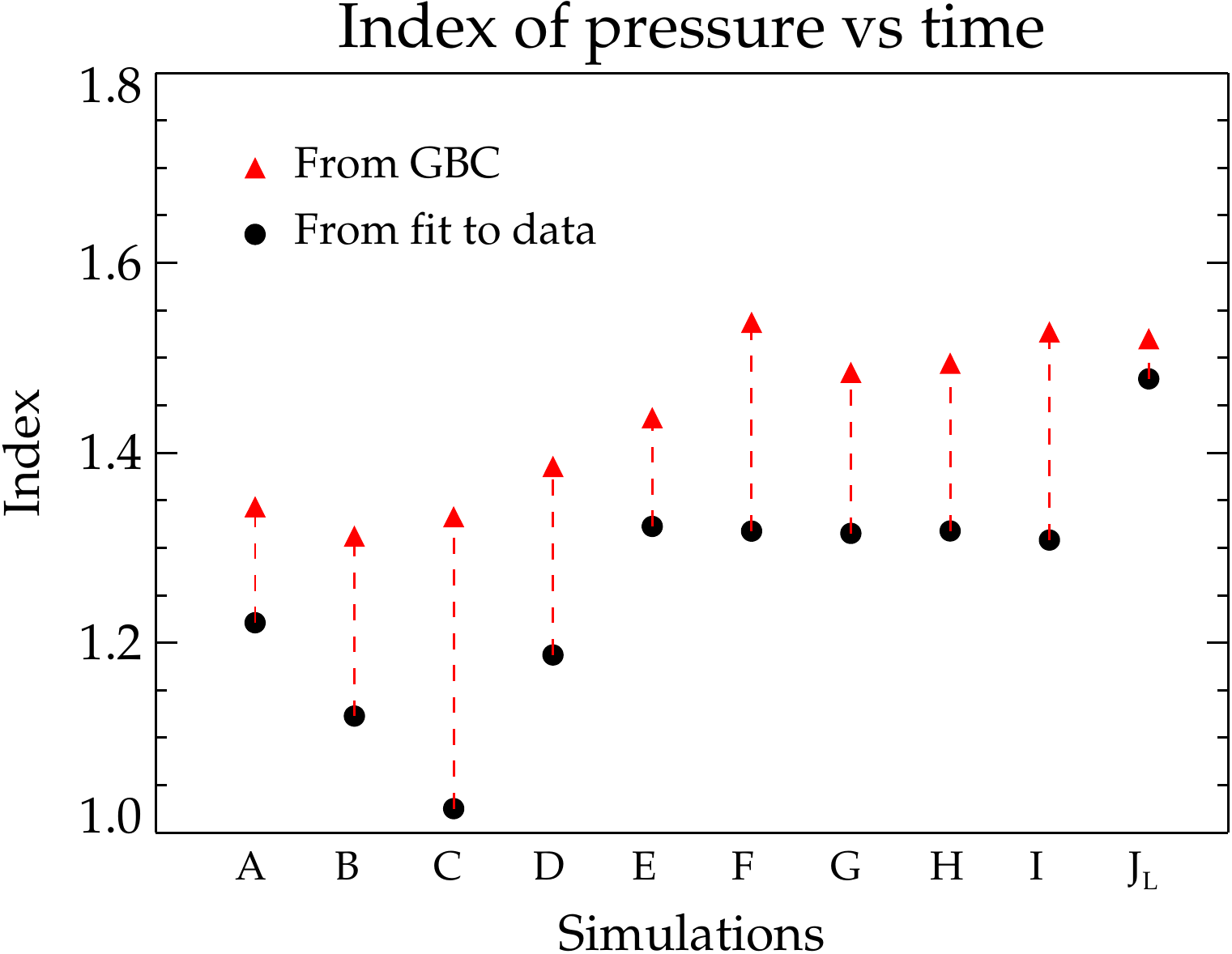}
	\caption{\small \textbf{Top}: Evolution of mean pressure in the cocoon with time for some simulations. The red lines show power-law fits to the pressure evolution. The blue line shows a fit to simulation J beyond a height of 15 kpc. \textbf{Bottom:} Comparison of the index of a power-law fit to the time evolution of the pressure with that predicted from GBC model (eq.~\ref{eq.pc4}).}
	\label{fig.prescomp}
\end{figure}
In the last panel of Fig.~\ref{fig.jetheight} we present the deceleration index $n$ derived from the fit coefficients. Low power jets and jets with lower magnetisation, which are more susceptible to instabilities (simulations A--D), have a mean deceleration index of $n \sim 0.6$. Faster jets which are not affected by instabilities have a lower deceleration index $n \sim 0.4$. The deceleration index and time scales obtained from the fit coefficients have been presented in Table~\ref{tab.gbcindex} in the Appendix~\ref{append.gbc}. The deceleration time scales were found to be approximately close to the time when the jet breaks out of the central core of $\sim 2$ kpc, which varies for different simulations depending on the jet advance speed. Stable jets have a slightly a higher value of deceleration time compared to unstable jets. Thus all jets show some deceleration from the onset, the degree of which depends on the jet stability, as inferred from the index.

The mean pressure in the cocoon evolves as a power-law in time at late times, with a slightly shallower slope at the very early times when the jet is just establishing a cocoon on injection. The pressure for some simulations are presented in the top panel of Fig.~\ref{fig.prescomp}. The pressure was fit with a function power-law in, time whose coefficient has then been compared to the value predicted by the GBC model (eq.~\ref{eq.pc4}), using the deceleration index $n$ derived from the fits to the jet length. For most of the simulations the index for the pressure was lower than predictions from GBC model by about $\sim 10-20\%$. Thus this demonstrates that the GBC model, overall, approximates well the expansion of the jet cocoon, although within $\sim 20\%$ margins. A more detailed model based on the momentum balance at the internal shocks as done in \citet{bromberg11a} may provide a closer match. However, given the various other uncertainties arising from complex developement of different MHD instabilities, we find the the present comparison with the simplified assumptions of the GBC model to be reasonable. 

Simulations A--C, with increasing $\sigma_B$, show a progressively poorer match with the theoretical values. This results from the stronger onset of instabilities (kink) with stronger magnetisation of the jet. Similarly, simulation D shows a poorer comparison than F, as D has more enhanced Kelvin-Helmholtz instabilities. Simulations with more stable jets (E--I) show nearly identical value of the exponent, implying that the pressure evolution is not much affected by the deceleration index of the jet. Simulation J shows a very good match for heights lower than $\sim 10$ kpc. At higher heights ($\gtrsim 15$ kpc) the lateral extent of the jet reaches the boundary of the domain with an outflow boundary condition. This makes the comparison of the mean pressure with the analytical models unreliable due to the loss of matter from the outflowing boundary condition; and hence excluded from the analysis. A comparison with the GBC model by evaluating the mean pressure will thus be misleading, and hence not presented here.

\subsubsection{Jet advance speed}
\begin{figure}
	\centering
	\includegraphics[width = 6.5 cm, keepaspectratio] 
	{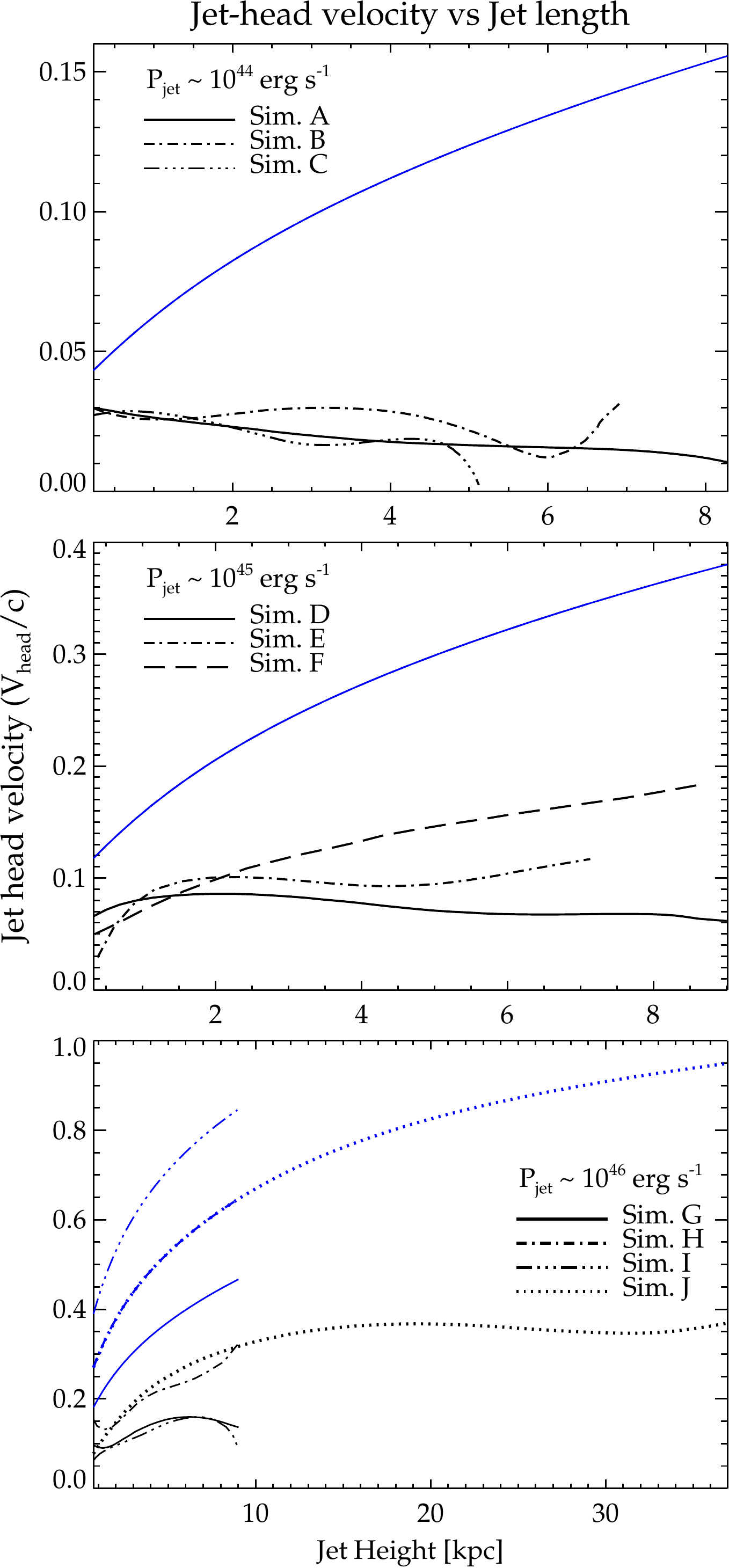}
	\caption{\small Speed of advance of the jet head as a function of jet height (see Sec.~\ref{sec.GBCcompare}). In blue is plotted the maximum velocity expected from a non-decelerating jet following eq.~\ref{eq.vhmarti} \citep[as derived in][]{marti97a}. }
	\label{fig.jetvel}
\end{figure}
In Fig.~\ref{fig.jetvel} we present the speed of advance of the jet head which is obtained by taking the derivative of a $6^{\rm th}$ order polynomial used to fit the evolution of the jet length with time (shown in Fig.~\ref{fig.jetheight}). In blue is plotted the maximum advance speed attainable for a non-decelerating jet following eq.~\ref{eq.vhmarti}. To compute the speed from eq.~\ref{eq.vhmarti} we assumed the jet parameters (velocity, pressure and density) to be the injected values. Firstly, the jet speeds (both theoretical and numerically computed), show an increase with distance. The apparent acceleration results from the jet expanding into a lower density medium that decreases as a power-law with distance beyond the core radius (as shown in eq.~\ref{eq.eqrho}). 

For simulations A, B and C with jet powers $\sim 10^{44} \ergs$ the jet advance speed mildly decreases with distance, being much lower than the maximum attainable value. This arises from the onset of kink like instabilities as discussed earlier in sec.~\ref{sec.kink} which result in strong deceleration of the jet. The jet head wobbles, spreading its energy over a much larger area and hence reducing the advance speed substantially. 

Simulations D and E show similar trend, which is distinctly different from that of simulation F. Although all three cases have nearly similar jet power of $\sim 10^{45} \ergs$, simulations D and E with lower magnetisation ($\sigma_B = 0.01$ and $0.05$ respectively) have unstable jets which show stronger mixing at the jet boundary and flaring of the jet axis as discussed earlier in Sec.~\ref{sec.moderate}. This causes the jets to decelerate which result in a flattening of the jet advance speed with distance. Simulation F on the other hand shows an increase in jet speed with a profile following more closely to the maximum theoretical line, although still lower. 

Simulations G--J show similar qualitative trends for the evolution of the jet speed, with a gradual increase with distance. At larger scales the ambient density may become comparable to the jet density, such that the earlier approximation of  $\eta_j f(\rbar)^{-1} \ll 1$ used in eq.~\ref{eq.vh1} (and later in Appendix~\ref{append.gbc}) is no longer valid. The jet head velocity will then become $v_h \sim v_j$, independent of the radial distance, as is seen in the last panel of Fig.~\ref{fig.jetvel}, showing a flattening of the theoretical curve for simulation J. The actual jet head speed computed numerically asymptotes more quickly to a constant value of $\sim 0.35c$ than the theoretical curve. This is likely due to a combination of added deceleration due to small scale instabilities resulting in lowering of the jet speed, besides the effect of entering into a low density ambient medium which results in constant jet advance speed. 

\begin{figure*}
	\centering
	\hspace{-0.1cm}\includegraphics[width = 9.cm, keepaspectratio] 
	{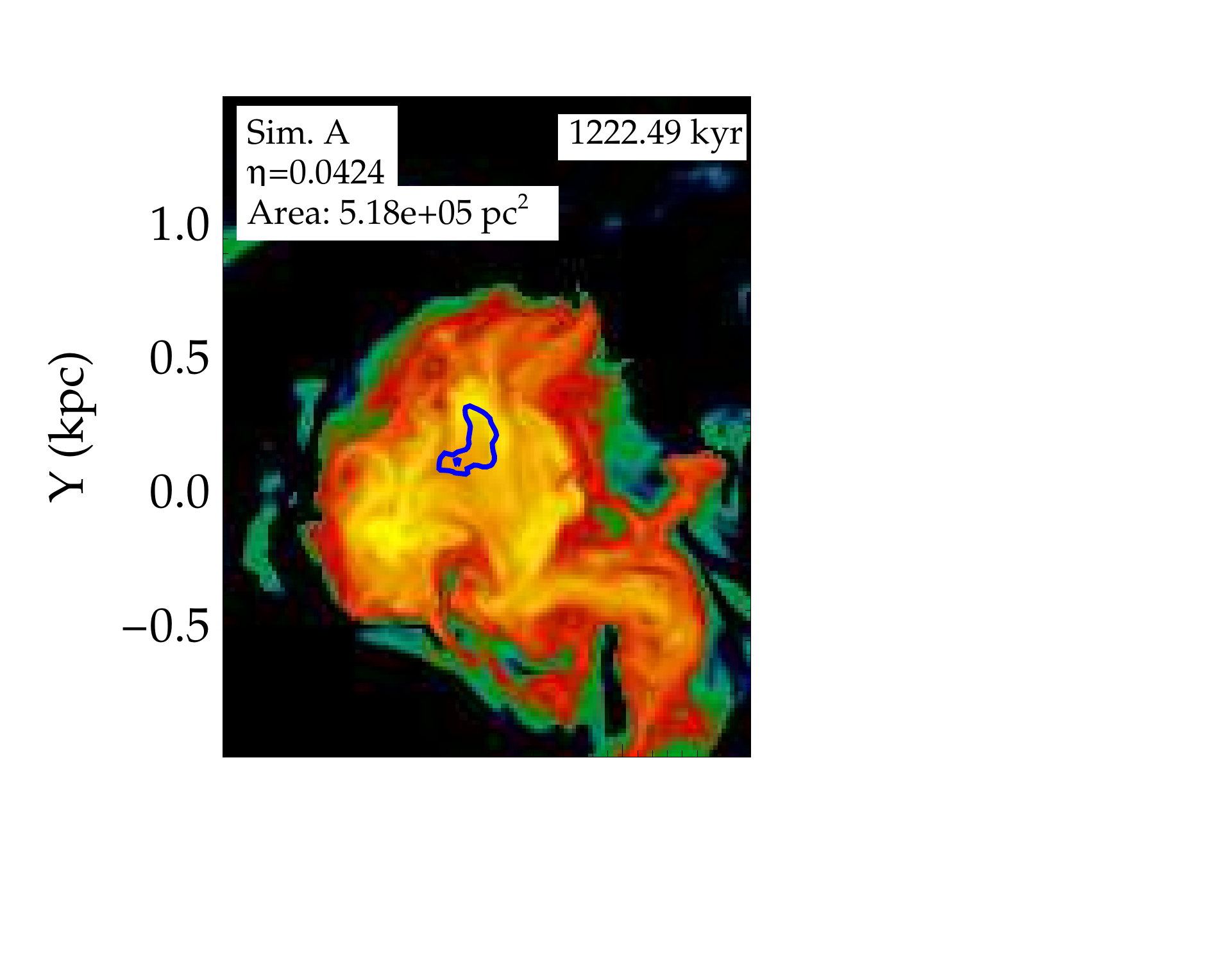}\vspace{-0.7cm}\hspace{-5.cm}
	\includegraphics[width = 9.cm, keepaspectratio] 
	{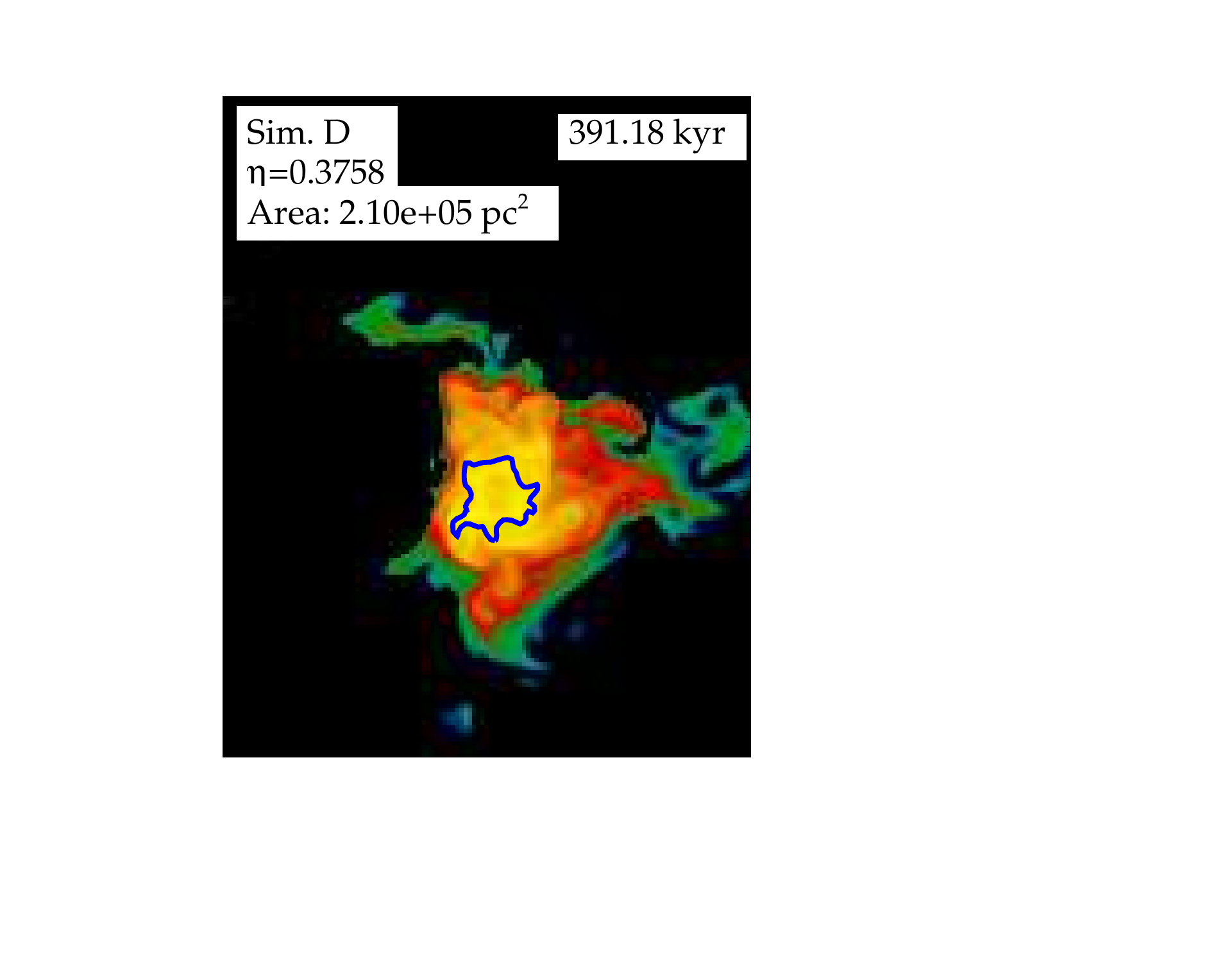}\vspace{-0.7cm}\hspace{-5.cm}
	\includegraphics[width = 9.cm, keepaspectratio] 
	{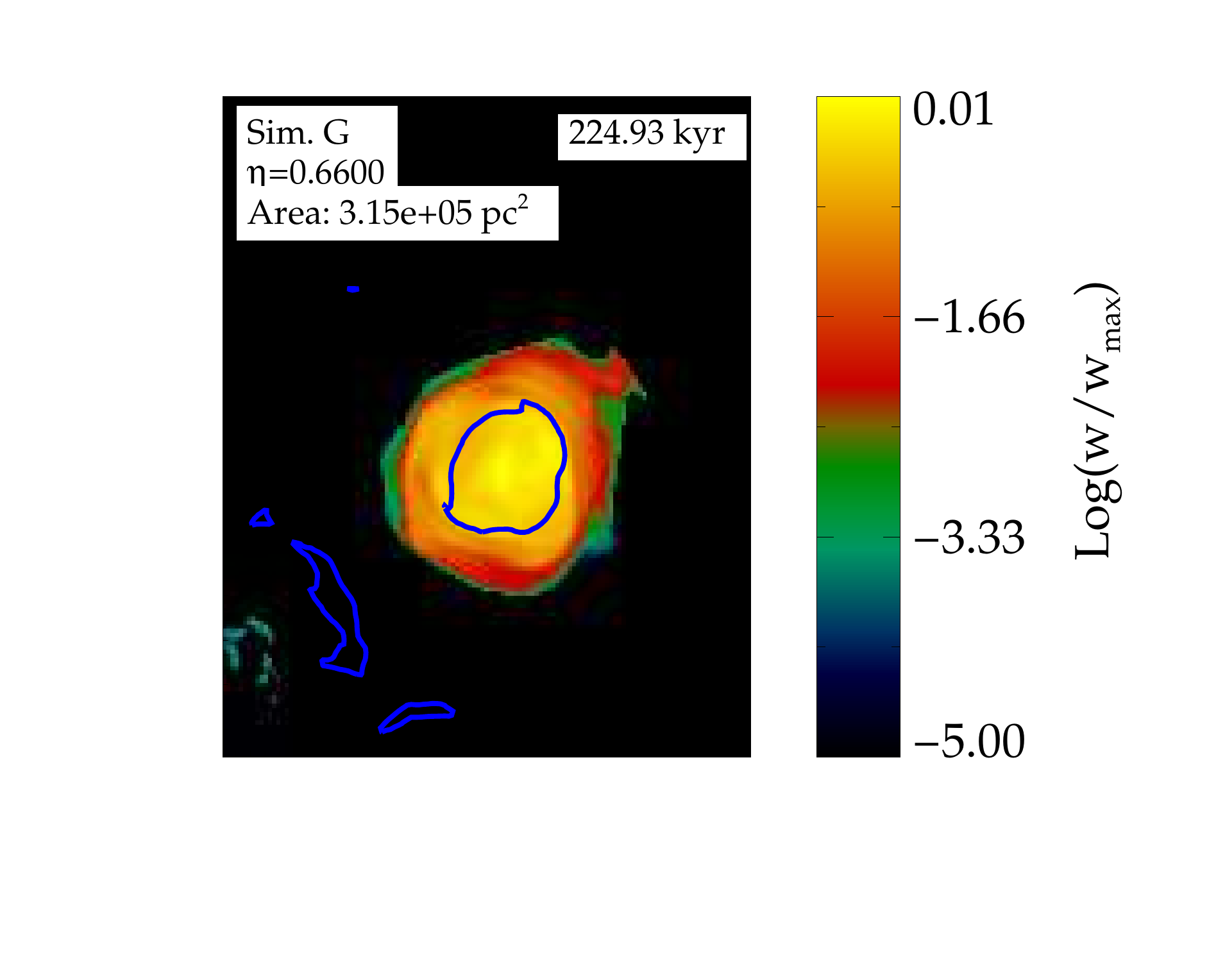}\vspace{-0.7cm}
        \includegraphics[width = 9.cm, keepaspectratio] 
	{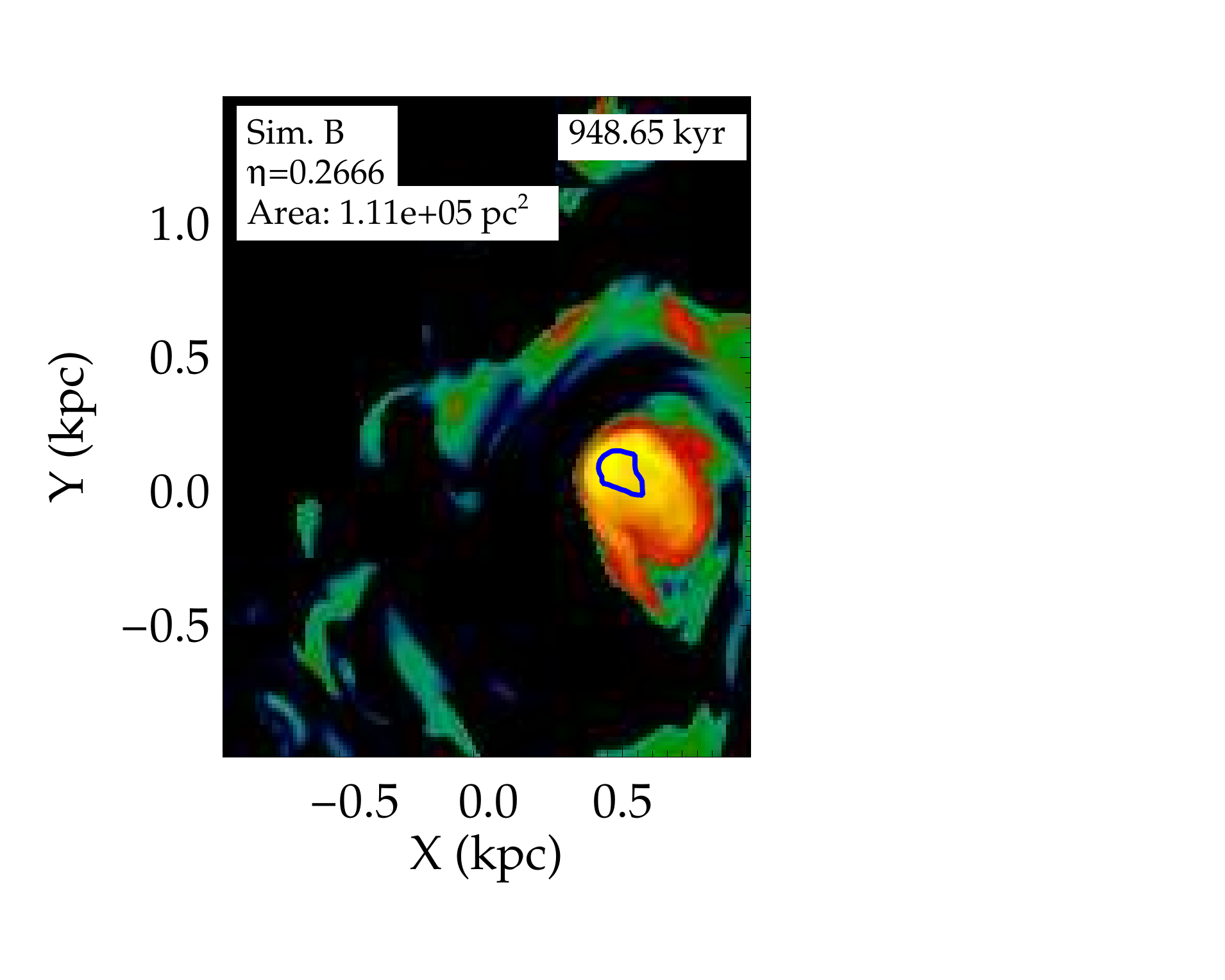}\hspace{-5.cm}
	\includegraphics[width = 9.cm, keepaspectratio] 
	{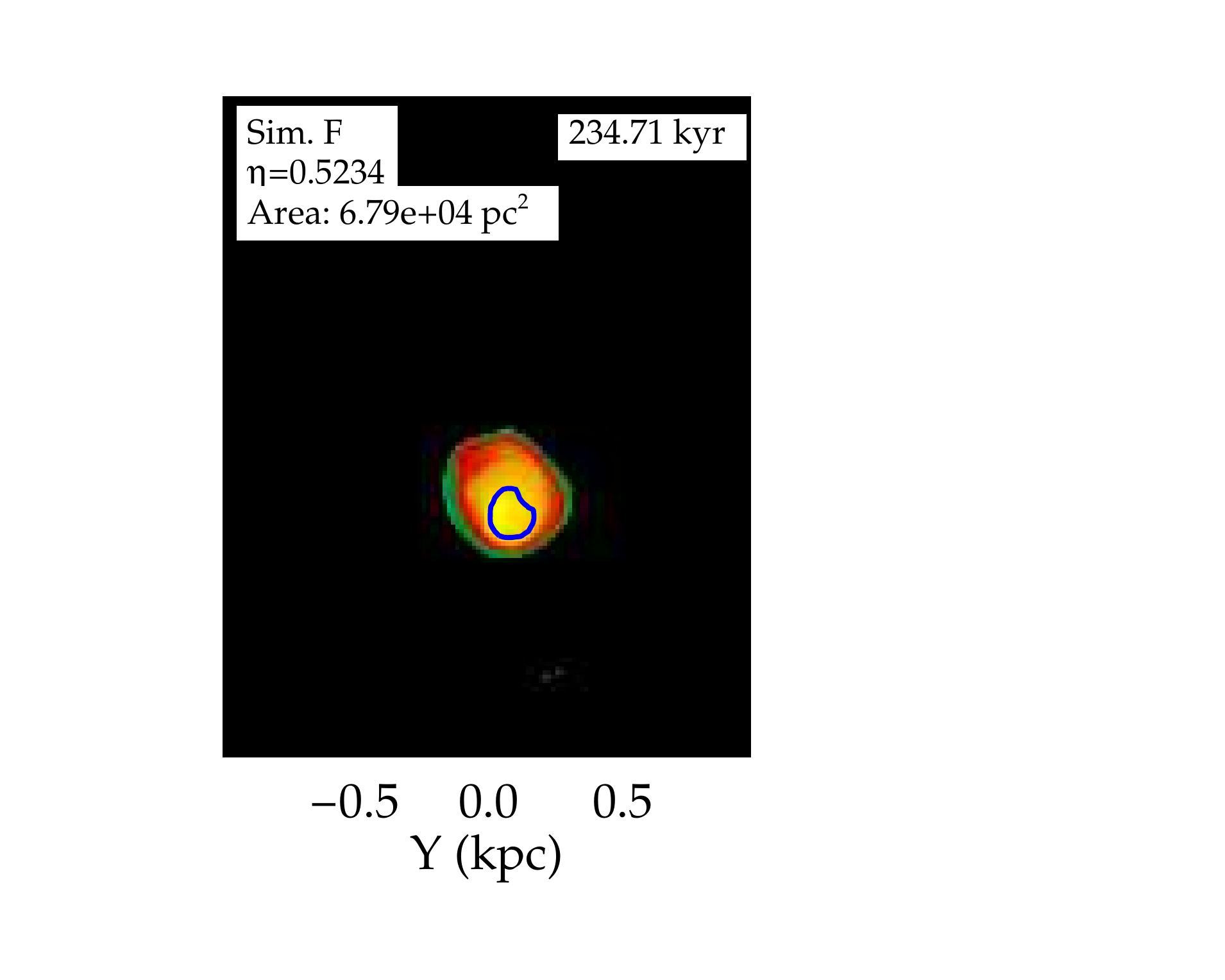}\hspace{-5.cm}
	\includegraphics[width = 9.cm, keepaspectratio] 
	{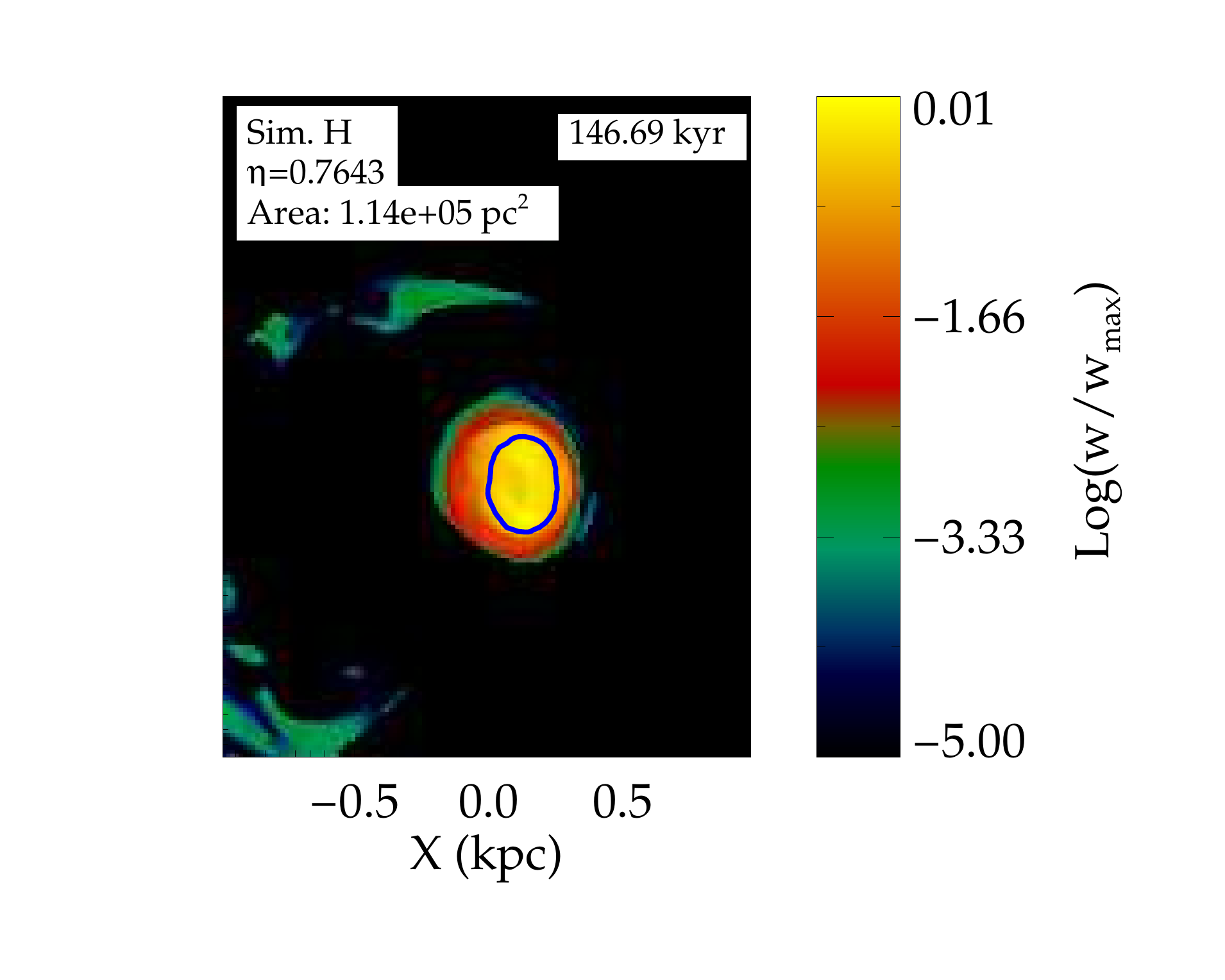}
	\caption{\small We present the cross-section of the positive jet enthalpy flux ($w = \gamma^2 \rho h v_z$) along the direction of the jet launch, normalised to the maximum enthalphy ($w_{\rm max}$). The figures are at a height of $\sim 5$ kpc. The blue contour represent a jet tracer level of 0.8. In each panel we present the ratio ($\eta$) of the enthalpy flux within a jet tracer contour of 0.8 to the total positive jet flux (with jet tracer value $\geq 10^{-2}$). We also present the jet area (in pc$^2$), computed as the total area with a flux value $w/w_{\rm max} > 0.01$.  The top panels depict simulations where the jets are unstable to Kelvin-Helmholtz modes due to either lower magnetisation (sim. A and D) or higher pressure (sim. G), resulting in wider and more distorted cross-section of the jet. Lower panels are jets with stronger magnetisation, where KH modes have lower growth rates with more compact jet core. Simulation B shows a shift in the peak of the flux from the central region (0,0) due to kink-mode instabilities that bend the jet away from its initial launch axis. } 
	\label{fig.crosssec}
\end{figure*}

\section{Discussion}\label{sec.discuss}
In this paper we discuss the dynamics and evolution of relativistic jets with different initial starting parameters, evolving into a hydrostatic atmosphere. The primary results of this work are two folds: a) demonstration of the onset of different MHD instabilities for different jet parameters that significantly affect the dynamics and growth of the jet, b) comparison of the dynamics of the jets with generalised extension of the analytical model (GBC) for FR-II jets proposed by \citet{begelman89a}. The nature of the growth and development of the instabilities affect the dynamics and evolution of the fluid variables inside the jet and its cocoon, leading to deviations from the GBC model. 

We would like to note here that the results of the simulations depend on the assumptions of some jet parameters such as jet radius, jet magnetisation (defined here as the ratio of Poynting to enthalpy flux) and the density and pressure contrasts with the ambient medium. Although, the jet parameters are chosen to be approximately consistent with realistic estimates inferred from observations, as argued in Sec.~\ref{sec.jetsetup}, the absolute choices of some, such as the magnetisation, density contrast etc., were empirical. Similarly, the need to achieve sufficient resolution of the jet injection limits our choice of the jet radius to $\sim 100-200$ pc, which may be unphysically large at the given injection height. However, the qualitative results comparing the behaviour of jet dynamics for different jet parameters presented here are not exepcted to be affected by this approximation.

The primary focus of this work has been to systematically study the difference in jet dynamics for the variation of some jet parmaters, with others remaining constant. This highlights in a qualitative way, the relative importance of different physical quantities when compared to each other, with regards to the jet stability and dynamics; even though the absolute values of the assumed parameters may be different for specific systems. In this following sections we summarise the main results and discuss the implications of the jet stability on the jet dynamics and its comparison with analytical models.

\subsection{Growth of unstable modes}\label{sec.unstablemodes}
The type of instabilities in our simulations can be broadly grouped into two categories based on jet magnetisation and power:
\begin{enumerate}
\item \emph{Large scale modes at higher magnetisation:}  Low power jets ($\sim 10^{44} \ergs$) in simulations B and C with stronger magnetisation were found to be susceptible to kink modes that result in substantial bending of the jet head. The growth rate was lower for simulation A with an order of magnitude lower magnetisation, which did not show substantial bending of the jet axis during the run time of the simulation. However such strongly disruptive kink modes were not seen in more powerful jets (sim. D--J) during the run time of the simulations. Simulation E shows some bending of the jet over much longer length scales ($\sim 1$ kpc) but not as disruptive as in the low power jets.

The above results are in broad agreement with the results from linear stability analysis of the growth of $m=1$ modes in relativistic MHD jets \citep{bodo13a}. Growth rate of current driven instabilities (CDI) is higher for higher magnetisation. In relativistic jets however, for the same central value of the magnetic pitch parameter, the growth rate of CDI is lower \citep[Im($\omega$) $\propto \gamma^{-4}$, ][]{bodo13a}. Hence the absence of strong disruptive kink modes in faster, powerful jets can be due to weaker growth rates of the  CDI, which may manifest only for larger size of the jet. However, even at larger distances, recent results of \citet{tchekhovskoy16a} have demonstrated that the jets may remain fairly stable as they propagate into steeper density profiles beyond the galaxy core. Thus higher power jets with faster Lorentz factors that efficiently drill through the galaxy's core can remain stable up to very large distances.

\item \emph{Small scale modes at lower magnetisation or higher internal pressure:}
In simulations with lower magnetisation, velocity shear driven Kelvin-Helmholtz (KH) modes lead to a higher level of turbulence both close to the jet axis and in the cocoon. Such KH modes are disruptive and result in substantial deceleration of the jet with a decollimation of the jet axis.

In Fig.~\ref{fig.crosssec} we present the cross-section of the jet enthalpy flux ($w = \gamma^2 \rho h v_z$, $\rho h$ being computed from eq.~\ref{eq.enth}) along the jet launch direction in the $X-Y$ plane, at a height of $\sim 5$ kpc for six different cases. The inner blue contour is for a value of the tracer equal to $0.8$. In the top row we have cases with low magnetisation, while the bottom row shows cases with high magnetisation; going from left to right, the simulations have an increase of the jet power and Lorentz factor. We also present in each panel the ratio $\eta$, of the positive jet enthalpy flux within a region with jet tracer $\geq 0.8$, to the total positive enthalphy flux (jet tracer $\geq 10^{-2}$). This quantity gives an approximate estimate of the compactness of the jet. A lower value of eta would represent a jet that is more spread out. Additionally, we also present in each panel the jet cross section area, defined as the area with $w/w_{\rm max} \geq 0.01$, $w_{\rm max}$ being the maximum enthalpy flux at the give height for each cross-section. 

The figure displays clearly the role of magnetic field and instabilities in determining the mixing properties for the  different cases. We can see that, in the top row, the jet cross-section is more deformed than in the bottom row. In particular, cases A (top left panel) and D (top middle panel) show very corrugatedand contours of the jet cross-section. This is indicative of the development of high $m$  KH modes that would favour mixing between jet and cocoon \citep[e.g.][]{rossi20a}. The unstable jets also contain a smaller fraction of the total enthalpy flux within a jet tracer of 0.8, as signified by the lower value of $\eta$ for the upper panels. Similarly, the jet cross section has a much larger area in the upper panels. All these indicate that the jet spine in cases with lower magnetisation are prone to KH mode instabilities resulting in deformed non-regular jet cross-section which is spread over a larger area. 

Case G (top right panel) has a higher Lorentz factor and is more stable than the lower $\gamma$ cases. However, as discussed earlier in Sec.~\ref{sec.highpower}, being hotter simulation G is more unstable than the other high $\gamma$ cases (e.g. simulation H in the lower panel). Correspondingly the jet cross-section is much less deformed than in cases A and D, but it shows an oval shaped deformation when compared to H, possibly indicating higher order modes. The cases in the bottom row have a higher magnetisation and the magnetic tension associated with the toroidal component of the magnetic field opposes the jet deformation and stabilises high $m$ KH modes and, correspondingly, the contours are less deformed.

Similar results have been presented in \citet{mignone10a} and \citet{rossi20a}, where the jet core for a relativistic hydrodynamic jet was found to be more diffuse and decollimated as compared to a jet with a magnetic field. The added magnetic field shields the inner core of the jet by suppressing the KH modes. Linear stability analysis \citep{bodo13a} suggest that for similar magnetic pitch, KH modes have slower growth rate at higher magnetisation.
\end{enumerate}

\subsection{Impact of instabilities on jet dynamics}
The MHD instabilities described above significantly affect the dynamics and evolution of the jet as well its morphology. We list below the major implications:
\begin{enumerate}
\item \emph{Jet deceleration:}
The low power jets (simulations A--C with $P_j \sim 10^{44} \ergs$) are strongly decelerated with mean advance speeds nearly an order of magnitude lower than the maximum possible values predicted by analytical estimates (see Fig.~\ref{fig.jetvel}). Although the nature of instabilities is different for the different simulations (kink modes for Sim. B and C, Kelvin-Helmholtz for Sim. A), all show strong deceleration with a high value of the deceleration index $n$ (eq.~\ref{eq.vh2}) as seen in Fig.~\ref{fig.jetheight}. Amongst the moderate power jets, simulation D with $\sigma_B=0.01$ also shows a flattening of the advance speed and a higher deceleration index than simulations E and F with higher magnetisation.

\item\emph{Self-similar expansion for unstable jets:}
Simulations which suffer strong deceleration (A--D) due to instabilities, evolve more close to a self-similar expansion. As described at the end of Appendix~\ref{append.gbc} for a density profile with $\alpha=1.166$ (eq.~\ref{eq.frho1}), a jet will evolve self-similarly for $n\simeq0.67$, close to the deceleration index for simulations A--D. The axis-ratio plots of simulations B, C and D show a flattening to a constant value beyond a certain time. A constant axis-ratio is indicative of a self-similar expansion of the jet-cocoon. The self-similar expansion likely results from the energy from the jet being more uniformly spread to a larger volume within cocoon. For more stable jets, the ram pressure at the jet head results in a stronger pressure at the mach disc which in turn leads to a larger advance speed than expansion rate for a self-similar jet. Hence the axis ratio of simulations E onwards show a steady increase with time resulting in more conical cocoon profiles.

There has been considerable debate in the literature over the nature of expansion of the jet-cocoon. Self-similar expansion is a convenient assumption for deriving analytical results \citep{falle91a,kaiser97a}.   Although \citet[][hereafter KF98]{komissarov98a} argue that for a jet with a half-opening angle of $\theta_i$, self-similar evolution is expected for length scales larger than the characteristic length of 
\begin{align}
l_c &= \left(\frac{2P_j}{\theta_i \pi \rho_a c^3}\right)^{1/2} \left\lbrack \frac{\gamma_j^2}{(\gamma_j - 1)(\gamma_j^2 -1)}\right\rbrack^{1/2} \\
 &\simeq 85 \pc \times \left(\frac{P_j}{10^{45} \ergs}\right)^{1/2} \left(\frac{\theta_i}{5^\circ}\right)^{-1/2} \left(\frac{n_a}{0.1 \cc}\right)^{-1/2} \nonumber\\ 
 & \times \left\lbrack\frac{\gamma_5^2}{(\gamma_5 - 1)(\gamma_5^2 -1)}\right\rbrack^{1/2} \,\,;\,\, \mbox{with } \gamma_5 = 5 ,
\end{align}
numerical simulations have not found this to be true for all cases. KF98 find that for some simulations, a self-similar phase is established only at late times \citep[similar to ][]{sheck02a,perucho07a,perucho19a}. The intermediate phase in KF98 was characterised by a nearly constant advance speed (in an uniform external medium) and increasing axis ratio, similar to predictions of \citet{begelman89a}, which is true for a collimated jet with $\theta_i = 0$, implying $l_c = \infty \gg l_j$. The above findings support the results of our simulations where the self-similar phase ensues after the onset of fluid instabilities that start to decelerate the jet, which otherwise remains well collimated and is not self-similar.

\end{enumerate}

\subsection{Magnetic field of the jet and cocoon}\label{sec.magfield}
\begin{figure*}
	\centering
	\includegraphics[width = 12 cm, keepaspectratio] 
	{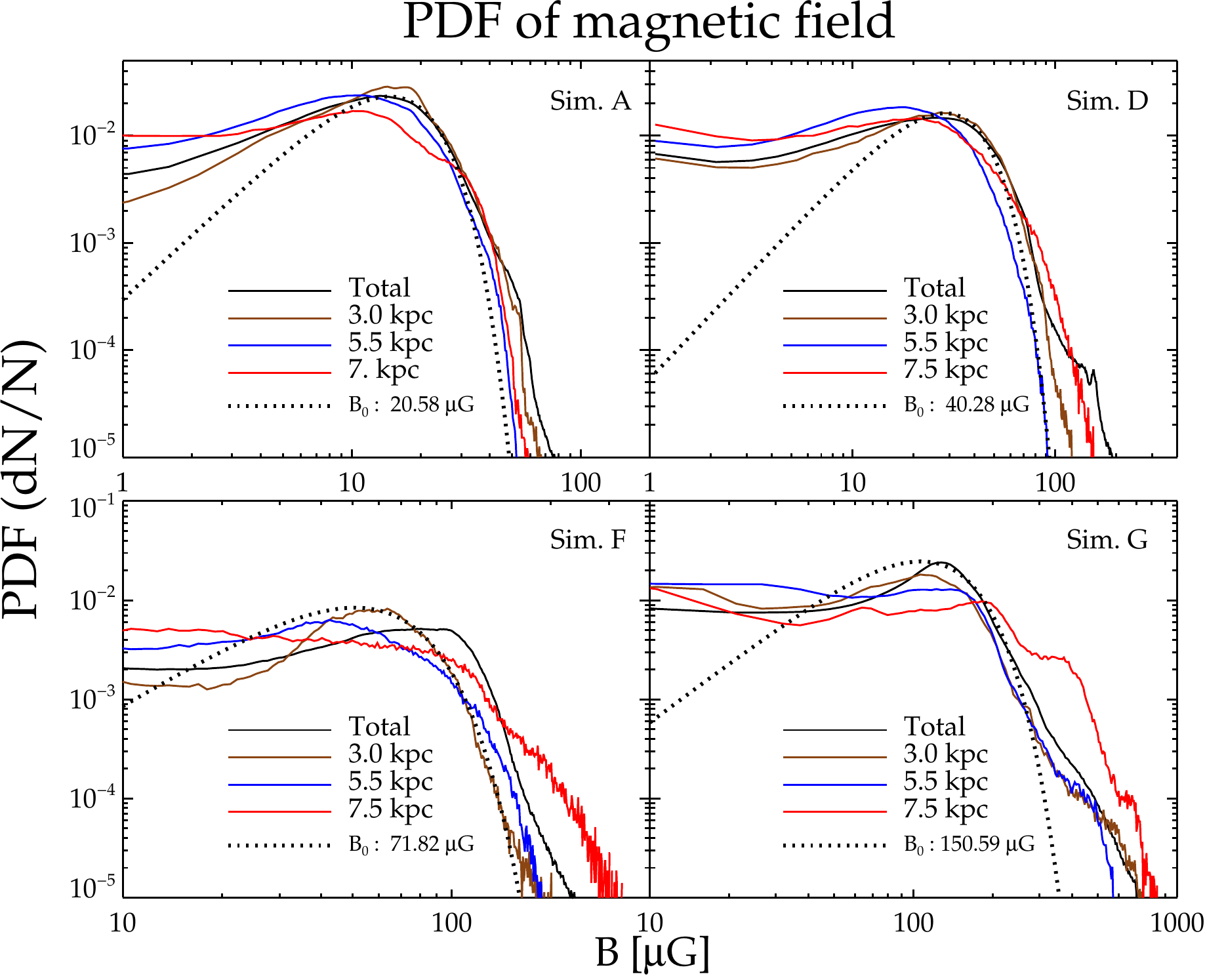}
	\caption{\small PDF of magnetic fields for different heights along the jet. Turbulent and unstable jets show near uniform distribution of magnetic fields at all heights, approximately described by a Maxwell-Boltzmann function (eq.~\ref{eq.mb}) presented in black dotted lines. The mean field strength $B_0$ for the  Maxwell-Boltzmann function is given for each figure. Non-turbulent jets show an extended tail at heights near the hotspot. The PDF are performed at times when the jet reaches the end of the simulation domain in the $Z$ axis.}
	\label{fig.magpdf}
\end{figure*}
\begin{enumerate}
\item \emph{Spatial distribution of magnetic field strength:}
The nature of the magnetic field distribution and its topology inside the cocoon depends on the jet dynamics. Turbulence in the jet cocoons for simulations with instabilities result in small scale magnetic fields varying over scales of $\Delta x-10\Delta x$, $\Delta x$ being the resolution of the simulation. This is demonstrated in Fig.~\ref{fig.sigcompare} and Fig.~\ref{fig.lpar} in  Sec.~\ref{sec.moderate}, where simulation D shows turbulent magnetic field over smaller length scales, whereas simulation F has ordered magnetic field over longer scales. Besides the intermittence in the scale of the magnetic fields, the jets with a turbulent cocoon have a more statistically homogenous distribution of magnetic field at different heights, as shown in Fig.~\ref{fig.magpdf} where the probability distribution function (PDF) of the strength of the magnetic field is presented at different heights. 

For a powerful FRII like jet, it is expected that the field near the jet head will have higher values due to the strong bow shock. As the magnetic field is carried downstream by the backflow and they fill up the adiabatically expanding cocoon, their values would decrease. The PDFs of simulations F and G demonstrate the above, with lower magnetic fields near the bottom and higher field strengths near the jet head. However in unstable jets, the shocks at the jet head are weaker due to the deceleration of the jet from the induced instabilities. This also results in more homogenous distribution of magnetic field inside the cocoon, although intermittent. Hence the turbulent jets in simulations A and D have nearly similar PDF at different heights, with a slight increase to higher magnetic fields at larger heights for simulation D. 

For a magnetic field whose individual components have a random Gaussian distribution with zero mean, the field strength is distributed as a Maxwell-Boltzmann (MB) function \citep{tribble91a,murgia04a,hardcastle13a}:
\begin{equation}
P(B) = \sqrt{\frac{54}{\pi}} \frac{B^2 \exp\left(-(3/2)(B/B_0)^2\right)}{B_0^3}. \label{eq.mb}
\end{equation}
Here $B_0$ is the field strength for the mean magnetic field energy density \citep{hardcastle13a}:
\begin{equation}
\int_0^\infty B^2 P(B) dB = B_0^2.
\end{equation}
In Fig.~\ref{fig.magpdf} representative Maxwell-Boltzmann (hereafter MB) plots have been presented in dotted-black lines, which were obtained from approximate fits to the total magnetic field distribution inside the cocoon. The lines are not exact fits, but are seen to well represent the PDFs of sim. A and D for $B\gtrsim 10$ $\mu$G, and similarly the PDFs of the magnetic fields at lower heights for simulations F and G beyond the peak. This shows that the turbulent fields in the cocoon of the jets were well approximated by a distribution with Gaussian random components of the magnetic fields. The PDFs at heights closer to the jet head for simulations F and G however show strong departure from the MB distribution with an extended power-law tail for simulation F and complex features for simulation G. These arise from the strong interaction of the jet fluid at the bow-shock where the field strengths are likely enhanced due to compression from the shocks.

\item\emph{Variation of magnetic field strength with time:}
\begin{figure}
	\centering
	\includegraphics[width = 8 cm, keepaspectratio] 
	{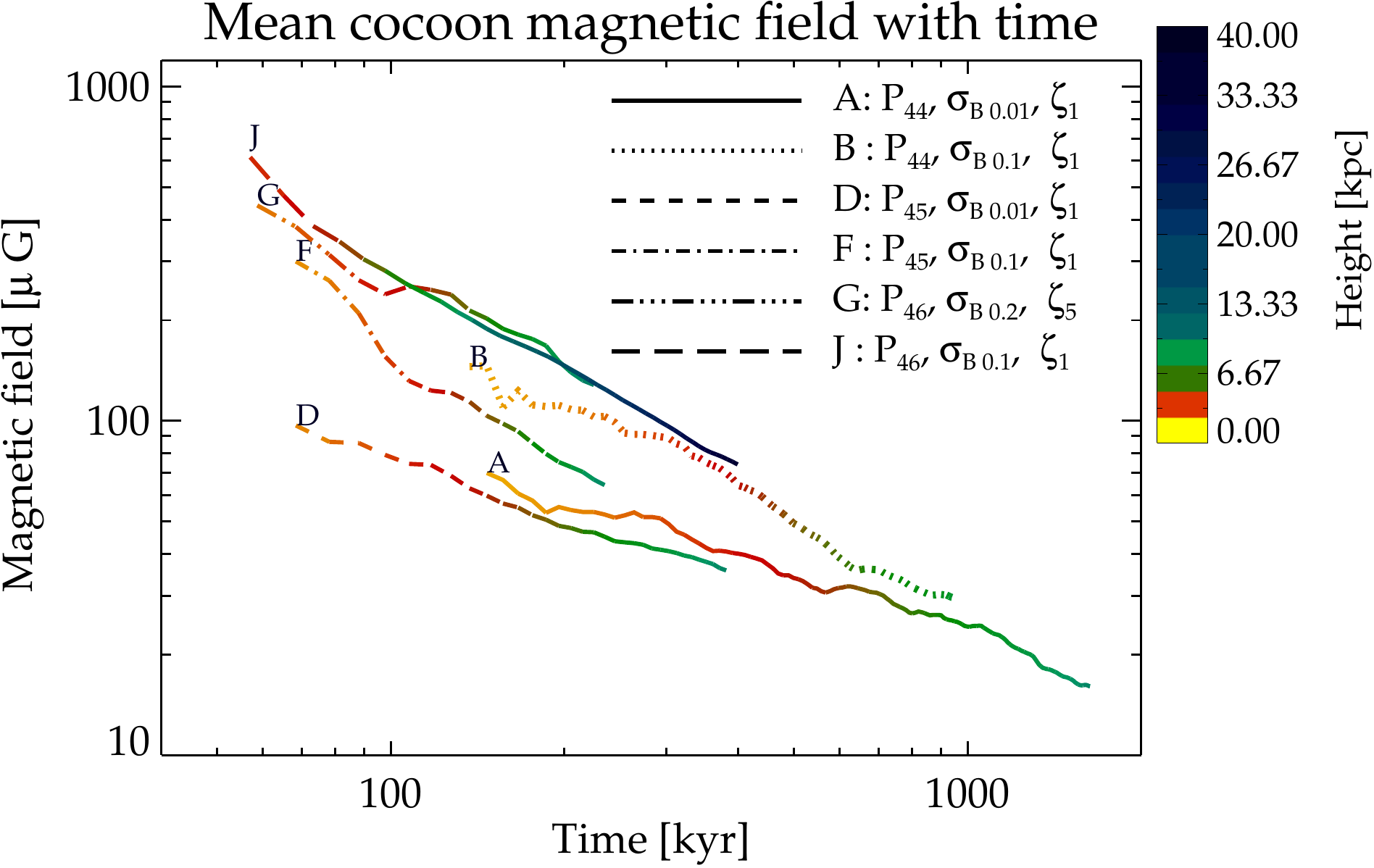}
	\includegraphics[width = 8 cm, keepaspectratio] 
	{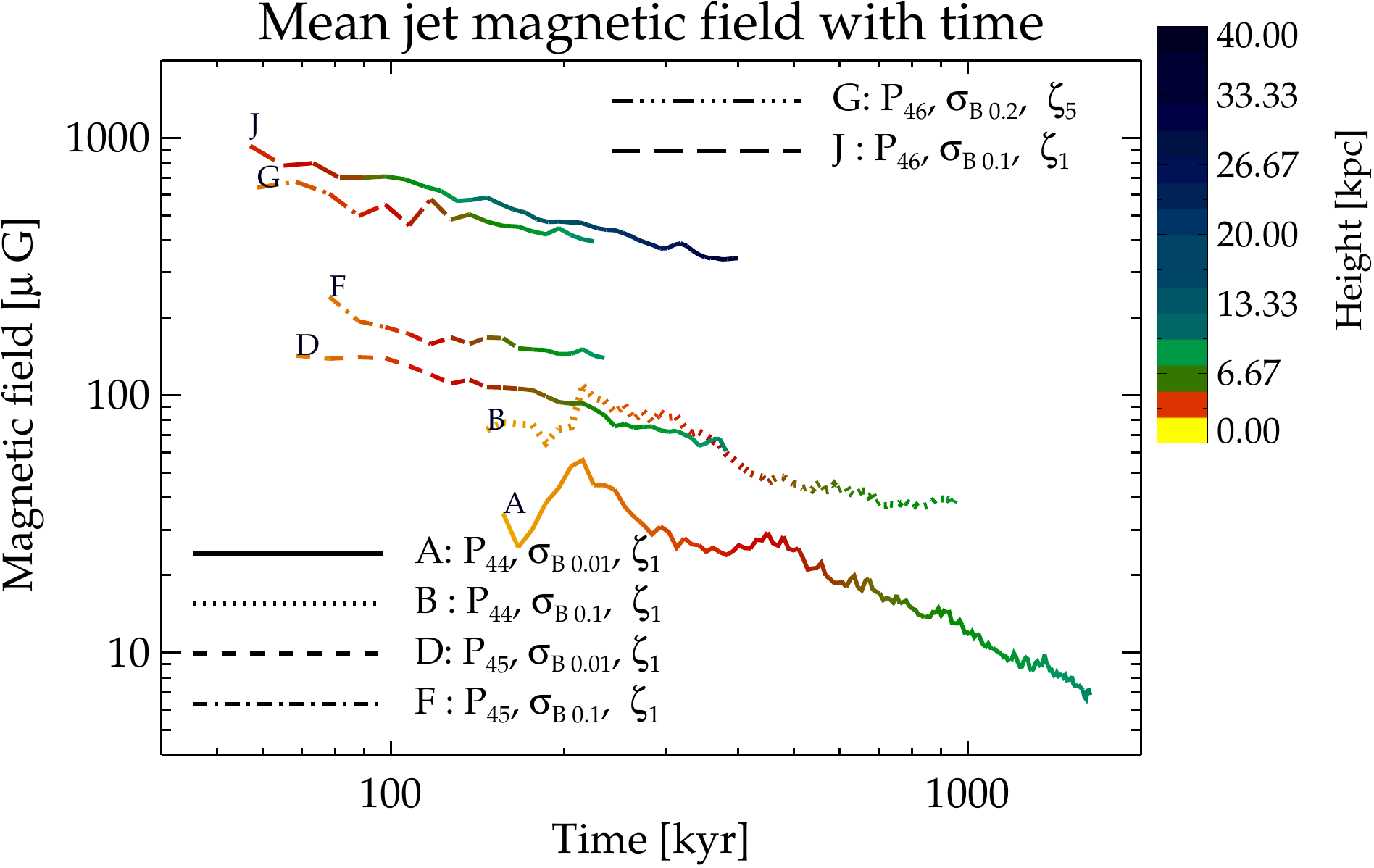}
	\caption{\small \textbf{Top:} Time evolution of the mean magnetic field in the cocoon for selected simulations with different initial parameters listed in the legends. The subscript to $P$ is the logarithm of the jet power, the value of jet magnetisation $\sigma_B$ and pressure-ratio are presented as sub-scripts as well. The beginning of each curve is marked with the initial for the simulation from the list in Table~\ref{tab.sims}. The lines are coloured according to the colourable on the right which denotes the height of the jet at that time.  See Sec.~\ref{sec.magfield} for details. \textbf{Bottom:} The mean magnetic field in the jet for the same simulations and similar legends as in the top panel. }
	\label{fig.bcmean}
\end{figure}
The magnetic field in the cocoon and the jet also evolve with time as the jet and its cocoon expand. In Fig.~\ref{fig.bcmean} we present the evolution of the mean magnetic field in the cocoon and jet separately. The regions with jet tracer: $10^{-7} < \mbox{ Tracer } < 0.9$ are identified as cocoon and those with $\mbox{Tracer } > 0.9$ are identified as jet material. The mean magnetic field in the cocoon decreases as a power-law with time due to the adiabatic expansion of the jet driven bubble. However, the rate of decrease depends on the nature of the simulation and onset of MHD instabilities. Simulations A and D with a lower magnetisation have a mean decay of $\propto t^{-0.6}$, whereas simulations B, the end phase of simulation F (for $t \gtrsim 100$ kyr and $Z \gtrsim 5$ kpc) and simulation J (for $t \gtrsim 100$ kyr and $Z \gtrsim 10$ kpc) show a power-law decline of $\propto t^{-1}$. 

The less steep decline in the field strength for the simulations with weaker magnetic fields could be due to onset of MHD instabilities discussed earlier in Sec.~\ref{sec.moderate}. Such instabilities result in a slower expansion of the jet which will result in a slower decline of mean field strength due to adiabatic expansion. Secondly, turbulence generated by the Kelvin-Helmholtz driven modes result in small scale fluctuation of the magnetic field, as shown in Fig.~\ref{fig.lpar}. This can result in moderate enhancement of the magnetic field  which may counteract the decrease of field strength due to stretching of the field lines. However, our current spatial resolution being limited, we cannot fully ascertain if such mode of field enhancement is dominant.

The field strength in the jet also follows a power-law evolution with time, which except for simulations F and A, have an index $\lesssim -0.6$. Simulation A follows a steeper decline at the later stages as $\propto t^{-1}$. The relatively steady power-law decline of the jet magnetic field with similar indices for different simulations imply that the jet core remains relatively steady. The rate of decline is slowest for simulation F ($\propto t^{-0.36}$) which does not show any signature of MHD instabilities. Simulation A has a sharper decline in the jet magnetic field as Kelvin-Helmholtz driven mixing of the jet lead to strong deceleration and decollimation of the jet (see Sec.~\ref{sec.kink}).
\end{enumerate}

\subsection{Implications for synchrotron emission}
The above results have several different implications for the nature non-thermal emission from jets which we list below.   
\begin{enumerate}
\item \emph{Morphology of emission:}
Powerful jets stable to fluid instabilities show the typical feature of a FRII jet with a strong pressure hotspot (see Fig.~\ref{fig.simG}) where the jet terminates, besides islands of enhanced pressure along the jet axis arising from recollimation shocks. The pressure at the hotspots is nearly two orders of magnitude higher than the mean pressure in the cocoon. These high pressure regions arising from shocks are expected to accelerate the electrons enhancing the synchrotron emission at the hotspot. Stable jets with higher magnetisation have conical shaped cocoons with narrower widths as the forward shock at the jet-head expands much faster due to very little deceleration. Jets with instabilities on the other hand show more wider cocoons with cylindrical shapes due to the deceleration of the jet. 

The simulations showing strong development of kink modes (simulations B and C) do not have prominent terminal hot-spot. Since the jet head swivels randomly in different direction due to the kink modes, the pressure at the jet head is spread evenly over a wider area. This results in a much wider cylindrical shaped cocoon with asymmetric features near the jet head due to changing orientation of the jet head. This may result in a wider diffuse emission at the top as the integrated emission will probe the whole volume where the shocked electrons are distributed. Emission at higher energies may however preferentially give weight to regions of strong shocks at the current location of the jet where the electrons are freshly accelerated. This may lead to a complex morphology of the emitting region at higher energies, which may differ from the emission dominated by low energy electrons.

\item \emph{Shock structures and emission profile:} Jets prone to instabilities have complex pressure profile at the jet head due to the motions of the jet head, which will result in multiple oblique shocks. This is in contrast to the standard model of an FRII jet with a single strong shock at the mach disc \citep{begelman89a,kaiser97a,falle91a}, which is often employed to calculate emission parameters and source ages \citep{pacholczyk70a,jaffe73a,murgia99a,harwood13a,harwood15a,harwood17a}. The complex shock structure with varying shock strengths will result in a wide variation of the energy distribution of the relativistic electrons being accelerated at these sites. Besides the stronger shocks at the hot spot, internal weak shocks develop inside the cocoon which may further accelerate the electrons as they flow across such shocks. Such multiple shock crossing will result in a variation of the resultant index of the power-law energy distribution, which is usually assumed to have a single value at low energies \citep{kardashev62a,harwood13a,harwood15a}. 

\item \emph{Cocoon magnetic field and electron ageing:} 
Models that estimate the time evolution of the synchrotron spectra assume a predefined distribution of the magnetic field \citep{harwood13a,harwood15a,harwood17a}. The simplest models such as by \citet[][ hereafter JP]{jaffe73a} assume a constant magnetic field. More recent sophisticated approaches have accounted for the turbulent nature of the magnetic field in the cocoon \citep{tribble91a,harwood13a,hardcastle13a}. In our simulations the magnetic field in the cocoon is well described by a Maxwell-Boltzmann distribution for the turbulent less powerful jets (as shown in Sec.~\ref{sec.magfield}), similar to the assumptions by \citet{tribble91a}.  For more powerful jets ($P_j \gtrsim 10^{46} \ergs$) however, the probability distribution function at heights near the jet-head have an extended tail beyond the mean Maxwell-Boltzmann profile. The nature of the field distribution significantly impacts the evolution of the spectra of electrons when they traverse through different magnetic fields, as demonstrated in \citet{harwood13a}. Such multiple shock crossings will subsequently affect the estimates of radiative ages of the synchrotron emitting sources. Besides the spatial distribution, our results show that the magnetic field in the cocoon show a steady decline with time as a power-law, as discussed earlier in Sec.~\ref{sec.magfield}. Such a secular decline of the magnetic field is also not considered in the analytical models of electron ageing, and will affect the break frequency of the synchrotron spectrum.

\end{enumerate}

We will discuss these in more quantitative detail in subsequent publications (Mukherjee et al. Paper II in prep) where we will discuss the results of some of the simulations presented here that have been performed with the new \textsc{lagrangian particle} module of \textsc{pluto} \citep{vaidya18a}. We will explore in detail the spectral evolution of the non-thermal electrons and the emission characteristics of synchrotron radiation at different wavelengths.

\section{Summary and conclusion}\label{sec.conclusion}
In conclusion, we can summarise our main results as:
\begin{enumerate}
\item We have performed simulations of relativistic jets of different powers and magnetisations up to a few tens of kilo parsec. One of the primary aims was to check for the growth of MHD instabilities as a function of different jet injection parameters.
\item MHD instabilities such as large-scale kink modes and small scale Kelvin-Helmholtz (KH) modes decelerate the jet, affecting its dynamics and morphology.
\item Large scale kink modes can result in global bending of the jet axis and significant deformation in the morphology of the jet and its cocoon.
\item Small scale KH modes cause turbulence in the jet cocoon, which in turn result in smaller length scales of the  magnetic field. Such modes disrupt the jet axis due to mixing with the cocoon plasma.
\item Small scale modes can also arise in jets with higher pressure or temperature (e.g. simulation G) due to smaller sound crossing times of perturbations, as predicted earlier by \citet{rosen99a}.
\item Low power jets ($P_{\rm jet} \sim 10^{44} \ergs$), with lower speeds and density contrasts, are susceptible to both modes. Jets with stronger magnetic fields (e.g. for a $\sigma_B \gtrsim 0.1$ which gives a peak central field of $B_0 \gtrsim 170 \, \mu$G ) are kink unstable, whereas those with lower magnetic fields show Kelvin-Helmholtz modes. 
\item Moderate power jets ($P_{\rm jet} \sim 10^{45} \ergs$) do not show appreciable disruption to kink instabilities up to 10 kpc. However, weakly magnetised jets ($\sigma_B \sim 0.01$ resulting in $B_0 \lesssim 150 \mu$G) show strong development of small scale KH modes.
\item Unstable jets show a greater resemblance to self-similar expansion of the jet and its cocoon.
\item Powerful jets ($P_j \sim 10^{46} \ergs$), with higher values of Lorentz factors and pressure or density contrasts, are less susceptible to instabilities (within the simulation run-times of this work). Such jets show a more closer match with the generalised Begelman-Cioffi \citep{begelman89a} relations (within $10\%-20\%$). Jets with instabilities show a poorer match with analytical predictions.
\item Jets less prone to instabilities show an increase in advance speed as they emerge into a radially falling ambient density field, asymptoting to a fraction of the maximum speed predicted by analytical relations. Unstable jets decelerate, resulting in either a constant advance speed at a value much slower than the maximum possible speeds, or show a decrease with distance and time.
\item The magnetic field distribution in the cocoon of unstable jets are well approximated by turbulent field distribution given by a Maxwell-Boltzmann (MB) function. For powerful stable jets, heights closer to the jet head show strong deviation from a standard MB form. Over-all the major volume of the cocoon shows a turbulent distribution of field strength, favouring the Tribble model \citep{tribble91a,hardcastle13a} for magnetic field distribution.
\item The mean magnetic field in the cocoon decays with time as the jet evolves, with unstable jets having a slower decay rate. 
\end{enumerate}

\section{Acknowledgement}
We thank the referees for their thorough scrutiny and constructive comments, which helped to improve the quality and clarity of the paper. We acknowledge support by CINECA through ISCRA and by the Accordo Quadro INAF-CINECA 2017 for the availability of high performance computing resources. The authors acknowledge support from the PRIN-MIUR project Multi-scale Simulations of High-Energy Astrophysical Plasmas (Prot. 2015L5EE2Y). The authors wish to acknowledge the UNITO Scientific Computing Competence Center for the availability of high-performance computing resources and support through the OCCAM cluster.

\section{Data Availability}
The derived data generated in this research will be shared on reasonable request to the corresponding author.

\appendix
\section{Comparison of jet density and pressure with analytical estimates}\label{sec.jetchi}
In this appendix we present a calculation to check for the consistency of the assumed choice of the density and pressure. We compute the ratio of the rest mass energy density to the sum of the internal energy and the jet pressure for an ideal gas using the parameters of our simulations. We compare the results to an approximate analytical calculation of the same parameter assuming the jet to be composed of non-thermal relativistic particles.  For an ideal gas, the ratio of the rest mass energy to the enthalpy without the rest mass can be expressed as \citep{komissarov96a,sutherland07a,wagner11a,mukherjee16a}
\begin{align}
\chi &= \frac{\rho c^2}{\rho h - \rho c^2} = \frac{\rho c^2}{\rho e + p} \\
     &= \left(\frac{\Gamma - 1}{\Gamma}\right) \frac{\rho c^2}{p} 
\end{align}
The parameter $\chi$ gives a relative estimate of whether the jet is enthalpy dominated or matter dominated, and can be used to estimate the density of an analogous classical jet with similar power, velocity and pressure as that of a relativistic jet \citep{komissarov96a,sutherland07a}. For the choice of density and pressure in our simulations, $\chi$ ranges between: $\sim 8.66 - 44.44$, which we obtain by assuming $\Gamma = 5/3$ and using the values of $\Theta_j$ in Table~\ref{tab.sims}. For the given ranges of $\Theta_j$, an ideal gas equation of state with $\Gamma = 5/3$ is a good approximation \cite{mignone07a}.

Alternatively, $\chi$ can be also be derived by assuming the jet to be composed of relativistic non-thermal particles (electrons) with a distribution function which is power-law in particle energy as: 
\begin{equation}
N_e(\gamma) d\gamma = K  \gamma_e^{-p} \quad \gamma \in (\gamma_1, \gamma_2)
\end{equation}  
The total number density ($n_e$) and energy densities ($\varepsilon$) of the particles are obtained by integrating over the distribution function as
\begin{align}
n_e &\simeq \frac{K m_e}{p-2} \gamma_1^{-(p-1)} \\
\varepsilon &\simeq \frac{K m_e c^2}{p-2} \gamma_1^{-(p-2)}
\end{align}
where we have assumed $p > 2$ \citep{worrall09a,hovatta14a} and $\gamma_2 \gg \gamma_1$ which is valid for synchrotron emitting sources as observations constrain the Lorentz factors to vary between $\gamma_1 \gtrsim 10 - 100$ \cite{wardle98a,godfrey09a} and $\gamma_2 \gtrsim 10^6-10^8$ \citep{worrall09a,croston09a,ghisellini14a,migliori20a}. Following the principle of equipartition, one can assume that the density and energy of the non-thermal particles are a fraction ($\eta$) of the total fluid values. Thus the parameter $\chi$ can be computed as
\begin{align}
\chi &= \frac{\rho c^2}{\rho e + p} = \frac{\eta f_{pi} n_e m_+ c^2}{(4/3) \eta \varepsilon} \\
     &= f_{pi}\frac{3(p-2)}{4(p-1)}\frac{m_+}{m_e}\gamma_1^{-1} 
\end{align}
where we have considered the fluid density to be $\rho = m_e n_e + m_+ n_+ =  f_{pi} m_+ n_e$ in a net charge neutral fluid ($n_e = n_+$). The pressure and internal energy densities of the highly relativistic non-thermal gas are related as $p = \varepsilon/3$. Here the + subscript denotes the positively charged particles which are positrons for a leptonic jet and ions for a hadronic jet. The factor $f_{pi} = 1$ for a hadronic jet ($m_p \gg m_e$) and $f_{pi} = 2$ for a leptonic composition of the jet. The above equation is similar to that derived in \citet{nawaz14a}.

For a $\gamma_1 \gtrsim 10 - 100$ \cite{wardle98a,godfrey09a}, and a spectral index value of $p=2.4$ \cite{cotton09a}, the parameter $\chi$ for a hadronic jet is $\chi \sim 4 - 39$. The above range is close to the values inferred from our choices of the simulation parameters. This demonstrates that the values of densities and pressure used in our simulations are consistent with a hadronic jet. Although, many models prefer an electron-positron jet, there are several counter examples of dominant hadronic components in jets and the debate on jet composition is not yet settled \citep{sikora2000a,celotti01a,sheck02a,worrall09a}.

\section{Generalised Begellman-Cioffi (GBC) relations}\label{append.gbc}
For a jet expanding into an ambient medium with a density profile 
\begin{equation}
n_a = n_0 f(\rbar) = \frac{n_0}{(1+\rbar)^\alpha} \,\,\,\mbox{ with } \rbar = \frac{r}{a},
\end{equation} 
$a$ being a scale length, the velocity of the jet head is given by eq.~\ref{eq.vh2}. For our simulations, the density profile obtained by numerically solving eq.~\ref{eq.halopres} was found to be described well by an approximate analytical expression in two different spatial regimes, as:
\begin{align}
n_a &= n_0 f(\rbar) \,\,\, ; \,\,\, \rbar = r/a \label{eq.eqrho}\\
f(\rbar) &= \frac{1}{\left(1 + \rbar\right)^{1.166}} \label{eq.frho1} \,\,\, r < 10 \kpc \\
f(\rbar) &= \rbar^{-0.829} \label{eq.frho3} \,\,\, r > 15 \kpc
\end{align}
with $n_0 = 0.103 \cc$ and $a = 0.63$ kpc.

In the equations that follow, length scales have been normalised with the length scale $a$ of the density profile (e.g. $\lbar = l/a$) and time with the deceleration time scale $\tau$ as $\tbar = t/\tau$. Thus the evolution of the jet length is given by:
\begin{align}
  \frac{d\lbar}{d\tbar} & = v_h^M g(\tbar) \left(\frac{\tau}{a}\right) = f(\lbar)^{-1/2} \frac{\Lbar}{\left(1 + \tbar\right)^n} \label{eq.apend1} \\
\mbox{ where }  &\Lbar =  \gamma_j v_j \left( \frac{\tau}{a} \right) \eta_0^{1/2}  \left\lbrack 1 + \frac{\Gamma p_j}{(\Gamma -1) \rho_j c^2} \right \rbrack^{1/2} \nonumber 
\end{align}
 Here $\Lbar = $ is a scale length normalised to the scale length of the density profile$a$, with typical value
\begin{align}
\Lbar &= 1.53 \left(\frac{\gamma_j}{5}\right) \left(\frac{\eta_0}{10^{-4}}\right)^{1/2}\left(\frac{v_j}{c}\right)\left(\frac{\tau}{100 \mbox{Kyr}}\right)\left(\frac{a}{1 \mbox{ kpc}}\right)^{-1} \nonumber \\
     & \times \left\lbrack \frac{1 + \left(\Gamma/(\Gamma-1)\right)\Theta_j}{1.038} \right\rbrack^{1/2} \label{eq.L0bar}
\end{align}
In the last term in eq.~\ref{eq.L0bar}, $\Gamma$ is the adiabatic index of the Ideal gas equation of state, that we have assumed to be $\Gamma = 5/3$, which is relevant for our simulations (as shown in Table~\ref{tab.sims}). The temperature parameter has typical values of  $\Theta_j = p_j/(\rho_j c^2) \sim 0.0152 $ (see Table~\ref{tab.sims}). This is obtained for a jet with density contrast $\eta=10^{-4}$, in pressure equilibrium with the environment, where the ambient gas has density $n \sim 0.1 \cc$, mean molecular weight $\mu \sim 0.6$ and temperature $T\sim 10^7$K. Overall the last term contributes a value close to unity. 

Assuming a density profile as in eq.~\ref{eq.eqrho}, eq.~\ref{eq.apend1} can be integrated for the two limiting cases as
\begin{align}
\lbar &= \frac{\Lbar}{(1-n)} \left(1+\tbar\right)^n - \frac{\Lbar}{(1-n)} \,\,\, \mbox{ for } \lbar \ll 1 \\
\lbar^{\frac{(2-\alpha)}{2}} &= \frac{(2-\alpha)\Lbar}{2(1-n)}\left(1 + \tbar\right)^{1-n} - \frac{(2-\alpha)\Lbar}{2(1-n)} \mbox{ for } \lbar \gg 1
\end{align}
The above equations can be further simplified for the two limiting cases of $t \ll \tau$ and $t \gg \tau$ to get
\begin{itemize}
\item $\lbar \ll 1$ and $\tbar \ll 1$:
\begin{equation}
\lbar = \Lbar \tbar \label{eq.lbar1}
\end{equation}
\item $\lbar \ll 1$ and $\tbar \gg 1$:
\begin{equation}
\lbar = \frac{\Lbar}{(1-n)} \tbar^{(1-n)} \label{eq.lbar2}
\end{equation}
\item $\lbar \gg 1$ and $\tbar \ll 1$:
\begin{equation}
\lbar = \left\lbrack\frac{(2-\alpha)\Lbar}{2}\right\rbrack^{\frac{2}{2-\alpha}} \tbar^{\frac{2}{2-\alpha}} \label{eq.lbar3}
\end{equation}
\item $\lbar \gg 1$ and $\tbar \gg 1$:
\begin{equation}
\lbar = \left\lbrack\frac{(2-\alpha)\Lbar}{2(1-n)}\right\rbrack^{\frac{2}{2-\alpha}} \tbar^{\frac{2(1-n)}{2-\alpha}} \label{eq.lbar4}
\end{equation}
\end{itemize}
Eq.~\ref{eq.lbar1} and eq.~\ref{eq.lbar2} refer to the jet evolution within the core of the density profile, whereas eq.~\ref{eq.lbar3} and eq.~\ref{eq.lbar4} are for larger scales where the density profile is approximately a power-law with radius. From eq.~\ref{eq.lbar4} we see that for a decelerating jet the, the jet evolves slower by a factor of $(1-n)$ as compared to a non-decelerating jet. Eq.~\ref{eq.lbar4} also implies that the deceleration coefficient $n$ should be less than unity ($n < 1$) to have non-imaginary values of $\lbar$. From the coefficients to the power-law fit to the evolution of jet height presented in Fig.~\ref{fig.jetheight} and assuming jet parameters at injection, we find the deceleration index and the deceleration time scale (from eq.~\ref{eq.L0bar} and eq.~\ref{eq.lbar4}), presented in Table~\ref{tab.gbcindex}. J$_L$ and J$_U$ refer to fits to the jet height of distances $l \leq 10$ kpc and $l \geq 15$ kpc respectively. Since the power-law index of the density profile changes around $r \sim 10$ kpc, different values of $\alpha$ defined in eq.~\ref{eq.frho1} and eq.~\ref{eq.frho3} have been used to compute the deceleration index $\tau$ from eq.~\ref{eq.lbar4}
 
\begin{table}
\caption{Deceleration index and deceleration time scale}\label{tab.gbcindex}
\begin{center}
\begin{tabular}{| l | c | c |}
\hline
Sim. &  n   & $\tau$ (kyr) \\
\hline
A    & 0.68 & 216 \\
B    & 0.72 & 210 \\
C    & 0.69 & 204 \\
D    & 0.62 & 55 \\
E    & 0.55 & 71 \\
F    & 0.4  & 119 \\
G    & 0.48 & 65 \\
H    & 0.47 & 39 \\
I    & 0.42 & 110 \\
J$_L$ & 0.43 & 204 \\
J$_U$ & 0.24 & 204 \\
\hline
\end{tabular}
\end{center}
\begin{tablenotes}
\item The above have been computed from coefficients of power-law fits to the jet length using  eq.~\ref{eq.L0bar} and eq.~\ref{eq.lbar4}.
\end{tablenotes}
\end{table}

Equating the cocoon pressure in eq.~\ref{eq.pc0} to the ram pressure of the ambient medium and assuming that the cocoon is over-pressured compared to the ambient gas, the rate of expansion of the cocoon radius can be obtained as:
\begin{align}
p_c &= (\Gamma - 1) \frac{P_j t}{(4/3) \pi a^3 \rbar^2_c \lbar} = \rho_a(\rcbar) v_c^2 \\
\rcbar &f(\rcbar)^{1/2} \frac{d\rcbar}{d\tbar} = G\left(\frac{\tbar}{\lbar}\right)^{1/2} \,\, ; \,\, G = \left\lbrack \frac{3(\Gamma-1) P_j \tau^3}{4\pi a^5\rho_0}\right\rbrack^{1/2} \label{eq.pc0b}
\end{align}
Here $G$ is a dimensionless constant whose typical value would be
\begin{align}
G &= 0.42 \left(\frac{P_j}{10^{45} \ergs}\right)^{1/2} \left(\frac{\tau}{100 \mbox{ Kyr}}\right)^{3/2} \nonumber \\
  & \times \left(\frac{a}{1 \kpc}\right)^{-5/2} \left(\frac{n_0}{0.1 \cc}\right)^{-1/2},
\end{align}
where we have assumed $\Gamma = 5/3$ (ideal EOS) and $\mu = 0.6$ for the mean molecular weight.
Eq.~\ref{eq.pc0b} can be integrated in the various limits as done in eq.~\ref{eq.lbar1}--\ref{eq.lbar4}, to find the time evolution of the cocoon radius and pressure:
\begin{itemize}
\item $\rcbar \ll 1$ and $\tbar \ll 1$:
\begin{align}
\rcbar &= \left(\frac{2 G}{\Lbar^{1/2}}\right)^{1/2} \tbar^{1/2}                \label{eq.rcbar1}  \\
p_c    &= \frac{3 P_j (\Gamma -1) \tau}{8 \pi a^3 G \sqrt{\Lbar}} \,\, \tbar^{-1}     \label{eq.pc1} 
\end{align}
\item $\rcbar \ll 1$ and $\tbar \gg 1$:
\begin{align}
\rcbar &= \left(\frac{4 G (1-n)}{\Lbar^{1/2}(n+2)}\right)^{1/2} \tbar^{(n+2)/4}               \label{eq.rcbar1}  \\
p_c    &= \left\lbrack\frac{3 P_j (\Gamma -1) (n+2) \tau}{16 \pi a^3 G \sqrt{\Lbar}}\right\rbrack \tbar^{-(2-n)/2}     \label{eq.pc2} 
\end{align}
\item $\rcbar \gg 1$ and $\tbar \ll 1$:
\begin{align}
\rcbar &= \left\lbrack \frac{G}{\Lbar^{1/(2-\alpha)}} \frac{(2-\alpha)(4-\alpha)}{(4-3\alpha)} \left(\frac{2}{2-\alpha}\right)^{1/(2-\alpha)}   \right\rbrack^{2/(4-\alpha)}  \nonumber\\
       & \times \tbar^{(4-3\alpha)/((2-\alpha)(4-\alpha))}               \label{eq.rcbar3}  \\
p_c    &=   \frac{3 P_j \tau (\Gamma -1)}{4 \pi a^3} \left\lbrack \frac{\sqrt{2}(4 - 3\alpha)}{G\sqrt{\Lbar}(2-\alpha)^{3/2}(4-\alpha)} \right\rbrack^{4/(4-\alpha)} \nonumber \\
       &\times \tbar^{-(4+\alpha)/(4-\alpha)}\label{eq.pc3}   
\end{align}
\item $\rcbar \gg 1$ and $\tbar \gg 1$:
\begin{align}
\rcbar &= G^{2/(4-\alpha)} \left\lbrack \frac{2(1-n)}{(2-\alpha)\Lbar}\right\rbrack^{2/((2-\alpha)(4-\alpha))} \nonumber \\
       &  \times \left\lbrack \frac{(2-\alpha)(4-\alpha)}{(4+2n-3\alpha)}\right\rbrack^{2/(4-\alpha)} \tbar^{(4+2n-3\alpha)/((2-\alpha)(4-\alpha))} \label{eq.rcbar4}\\
p_c &= \frac{3 P_j \tau (\Gamma -1)}{4 \pi a^3} \left\lbrack \frac{(4+2n-3\alpha)\sqrt{2(1-n)}}{G\sqrt{\Lbar}(2-\alpha)^{3/2}(4-\alpha)} \right\rbrack^{4/(4-\alpha)} \nonumber \\
&\times \tbar^{-(4+\alpha-2n)/(4-\alpha)}  \label{eq.pc4}
\end{align}
\end{itemize}
The exponent of time in eq.~\ref{eq.lbar4} and eq.~\ref{eq.pc4} is identical to that derived earlier in \citet{perucho07a}. Note that for $n = \frac{(4+\alpha)}{2(5-\alpha)}$, the exponent of time for jet length in eq.~\ref{eq.lbar4} is $\lbar \propto t^{3/(5-\alpha)}$ and cocoon pressure in eq.~\ref{eq.pc4} is $p_c \propto t^{-(4+\alpha)(5-\alpha)}$. This is identical to the solutions for a self-similar evolution of the jet cocoon derived earlier in \citep{kaiser97a,falle91a}. 

\def\apj{ApJ}%
\def\mnras{MNRAS}%
\def\aap{A\&A}%
\def\apjl{ApJ}
\def\physrep{PhR}
\def\apjs{ApJS}
\def\pasa{PASA}
\def\pasj{PASJ}
\def\nat{Nature}
\def\memsai{MmSAI}
\def\aj{AJ}%
\def\aaps{A\&AS}%
\def\iaucirc{IAU~Circ.}%
\def\sovast{Soviet~Ast.}%
\def\apss{Ap\&SS}

\bibliographystyle{mnras}
\bibliography{dmrefs}


\end{document}